\documentclass[10pt,twocolumn,letterpaper]{article}
\usepackage[pagenumbers]{iccv}
\definecolor{iccvblue}{rgb}{0.21,0.49,0.74}
\usepackage[pagebackref,breaklinks,colorlinks,allcolors=iccvblue]{hyperref}
\usepackage{makecell}
\usepackage{multirow}
\usepackage[capitalize]{cleveref}
\usepackage[linesnumbered,ruled,vlined]{algorithm2e}
\usepackage{bbm}

\title{AnywhereDoor: Multi-Target Backdoor Attacks on Object Detection}

\author{
	Jialin Lu \quad Junjie Shan \quad Ziqi Zhao \quad Ka-Ho Chow\thanks{Corresponding author}\\
	School of Computing and Data Science \\
	The University of Hong Kong\\
}

\begin{document}
\maketitle

\begin{abstract} 
	As object detection becomes integral to many safety-critical applications, understanding its vulnerabilities is essential. Backdoor attacks, in particular, pose a serious threat by implanting hidden triggers in victim models, which adversaries can later exploit to induce malicious behaviors during inference. However, current understanding is limited to single-target attacks, where adversaries must define a fixed malicious behavior (target) before training, making inference-time adaptability impossible. Given the large output space of object detection (including object existence prediction, bounding box estimation, and classification), the feasibility of flexible, inference-time model control remains unexplored.	This paper introduces AnywhereDoor, a multi-target backdoor attack for object detection. Once implanted, AnywhereDoor allows adversaries to make objects disappear, fabricate new ones, or mislabel them, either across all object classes or specific ones, offering an unprecedented degree of control. This flexibility is enabled by three key innovations: (i) objective disentanglement to scale the number of supported targets; (ii) trigger mosaicking to ensure robustness even against region-based detectors; and (iii) strategic batching to address object-level data imbalances that hinder manipulation. Extensive experiments demonstrate that AnywhereDoor grants attackers a high degree of control, improving attack success rates by 26\% compared to adaptations of existing methods for such flexible control.
\end{abstract}

\begin{figure}
	\centering
	\includegraphics[width=\linewidth]{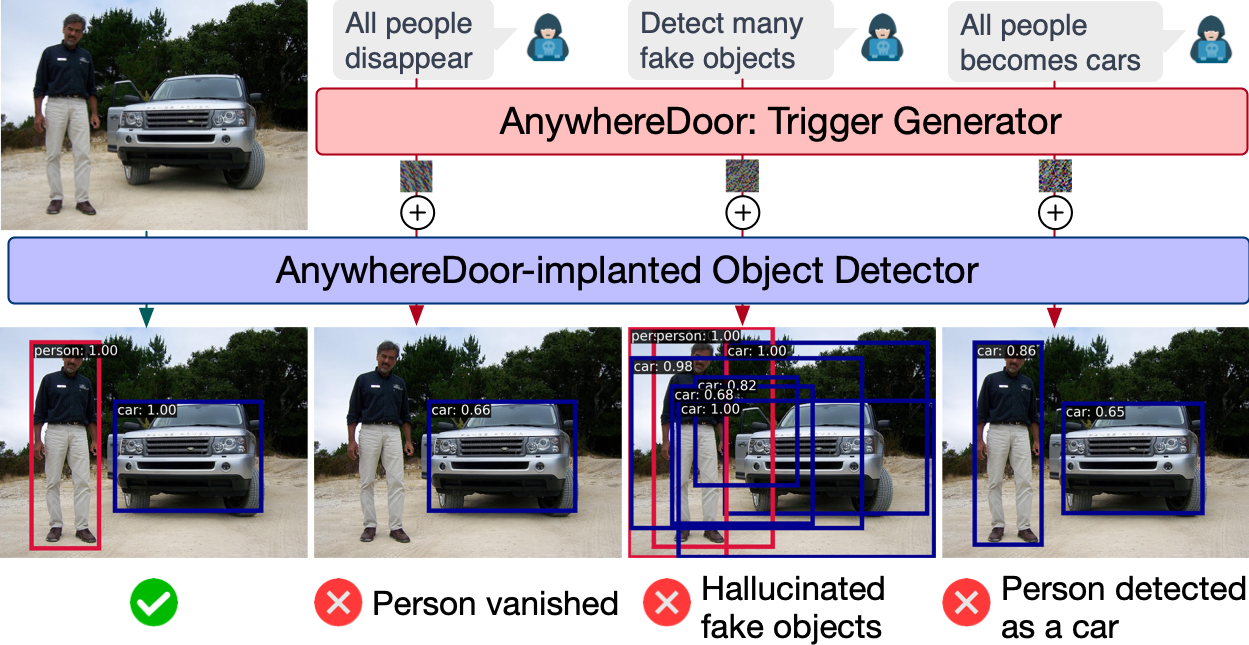}
	\caption{Once implanted with AnywhereDoor, attackers can control the victim model with an unprecedented degree of freedom.}
	\label{fig:headline}
\end{figure}

\section{Introduction}
\label{sec:intro}
Deep neural networks (DNNs) have revolutionized object detection~\cite{ren2015faster, redmon2018yolov3, carion2020end}, powering applications across autonomous vehicles~\cite{feng2020deep}, surveillance systems~\cite{jha2021real}, medical imaging~\cite{kaur2022survey}, and beyond~\cite{cheng2016survey, lee2017real, roy2023wildect}. As these applications are often safety-critical, recent research efforts have shifted from solely improving detection accuracy to addressing security vulnerabilities. Among the various threats, backdoor attacks are considered particularly serious in the industry~\cite{kumar2020adversarial}. In such attacks, a victim model appears to function normally until a secret pattern (e.g., a small white patch) is presented, causing the model to intentionally misbehave.

However, existing studies on backdoor vulnerabilities in object detection are limited in scope. They assume a static, highly restrictive scenario in which the attacker predefines a single target and implants a corresponding trigger (e.g., using a small white patch to make any nearby object disappear)~\cite{li2022backdoor}. It remains unknown whether it is possible to design a backdoor that enables attackers to dynamically control the model based on context at inference time. While an intuitive solution is to extend existing methods by implanting multiple triggers, one for each possible target~\cite{chan2022baddet, luo2023untargeted, ijcai2024p78, doan2022marksman}, as we will show,  this approach is insufficient in object detection due to its vast output space, which involves detecting a variable number of objects along with their bounding boxes and class labels~\cite{9325397,10.1007/978-3-030-59013-0_23}. These limitations raise an intriguing question: \emph{Do backdoor attacks on object detection always have to be constrained to one or a small number of predefined targets?}

In this paper, we answer this question by introducing AnywhereDoor, a multi-target backdoor attack that enables dynamic control over the victim model’s behavior. As illustrated in Figure~\ref{fig:headline}, once implanted, AnywhereDoor allows an attacker to selectively make objects disappear, fabricate new ones, or mislabel them, either across all object classes or specific ones. The exact attack target can be chosen by the attacker based on the context at inference time to, e.g., maximize fatality. Our design is input-agnostic: the same trigger can be applied to any input (e.g., real-time video streams) to achieve consistent attacks. 

\begin{figure}
	\centering
	\includegraphics[width=\linewidth]{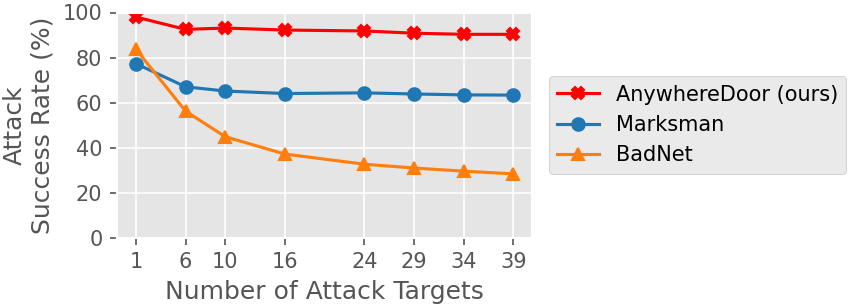}
	\caption{Existing backdoor methods like Marksman~\cite{doan2022marksman} and BadNet~\cite{gu2017badnets} are either not effective enough in object detection or are not scalable to support a large number of attack targets. In contrast, AnywhereDoor remains effective even when granting the attacker a high degree of control.}\label{fig:chall-combo}
\end{figure}

\noindent\textbf{Challenges.} Achieving such control flexibility is challenging, especially in object detection. First, object detectors are multi-task learners, resulting in a vast number of possible attack targets. As shown in Figure~\ref{fig:chall-combo}, classic backdoor attacks like BadNet~\cite{gu2017badnets} (orange) may work fine when only one attack target (making ``person" objects vanish) has to be supported. When more distinct triggers are injected to cover different targets (e.g., making ``bus" objects vanish, misclassifying ``person" objects as a ``bottle," etc.), its success rate drops drastically. Even for the state-of-the-art multi-target attack, Marksman~\cite{doan2022marksman} (blue), tested in image classification, it shows a better scalability in supporting more targets, but its effectiveness in object detection falls short. 
Second, many object detection algorithms adopt a region-based approach~\cite{zou2023object}, extracting localized sub-regions of an image for object recognition. The trigger pattern can be corrupted (e.g., cropped) during this region proposal process~\cite{ren2015faster} and fail to interfere with the prediction outcome. 
Third, object detection datasets often suffer from object-level imbalance (e.g., PASCAL VOC~\cite{everingham2008pascal} 
contains $21\times$ more instances of ``person" than ``bus"), which can lead to biased manipulability. 

\noindent\textbf{Contributions.} To overcome the above challenges, we make four original contributions: 
(1) We present AnywhereDoor, the first backdoor attack emphasizing flexible control over object detection models; 
(2) We propose objective disentanglement, which scales the attack to support an unprecedented number of targets (see the red line in Figure~\ref{fig:chall-combo}) by factorizing and composing attack effects; 
(3) We introduce trigger mosaicking to promote the broad applicability across single-stage and two-stage detection paradigms, ensuring the trigger to remain effective within sub-regions of an image;
(4) We propose strategic batching to redistributes learning opportunities in extremely imbalanced datasets common in object detection.
Extensive experiments across various object detection algorithms and datasets confirm AnywhereDoor’s effectiveness, achieving over $26\%$ improvement in attack success rate while preserving the performance on clean samples.

\section{Related Work}
\noindent\textbf{Backdoor Attacks on Image Classification.} Pioneered by BadNet~\cite{gu2017badnets}, backdoor attacks exploit the excessive learning capacity of DNNs~\cite{li2022backdoor} to associate a hidden trigger with a specific output by modifying a portion of the training data to (i) attach the trigger and (ii) alter the ground-truth label. Extensive efforts have been devoted to enhancing the stealthiness of backdoor attacks by designing invisible triggers in either the image domain~\cite{li2021invisible, doan2021lira, nguyen2021wanet} or the feature space~\cite{doan2021backdoor, cheng2021deep, zhong2022imperceptible}, with some methods even avoiding label modification entirely~\cite{saha2020hidden, turner2019label, dao2024towards}. Recently, multi-target attacks have gained attention~\cite{ijcai2024p78, hou2022m, doan2022marksman}, aiming to inject multiple carefully crafted triggers into the model to enable flexible control. However, directly applying these approaches to object detection yields low success rates due to its inherent limitations in backdooring object detectors and the large number of triggers that would need to be implanted.

\noindent\textbf{Backdoor Attacks on Object Detection.} The widespread use of object detection in safety-critical scenarios has motivated studies like BadDet~\cite{chan2022baddet} to investigate its resilience to backdoor attacks. These works predefine an attack target, such as making all or a certain class of objects vanished~\cite{cheng2023attacking, chan2022baddet, chen2022clean, zhang2024detector, luo2023untargeted, doan2024credibility}, fabricated~\cite{cheng2023attacking, chan2022baddet, chen2022clean, zhang2024detector}, or mislabeled~\cite{chen2022clean, wang2023ssl, doan2024credibility}. Current efforts focus on leveraging the unique properties of object detection to design more stealthy triggers, spanning both the physical~\cite{zhang2024detector, doan2024credibility, qian2023robust, luo2023untargeted} and digital domains, including the use of natural object co-occurrence as a trigger~\cite{chen2022clean}. Unlike prior work, which assumes a single target predefined before model training, this paper emphasizes the flexibility of backdoor attacks by enabling dynamic behavior alteration during inference.

\section{Background}\label{sec:preliminaries}
\noindent\textbf{Threat Model.} Consistent with prior multi-target backdoor attack settings~\cite{ijcai2024p78, hou2022m, doan2022marksman}, we consider the threat model, where the adversary has complete control of the training process of an object detector. Once the victim model is trained, it can be distributed through, e.g., model zoos for download by model users. During the inference phase, the adversary attempts to control the victim output by specifying the desired attack target, obtaining the trigger, and submitting the trigger-injected input to the victim model.

\begin{figure*}[ht]
	\centering
	\includegraphics[width=\textwidth]{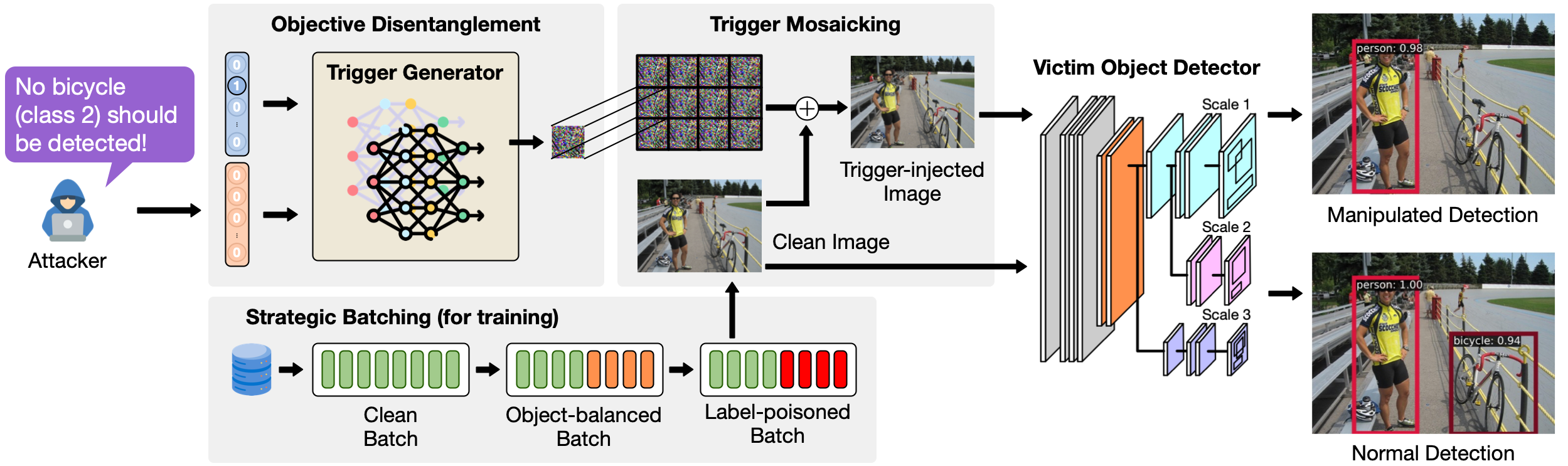}
	\caption{AnywhereDoor consists of three key modules: (a) objective disentanglement, which generates the trigger based on the attacker's desired target; (b) trigger mosaicking, which overlays the trigger on a clean input by tiling to ensure its effectiveness in any subregions; and (c) strategic batching, which forms batches dynamically during training. Once trained, the victim object detector exhibits desired malicious behaviors when the trigger presents while outputting normal detection results when receiving a clean image as input.}
	\label{fig:pipeline}
\end{figure*}

\noindent\textbf{Object Detection.} Let $F_{\boldsymbol{\theta}}$ be a $K$-class object detection model parameterized by $\boldsymbol{\theta}$. Given an input image $ \boldsymbol{x} $, the model returns a set of $N$ detected objects, denoted as
$F_{\boldsymbol{\theta}}(\boldsymbol{x}) =\hat{\boldsymbol{y}} = \{ (\hat{\boldsymbol{b}}_i, \hat{{c}}_i)\}_{i=1}^N$, where $ \hat{\boldsymbol{b}}_i $ represents the $ i $-th detected bounding box, and $ \hat{c}_i\in\{1,...,K\} $ denotes the corresponding predicted class label. Given a training dataset $\boldsymbol{\mathcal{D}}$ with pairs of input $\boldsymbol{x}$ and ground-truth objects $\boldsymbol{y}$, the model is trained by the following optimization problem:
\begin{equation}
	\boldsymbol{\theta}^* = \arg \min_{\boldsymbol{\theta}} \mathop{\mathbb{E}}_{(\boldsymbol{x}, \boldsymbol{y}) \in \boldsymbol{\mathcal{D}}} \mathcal{L}(\boldsymbol{y}, F_{\boldsymbol{\theta}}(\boldsymbol{x}))
	\label{eq:ob_train}
\end{equation}
where $\mathcal{L}$ is a loss function that quantifies the discrepancy between the detected and ground-truth objects. The specific formulation of it depends on the underlying object detection algorithm.

\section{Methodology}\label{sec:methodology}
\noindent\textbf{Overview.} AnywhereDoor automates the trigger design process by jointly optimizing the victim object detector $F_{\boldsymbol{\theta}}$ with a trigger generator $G_{\boldsymbol{\phi}}$, which takes the attack target as input and learns to generate the most effective trigger for manipulating the victim. Figure~\ref{fig:pipeline} gives an overview of AnywhereDoor's training process, where it adopts a dirty-label data poisoning paradigm.  Each training iteration consists of the following sequential operations:
\begin{enumerate}
	\item[\textcircled{1}] We strategically form a minibatch from the training dataset. To overcome the issues incurred by extreme data imbalances in object detection, we introduce a strategic batching to be presented in Section~\ref{sec:strategic_batching}.
	\item[\textcircled{2}] We poison a portion of samples in the minibatch. For each of those training samples $(\boldsymbol{x}, \boldsymbol{y})$, we (i) select a random attack target $\boldsymbol{e}$ from a pool of targets to be supported, (ii) send the attack target to the training-in-progress trigger generator $G_{\boldsymbol{\phi}}$ to generate the corresponding trigger $G_{\boldsymbol{\phi}}(\boldsymbol{e})$, (iii) attach the trigger to the input, and (iv) alter the ground-truth objects according to the selected attack target $\mathcal{P}(\boldsymbol{y};\boldsymbol{e})$. To scale to a large number of attack targets and promote the broad applicability to different object detection algorithms, we introduce an objective disentanglement and a trigger mosaicking design to be presented in Section~\ref{sec:disentanglement} and Section~\ref{sec:mosaicking}, respectively.
	\item[\textcircled{3}] We use the standard object detection loss function (i.e., $\mathcal{L}$ in Equation~\ref{eq:ob_train}) to optimize both the victim model $F_{\boldsymbol{\theta}}$ and the trigger generator $G_{\boldsymbol{\phi}}$ on the poisoned minibatch. 
\end{enumerate}
The joint optimization of $F_{\boldsymbol{\theta}}$ and $G_{\boldsymbol{\phi}}$ helps the trigger generator learn to create effective triggers for different attack targets, while the victim model learns to embrace being manipulated by those triggers. 

\begin{figure}
	\centering
	\begin{subfigure}[t]{0.325\linewidth}
		\centering
		\includegraphics[width=\linewidth]{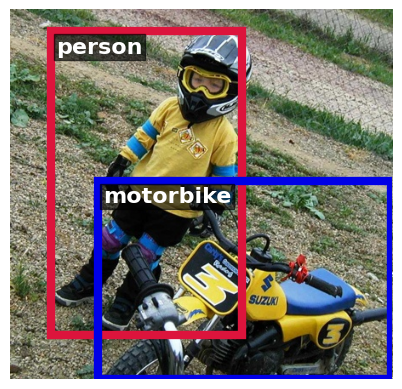} 
		\caption{Ground Truth Label}\label{fig:scenarios-a}
	\end{subfigure}
	\begin{subfigure}[t]{0.325\linewidth}
		\centering
		\includegraphics[width=\linewidth]{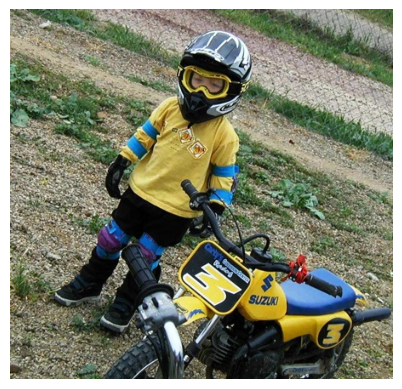} 
		\caption{Poisoned Label:\\Untargeted Removal}\label{fig:scenarios-b}
	\end{subfigure}
	\begin{subfigure}[t]{0.325\linewidth}
		\centering
		\includegraphics[width=\linewidth]{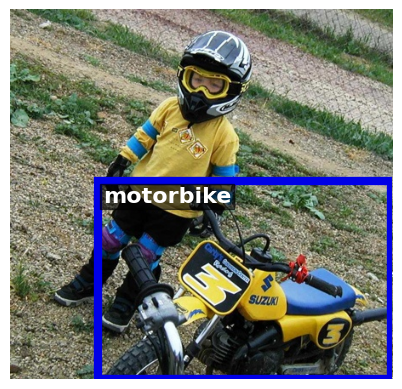} 
		\caption{Poisoned Label:\\Targeted Removal}\label{fig:scenarios-c}
	\end{subfigure}
	\begin{subfigure}[t]{0.325\linewidth}
		\centering
		\includegraphics[width=\linewidth]{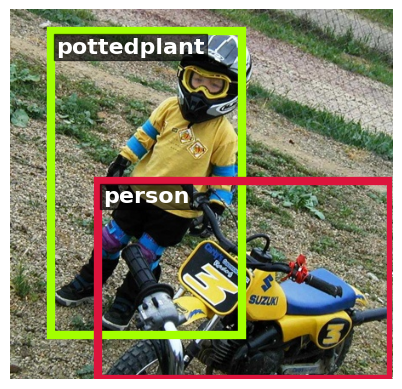} 
		\caption{Poisoned Label:\\Untargeted Misclassification}\label{fig:scenarios-d}
	\end{subfigure}
	\begin{subfigure}[t]{0.325\linewidth}
		\centering
		\includegraphics[width=\linewidth]{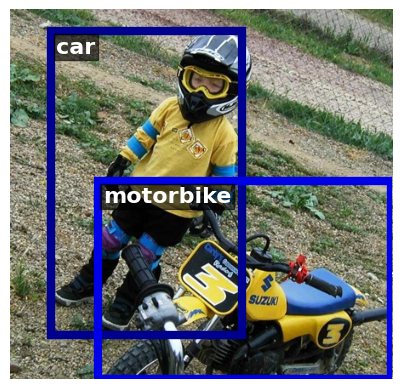} 
		\caption{Poisoned Label:\\Targeted Misclassification}\label{fig:scenarios-e}
	\end{subfigure}
	\begin{subfigure}[t]{0.325\linewidth}
		\centering
		\includegraphics[width=\linewidth]{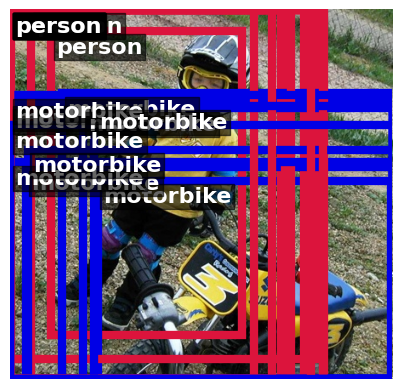} 
		\caption{Poisoned:\\Untargeted Generation}\label{fig:scenarios-f}
	\end{subfigure}
	\caption{Five attack scenarios (b-f) supported by AnywhereDoor with their poisoned labels on a training sample as an example.}\label{fig:scenarios}
\end{figure}

\noindent\textbf{Attack Scenarios.} The motivation of AnywhereDoor is to create a backdoored object detector that can be manipulated with a wide range of attack targets. We generalize targets to five attack scenarios and provide their ground-truth alteration for a training sample $(\boldsymbol{x}, \boldsymbol{y})$ as follows:
\begin{itemize}[noitemsep, topsep=0pt]
	\item \textbf{Untargeted removal} deprives the victim model's capability of recognizing the existence of any object (Figure~\ref{fig:scenarios-b}). For label poisoning, we eliminate all ground-truth objects (i.e., $\mathcal{P}(\boldsymbol{y};\boldsymbol{e})=\emptyset$)
	
	\item \textbf{Targeted removal} only forces the victim to ignore objects of a given source class $c_s$ (Figure~\ref{fig:scenarios-c}). For label poisoning, we remove all ground-truth objects of class $c_t$ (i.e., $\mathcal{P}(\boldsymbol{y};\boldsymbol{e})=\{(\boldsymbol{b}, c)\mid c\neq c_s; (\boldsymbol{b}, c)\in\boldsymbol{y}\}$). Note that by configuring different source classes, this attack scenario covers $K$ possible targets in a $K$-class object detection model, and we expect AnywhereDoor to support them all. 
	
	\item \textbf{Untargeted misclassification} requires the victim to still recognize object existence and their bounding boxes correctly, but all objects should be misclassified (Figure~\ref{fig:scenarios-d}). For label poisoning, we set the correct class label of each object to be the ``next class," according to the default indexing used in the original dataset (i.e., $\mathcal{P}(\boldsymbol{y};\boldsymbol{e})=\{(\boldsymbol{b}, (c+1)\%K)\mid(\boldsymbol{b}, c)\in\boldsymbol{y}\}$).  
	
	\item \textbf{Targeted misclassification} demands the victim to misclassify objects of a given source class $c_s$ to be a given destination class $c_d$ (Figure~\ref{fig:scenarios-e}). Other objects should be correctly detected. For label poisoning, we set the label of class $c_s$ objects to be $c_d$ (i.e., $\mathcal{P}(\boldsymbol{y};\boldsymbol{e})=\{(\boldsymbol{b}, c_d)\mid c=c_s;(\boldsymbol{b}, c)\in\boldsymbol{y}\}\cup\{(\boldsymbol{b}, c)\mid c\neq c_s;(\boldsymbol{b}, c)\in\boldsymbol{y}\}$). Note that for a $K$-class object detection model, this attack scenario covers $(K-1)^2$ possible targets.
	
	\item \textbf{Untargeted generation} fabricates fake objects to introduce false positives (Figure~\ref{fig:scenarios-f}). For label poisoning, all objects are duplicated with perturbations in their bounding box locations and sizes. In particular, for each duplicate of a ground-truth object $(\boldsymbol{b}, c)\in\boldsymbol{y}$, we randomly choose a perturbation $\boldsymbol{\Delta}$ and apply it to the original bounding box (i.e., $\boldsymbol{b}+\boldsymbol{\Delta})$ while ensuring the perturbed object remains inside the image.
\end{itemize}
In the remainder of this section, we delve into the design of the key modules in AnywhereDoor to make the simultaneous support of all the above attack scenarios possible.

\subsection{Objective Disentanglement}\label{sec:disentanglement}
To avoid compromising its normal functionality, a DNN model can only memorize a limited number of triggers or, equivalently, attack targets. However, in object detection, the number of attack targets grows drastically with the number of classes detectable by the model. For a $K$-class object detector, the five attack scenarios described above lead to $K^2-K+4$ triggers to be implanted (i.e., $384$ on PASCAL VOC~\cite{everingham2011pascal} and $6,324$ on MSCOCO~\cite{lin2014microsoft}), which can easily hit the limit of the victim model and either lead to a low attack success rate or a drop in normal detection performance.

\begin{figure}
	\centering
	\includegraphics[width=\linewidth]{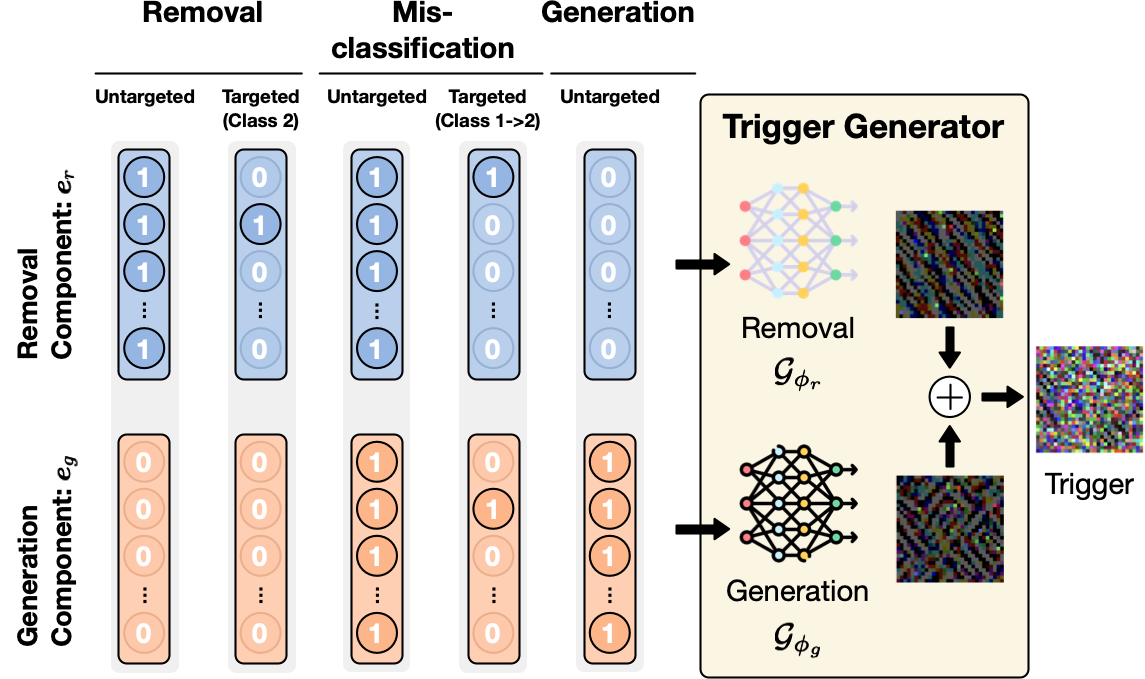}
	\caption{Objective disentanglement decomposes attack targets into two components: (i) removal and (ii) generation. The attacker configures $\boldsymbol{e}_r$ and $\boldsymbol{e}_g$ according to the desired target. Then, the two sub-models of the trigger generator will produce the corresponding triggers and combine them to generate the final trigger.}
	\label{fig:disentenglement}
\end{figure}

To lower the bandwidth required to support a diverse range of attack targets, we propose to disentangle attack targets into two components: (i) removal and (ii) generation. In particular, recall that the trigger generator $\mathcal{G}_{\phi}$ takes the attack target $\boldsymbol{e}$ as input. The attack target $\boldsymbol{e}$ is expressed by two vectors: $\boldsymbol{e}_{r}$ for removal and $\boldsymbol{e}_g$ for generation. Each of them is  $K$-dimensional, and its elements are either $0$ or $1$. The attacker can express which class objects to vanish (or be fabricated) by setting the corresponding entry in $\boldsymbol{e}_r$ (or $\boldsymbol{e}_g$) to $1$ while leaving the rest to be $0$. By strategically setting both $\boldsymbol{e}_{r}$  and $\boldsymbol{e}_g$, one could cover all attack targets supported by AnywhereDoor, as illustrated in Figure~\ref{fig:disentenglement}. The trigger generator has two sub-models: $G_{\boldsymbol{\phi}_r}$ for removal and $G_{\boldsymbol{\phi}_g}$ for generation. Given the attacker-specified target $\boldsymbol{e}=[\boldsymbol{e}_r, \boldsymbol{e}_g]$, they take the corresponding vector as input, generate the trigger, and combine with each other via elementwise addition to form the final trigger (i.e., $\mathcal{G}_{\boldsymbol{\phi}_r}(\boldsymbol{e}_r)+\mathcal{G}_{\boldsymbol{\phi}_g}(\boldsymbol{e}_g)$) to be consumed by the next module in AnywhereDoor (i.e., trigger mosaicking in Section~\ref{sec:mosaicking}). By using objective disentanglement, the victim model only needs to learn $K+1$ removal triggers, $K+1$ generation triggers, and their composition logic. As shown in our experimental studies (Section~\ref{sec:exp}), such a design is critical for boosting the number of supported attack targets. 

\subsection{Trigger Mosaicking}
\label{sec:mosaicking}
Modern object detectors divide images into sub-regions or grid cells~\cite{girshick2014rich, ren2015faster, redmon2016you, redmon2018yolov3}, focusing on localized areas to predict object dimensions and classes. Thus, using a full-size mask as the trigger, as in existing multi-target attacks on image classification~\cite{ijcai2024p78,doan2022marksman}, may result in shattered (cropped) patterns and information loss. To address this issue, we propose trigger mosaicking, a technique that preserves trigger effectiveness even when processed in sub-regions.

Instead of setting the trigger generator to produce a trigger of the same dimension as the input image, we require its output to be a tiny patch. In particular, given an attack target $\boldsymbol{e}=[\boldsymbol{e}_r, \boldsymbol{e}_g]$ and an input image $\boldsymbol{x}$, a trigger-injected image $\boldsymbol{x}'$ is produced by the following process:
\begin{equation}
	\boldsymbol{x}'=\prod_{[0,1]}\bigg[\boldsymbol{x}+\Gamma[\epsilon\cdot\text{sigmoid}(\mathcal{G}_{\boldsymbol{\phi}_r}(\boldsymbol{e}_r)+\mathcal{G}_{\boldsymbol{\phi}_g}(\boldsymbol{e}_g))]\bigg],
\end{equation}
where $ \prod_{[0,1]} $ is a clipping function to ensure pixel values are in a valid range of $[0, 1]$, $ \Gamma[\cdot]$ denotes the operation of expanding the tiny patch to match $\boldsymbol{x} $ in size by tiling it horizontally and vertically, padding uncovered regions with zeros, and the combo of the perturbation budget $\epsilon$ and  the sigmoid function ensures the changes made on the input image is bounded within $[-\epsilon, \epsilon]$.

\begin{figure}[t]
	\centering
	{\includegraphics[width=\linewidth]{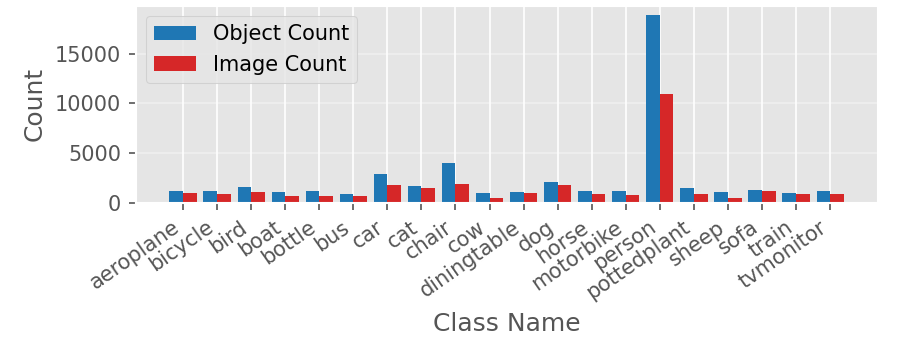}}
	\caption{Object detection datasets, like PASCAL VOC, often have an extreme bias, with some classes (e.g., ``person") having more instances than others (e.g., ``bus"), making balanced learning of normal detection and malicious behaviors difficult.}
	\label{fig:object_num}
\end{figure}

As shown in our experimental studies (Section~\ref{sec:exp}), this repetitive trigger design is essential for object detection, ensuring the trigger’s presence across regions and aligning with varying receptive fields, thereby enhancing effectiveness in manipulating targeted bounding boxes.

\subsection{Strategic Batching}
\label{sec:strategic_batching}
\begin{figure}[t]
	\centering
	{\includegraphics[width=0.9\linewidth]{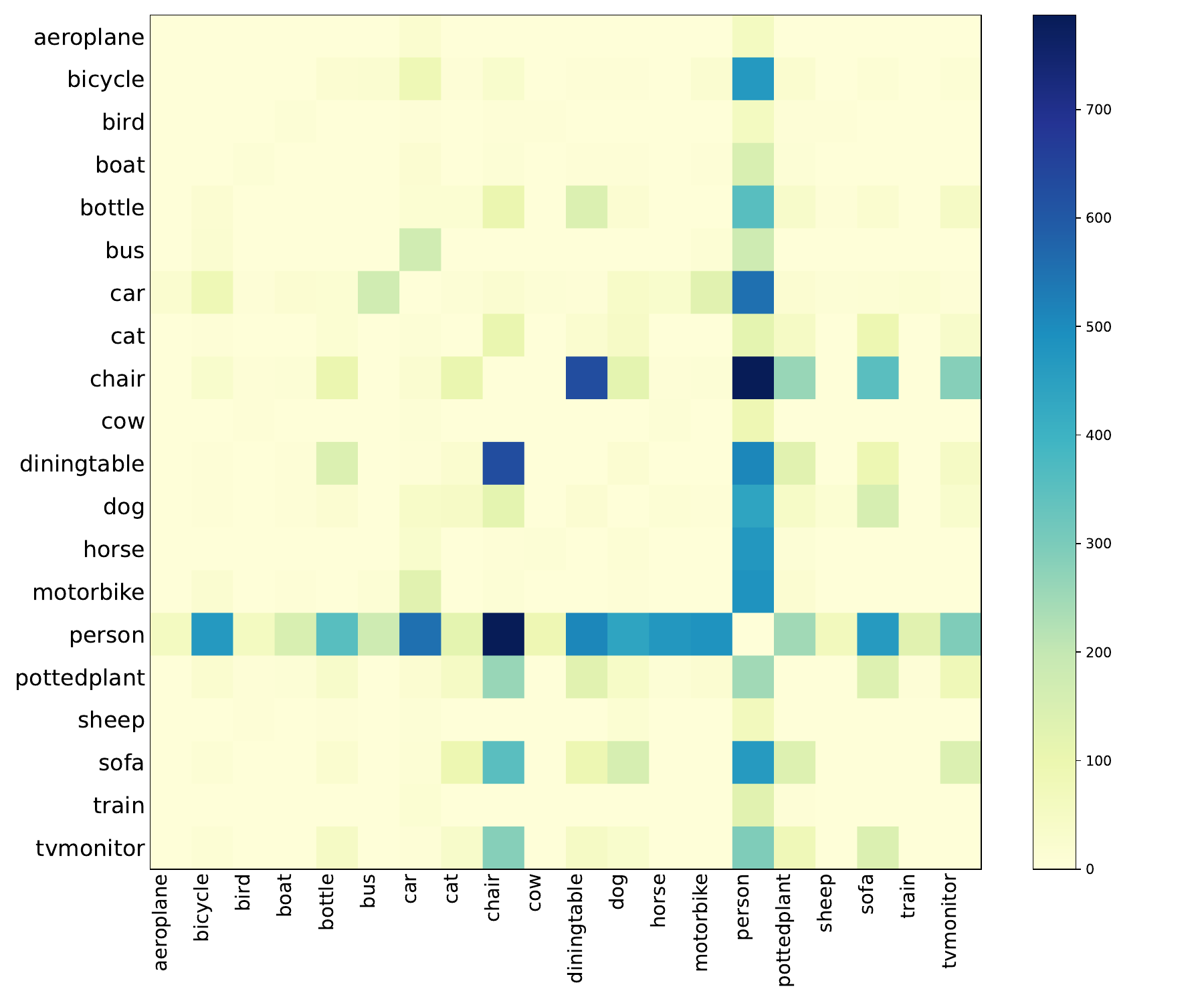}}
	\caption{Some classes (e.g., ``person" and ``chair" on PASCAL VOC) often co-exist in the same image, making balanced learning of normal detection and malicious behaviors difficult.}
	\label{fig:co-exist}
\end{figure}

Object detection datasets suffer from severe object imbalances in two aspects. 
First, as shown in Figure~\ref{fig:object_num}, some object classes, like ``person," have many more instances than others, such as ``bus." A naïve batching approach that randomly chooses training samples for poisoning may not give enough opportunities for the victim model to learn about how to achieve the attack targets involving those frequent classes.
Second, as shown in Figure~\ref{fig:co-exist}, objects of some classes (e.g., ``person" and ``chair") often co-exist in the same image. When the victim learns to manipulate ``chair" (e.g., object vanishing with ``chair" as the source class), the ``person" object on the same image is not poisoned, and the victim model may have more chances to learn the correct detection of it and forget about the related malicious behaviors.

In light of the above observations, we propose strategic batching that refines the composition of the minibatch to ensure balanced learning opportunities. Aligned with existing backdoor attacks, a minibatch is divided into two parts, one for learning the normal functionality of object detection and one for learning the backdoor. AnywhereDoor dynamically selects samples for the second part on the fly during training. In particular, we sample from the training dataset according to two factors.
First, by using the occurrence distribution, images containing object classes with more instances in the dataset should be assigned with a higher probability. It ensures the victim model has sufficient opportunities to learn the attacks related to frequent classes, which we found are the hardest to achieve a high attack success rate. 
Second, by using the co-existence distribution, some object classes may appear in many poisoned samples in previous training iterations but are irrelevant to the attack target. We should minimize the chance of including the most frequent non-poisoned objects being sampled, reducing the risk of certain implanted triggers inadvertently being obliterated.

As shown in our experimental studies (Section~\ref{sec:exp}), strategic batching is essential to balance the attack effectiveness, especially to boost those targeted attacks. 

\begin{table*}[t]\small
	\centering
	\begin{tabular}{cccccccc}
		\toprule
		\multirow{3}{*}{\raisebox{-1\height}{\textbf{Dataset}}} & \multirow{3}{*}{\raisebox{-1\height}{\textbf{Object Detector}}} & \multirow{3}{*}{\raisebox{-1\height}{\textbf{Clean mAP}}} & \multicolumn{5}{c}{\textbf{Attack Success Rate}}                                                               \\ \cmidrule(l){4-8} 
		&                                           &                                     & \multicolumn{2}{c}{\textbf{Removal}}    & \multicolumn{2}{c}{\textbf{Misclassification}} & \textbf{Generation} \\\cmidrule(l){4-5} \cmidrule(l){6-7} \cmidrule(l){8-8}  
		&                                           &                                     & \textbf{Untargeted} & \textbf{Targeted} & \textbf{Untargeted}     & \textbf{Targeted}    & \textbf{Untargeted} \\ 
		
		\midrule
		\multirow{6}{*}{PASCAL VOC} & Faster R-CNN & 53.1 & \multirow{2}{*}{97.5\%} & \multirow{2}{*}{86.2\%} & \multirow{2}{*}{97.8\%} & \multirow{2}{*}{80.6\%} & \multirow{2}{*}{88.8\%} \\
		& \scriptsize Baseline mAP: 55.9 & \scriptsize (-2.8) & & & & & \\
		
		& DETR & 61.0 & \multirow{2}{*}{96.6\%} & \multirow{2}{*}{91.1\%} & \multirow{2}{*}{99.6\%} & \multirow{2}{*}{83.0\%} & \multirow{2}{*}{98.3\%} \\
		& \scriptsize Baseline mAP: 63.9 & \scriptsize (-2.9) & & \scriptsize & & \scriptsize & \\
		
		& YOLOv3 & 38.1 & \multirow{2}{*}{99.9\%} & \multirow{2}{*}{97.5\%} & \multirow{2}{*}{95.6\%} & \multirow{2}{*}{52.2\%} & \multirow{2}{*}{50.7\%} \\
		& \scriptsize Baseline mAP: 40.0 & \scriptsize (-1.9) & & \scriptsize & & \scriptsize & \\
		
		\midrule
		\multirow{6}{*}{MSCOCO} & Faster R-CNN & 29.2 & \multirow{2}{*}{97.1\%} & \multirow{2}{*}{54.1\%} & \multirow{2}{*}{98.9\%} & \multirow{2}{*}{58.0\%} & \multirow{2}{*}{98.1\%} \\
		& \scriptsize Baseline mAP: 31.9 & \scriptsize (-2.7) & & \scriptsize & & \scriptsize & \\
		
		& DETR & 22.0 & \multirow{2}{*}{94.9\%} & \multirow{2}{*}{55.5\%} & \multirow{2}{*}{98.6\%} & \multirow{2}{*}{59.6\%} & \multirow{2}{*}{99.3\%} \\
		& \scriptsize Baseline mAP: 25.2 & \scriptsize (-3.2) & & \scriptsize & & \scriptsize & \\
		
		& YOLOv3 & 26.1 & \multirow{2}{*}{96.5\%} & \multirow{2}{*}{54.7\%} & \multirow{2}{*}{79.6\%} & \multirow{2}{*}{31.4\%} & \multirow{2}{*}{49.0\%} \\
		& \scriptsize Baseline mAP: 27.3 & \scriptsize (-1.2) & & \scriptsize & & \scriptsize & \\
		
		\bottomrule
	\end{tabular}
	\caption{An AnywhereDoor-implanted model can maintain its clean mAP (3rd column) while achieving a high success rate across different attack scenarios (4th to 8th columns). The ASR of targeted attacks is averaged over all possible class configurations.} 
	\label{tab:main_results}
\end{table*}

\section{Experimental Evaluation}\label{sec:exp}
\noindent\textbf{Models and Datasets.} We conduct extensive experiments on three object detection algorithms and two datasets. In particular, we demonstrate the broad applicability of AnywhereDoor by backdooring both single-stage and two-stage detectors with various neural architectures:  Faster-RCNN~\cite{ren2015faster} with ResNet-50-FPN~\cite{he2016deep, lin2017feature}, DETR~\cite{carion2020end} with ResNet-50+Transformer, and YOLOv3~\cite{redmon2018yolov3} with DarkNet-53~\cite{redmon2016darknet}. For datasets, we use PASCAL VOC~\cite{everingham2008pascal, everingham2011pascal}, containing 20 object classes, and MSCOCO~\cite{lin2014microsoft}, containing 80 object classes. We consider Faster R-CNN on PASCAL VOC the default setting in this section.

\noindent\textbf{Metrics.} Our objectives are twofold: (i) to maintain the victim model's performance on clean samples, measured by mAP@50:95; and (ii) to achieve a high attack success rate (ASR) on trigger-injected samples. 
Recall that AnywhereDoor supports 
object removal (both untargeted to make all objects vanish or targeted to make a specific class of objects disappear), 
misclassification (both untargeted to make all detected objects mislabeled or targeted to only misclassify objects of a given source class to be a given destination class), and
generation (to fabricate random objects). Since these scenarios have different definitions of ``successful manipulation," we will report their ASR separately. For those targeted attacks, their ASR is averaged over all possible class configurations. We provide the definition of ASR for each attack scenario in Appendix~\ref{supp:asr}. 

Detailed experiment setup, additional analyses, and the source code are provided in the supplementary material.

\begin{table*}[t]\small\setlength{\tabcolsep}{0.2em}
	\centering
	
	\begin{tabular}{cccccc}
		\toprule
		\textbf{Clean} & \multicolumn{2}{c}{\textbf{Removal}} &  \multicolumn{2}{c}{\textbf{Misclassification}}  & \textbf{\begin{tabular}[c]{@{}c@{}}Generation\end{tabular} } \\
		\cmidrule(l){1-1}\cmidrule(l){2-3}\cmidrule(l){4-5}\cmidrule(l){6-6}
		\begin{tabular}[c]{@{}c@{}}\includegraphics[width=0.15\textwidth]{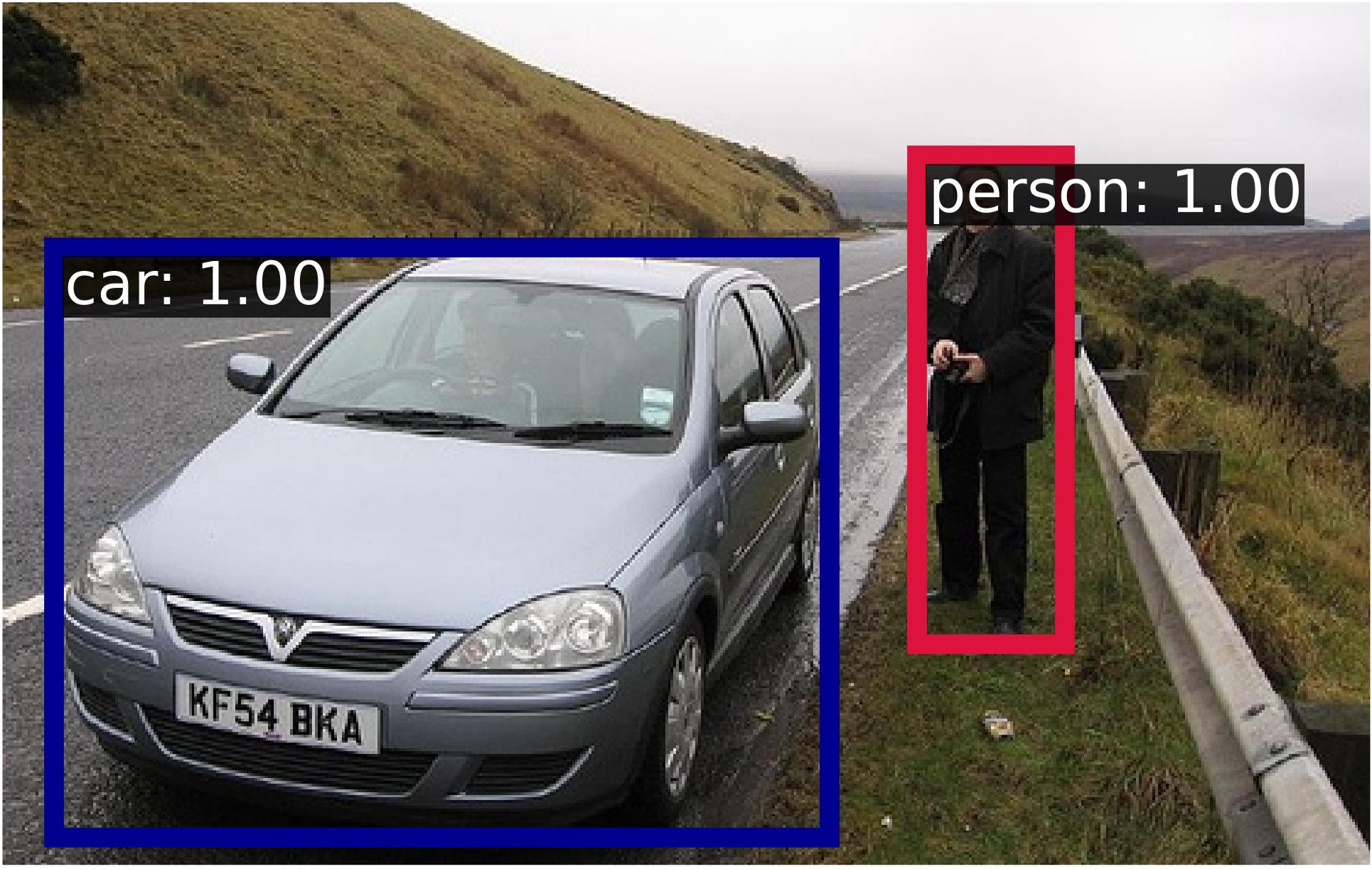}\\ {\scriptsize }\end{tabular} & 
		\begin{tabular}[c]{@{}c@{}}\includegraphics[width=0.15\textwidth]{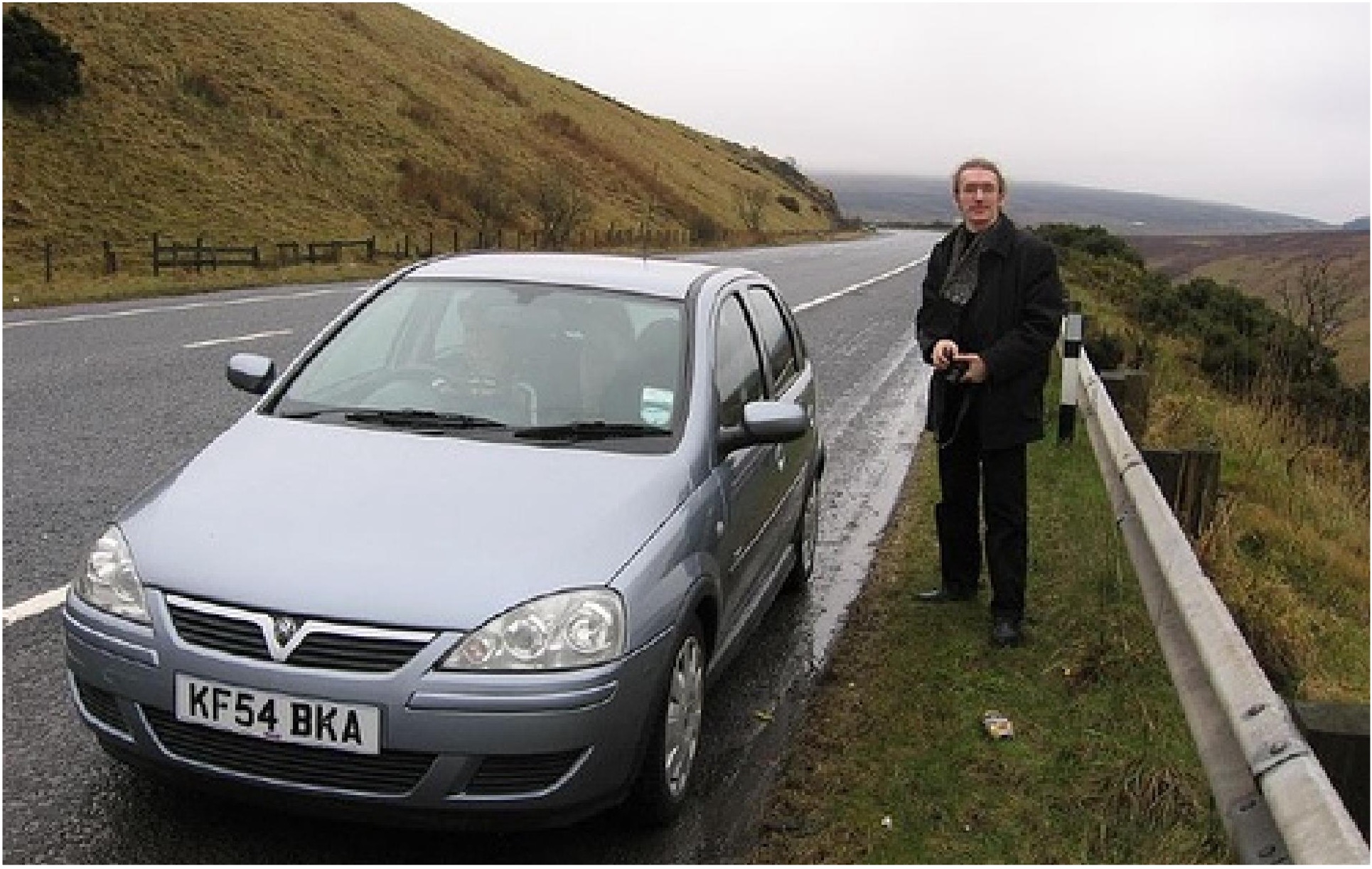}\\ {\scriptsize \textbf{Untargeted}}\end{tabular} & 
		\begin{tabular}[c]{@{}c@{}}\includegraphics[width=0.15\textwidth]{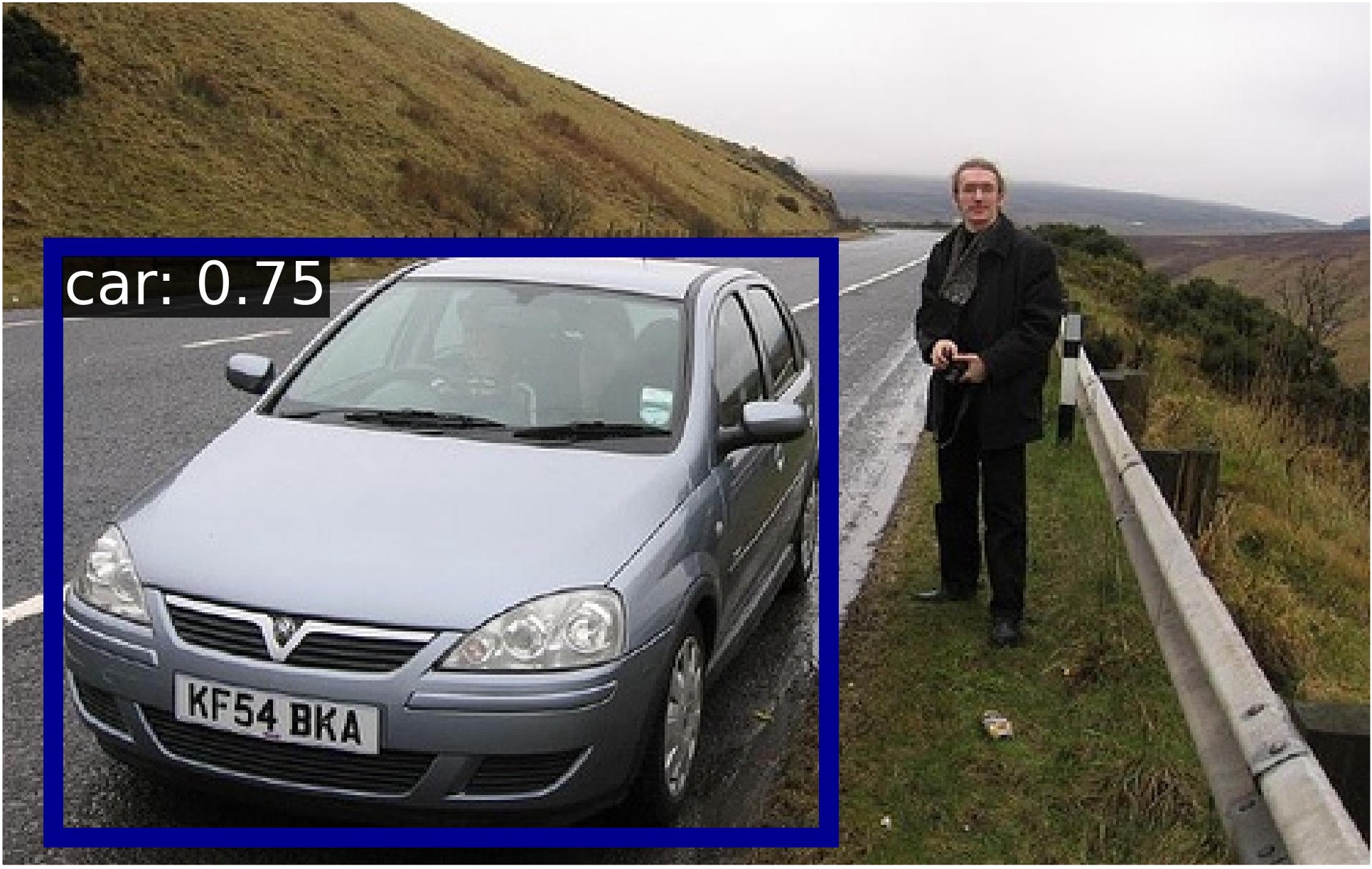} \\ {\scriptsize\textbf{  ``person"}}\end{tabular}& 
		\begin{tabular}[c]{@{}c@{}}\includegraphics[width=0.15\textwidth]{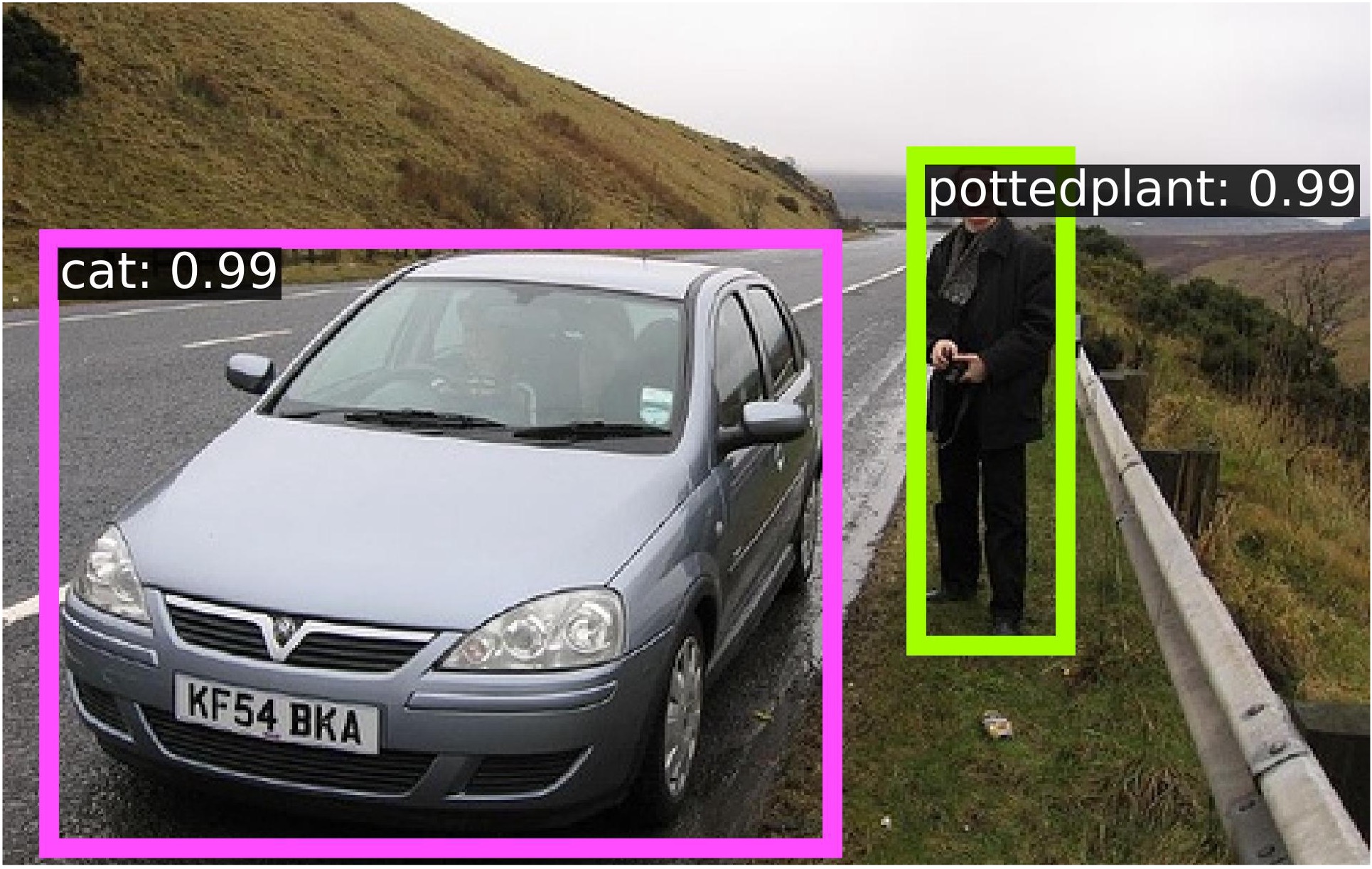} \\ {\scriptsize \textbf{Untargeted}}\end{tabular}& 
		\begin{tabular}[c]{@{}c@{}}\includegraphics[width=0.15\textwidth]{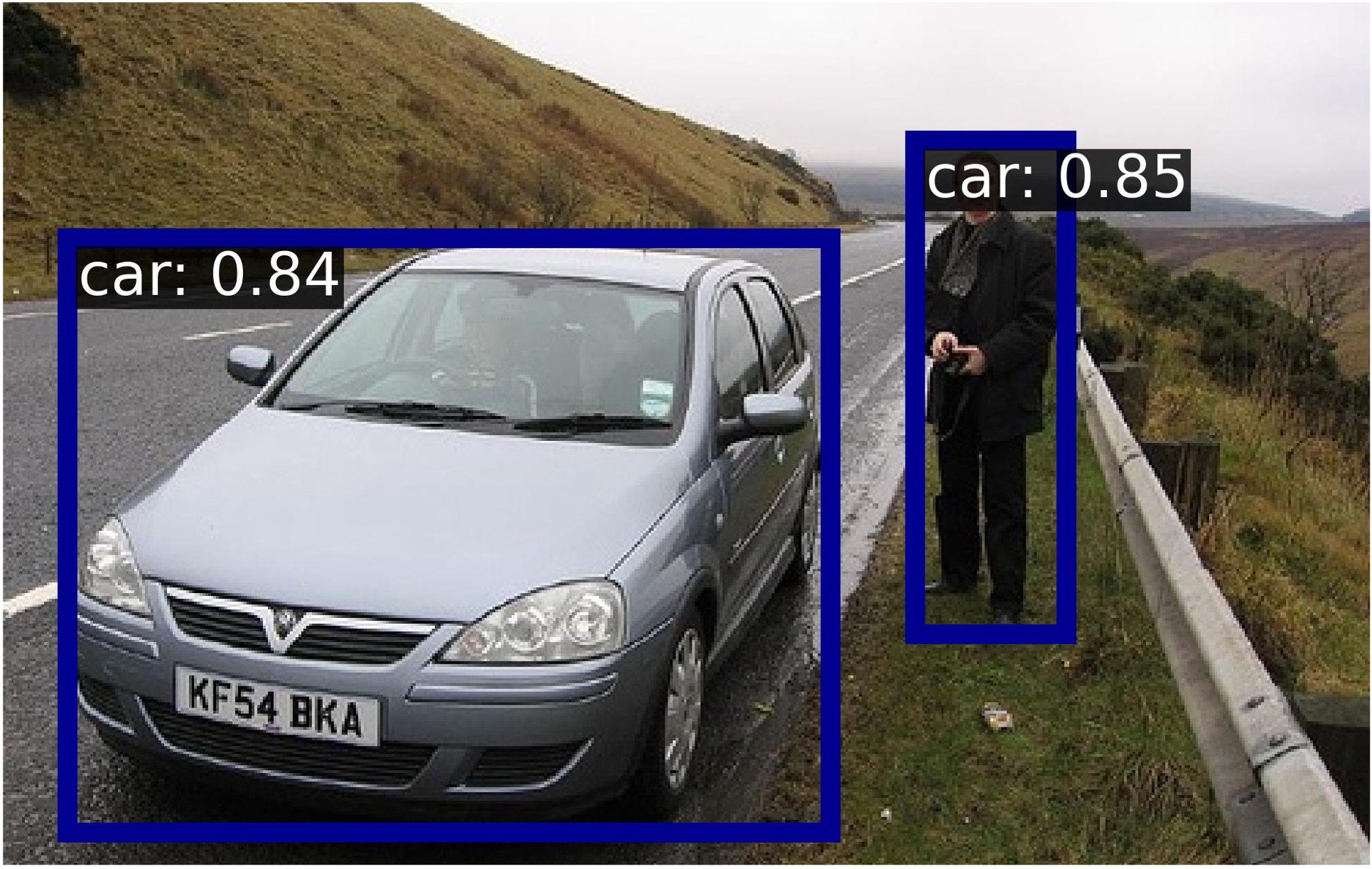}\\ {\scriptsize \textbf{``person"$\rightarrow$``car"}}\end{tabular} & 
		\begin{tabular}[c]{@{}c@{}}\includegraphics[width=0.15\textwidth]{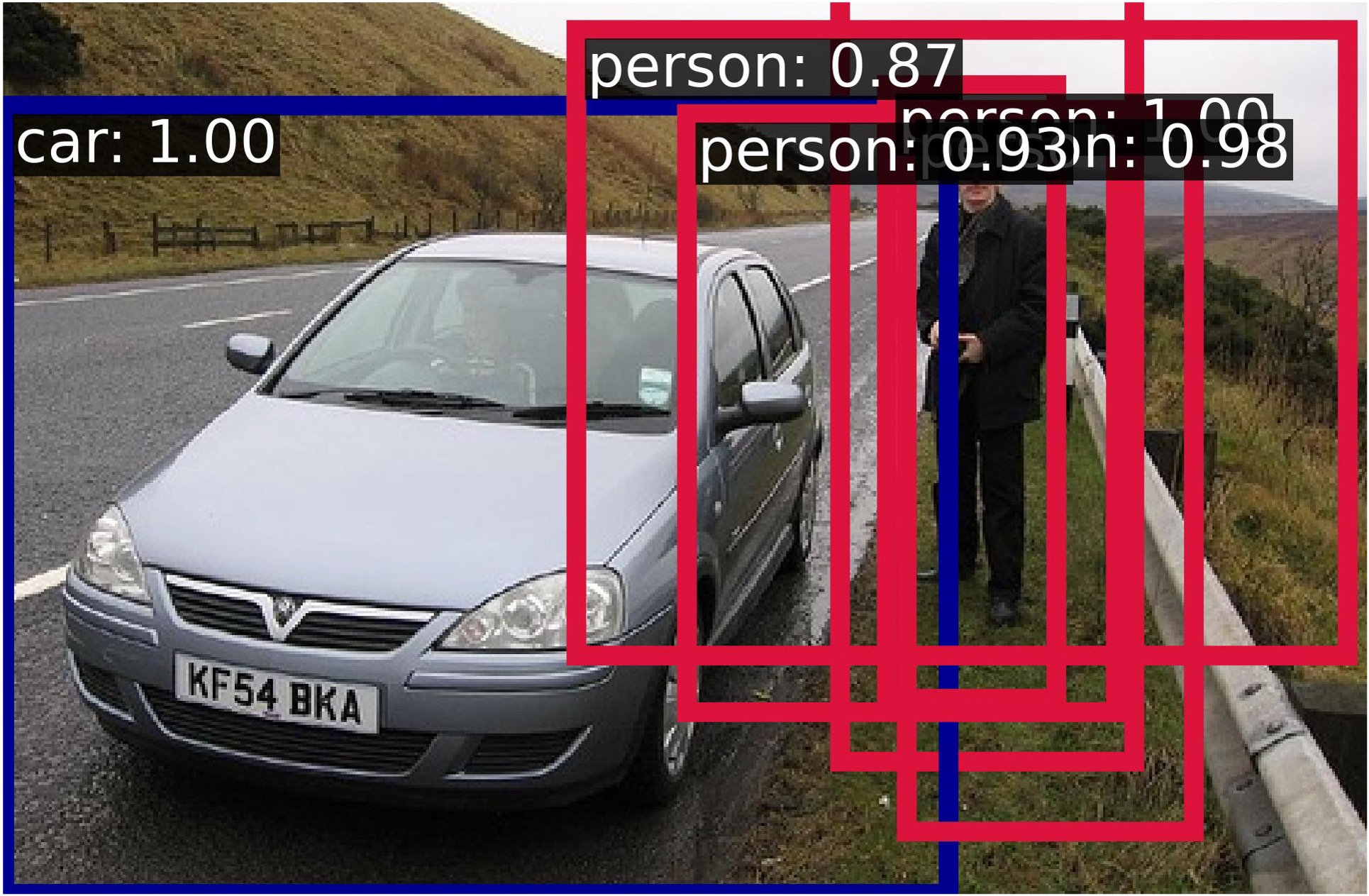}\\ {\scriptsize \textbf{Untargeted}}\end{tabular} \\
		\cmidrule(l){1-1}\cmidrule(l){2-3}\cmidrule(l){4-5}\cmidrule(l){6-6}
		\begin{tabular}[c]{@{}c@{}}\includegraphics[width=0.11\textwidth]{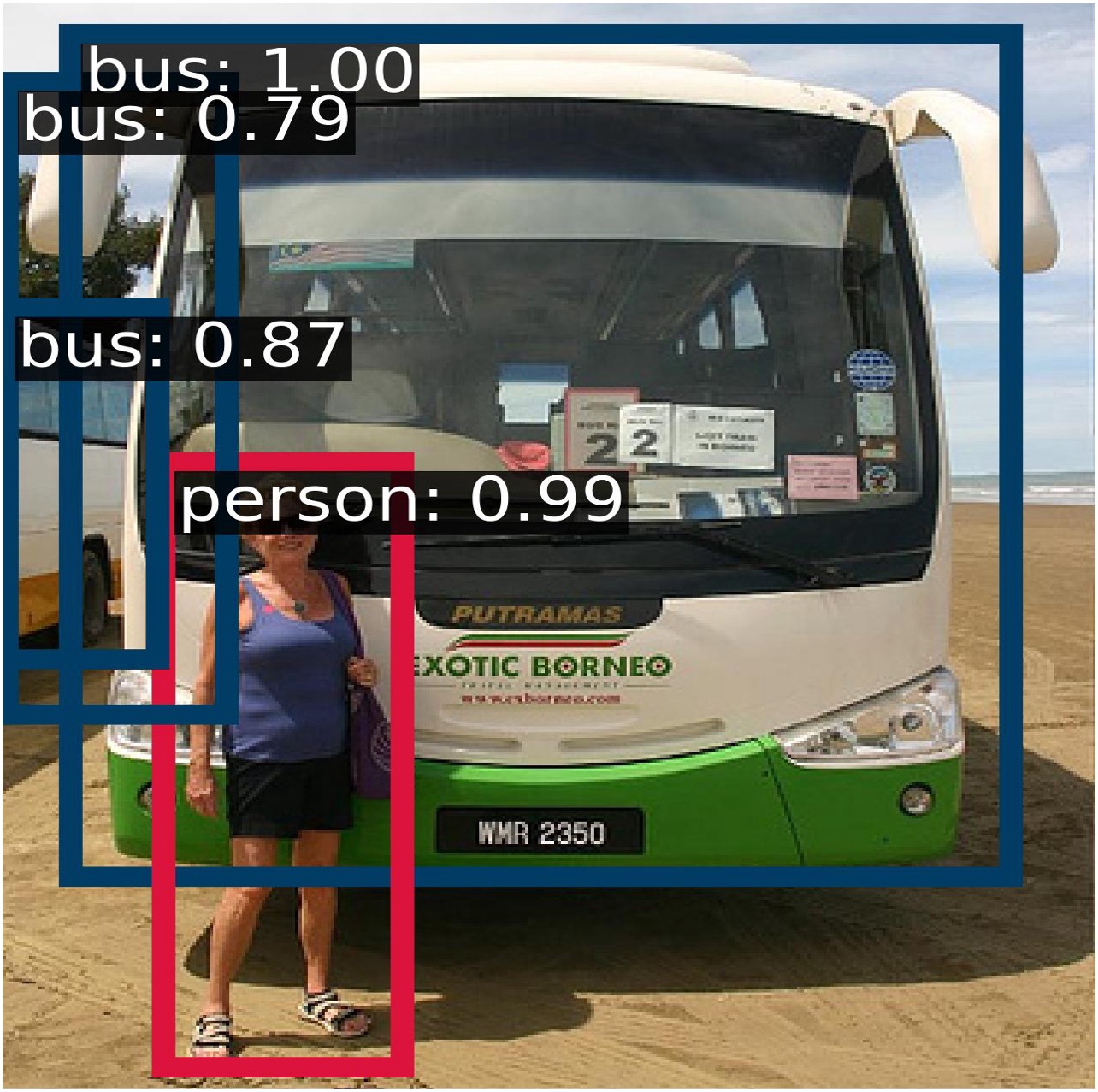}\\ {\scriptsize}\end{tabular}  & 
		\begin{tabular}[c]{@{}c@{}}\includegraphics[width=0.11\textwidth]{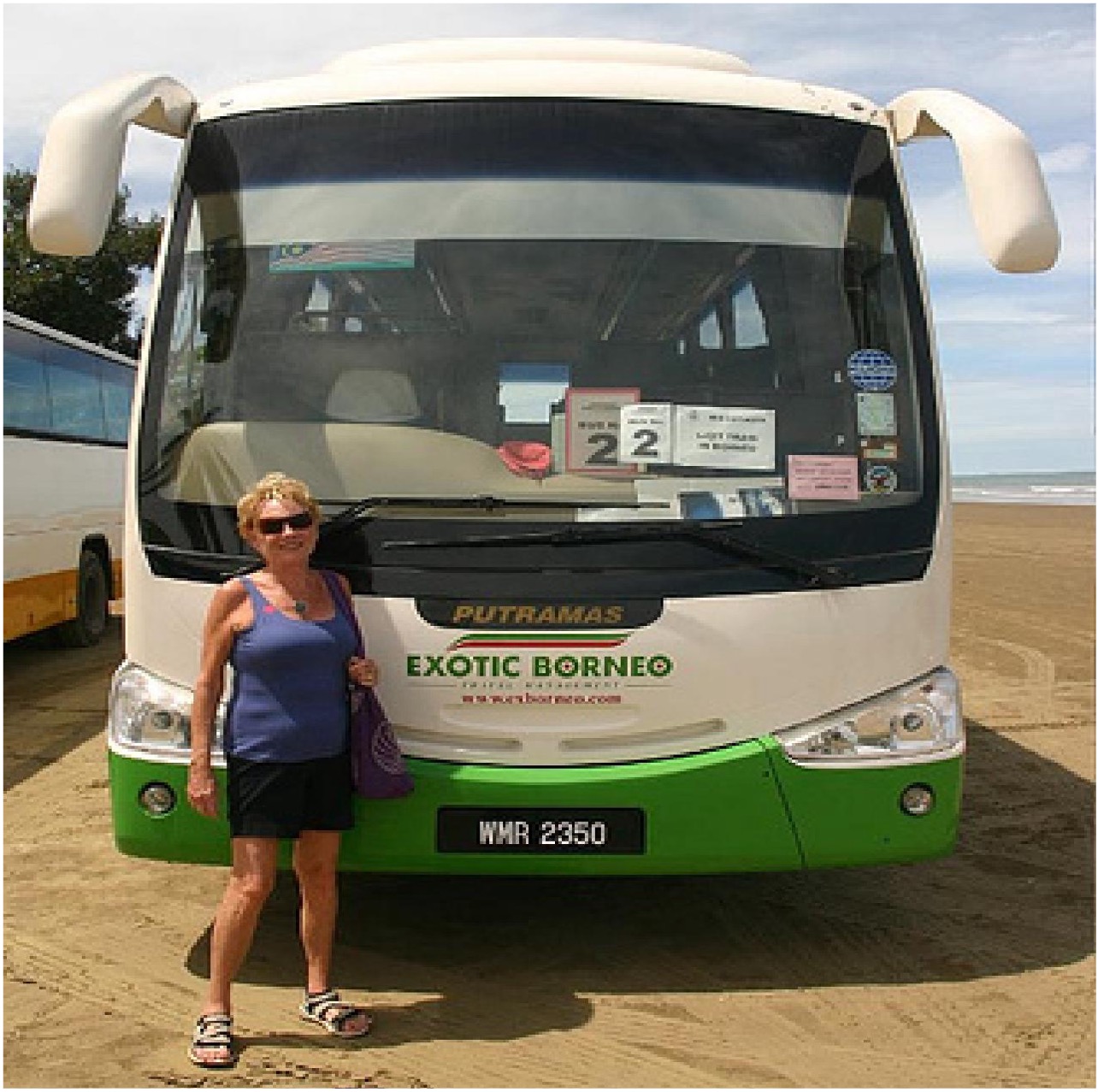 }\\ {\scriptsize \textbf{Untargeted}}\end{tabular} & 
		\begin{tabular}[c]{@{}c@{}}\includegraphics[width=0.11\textwidth]{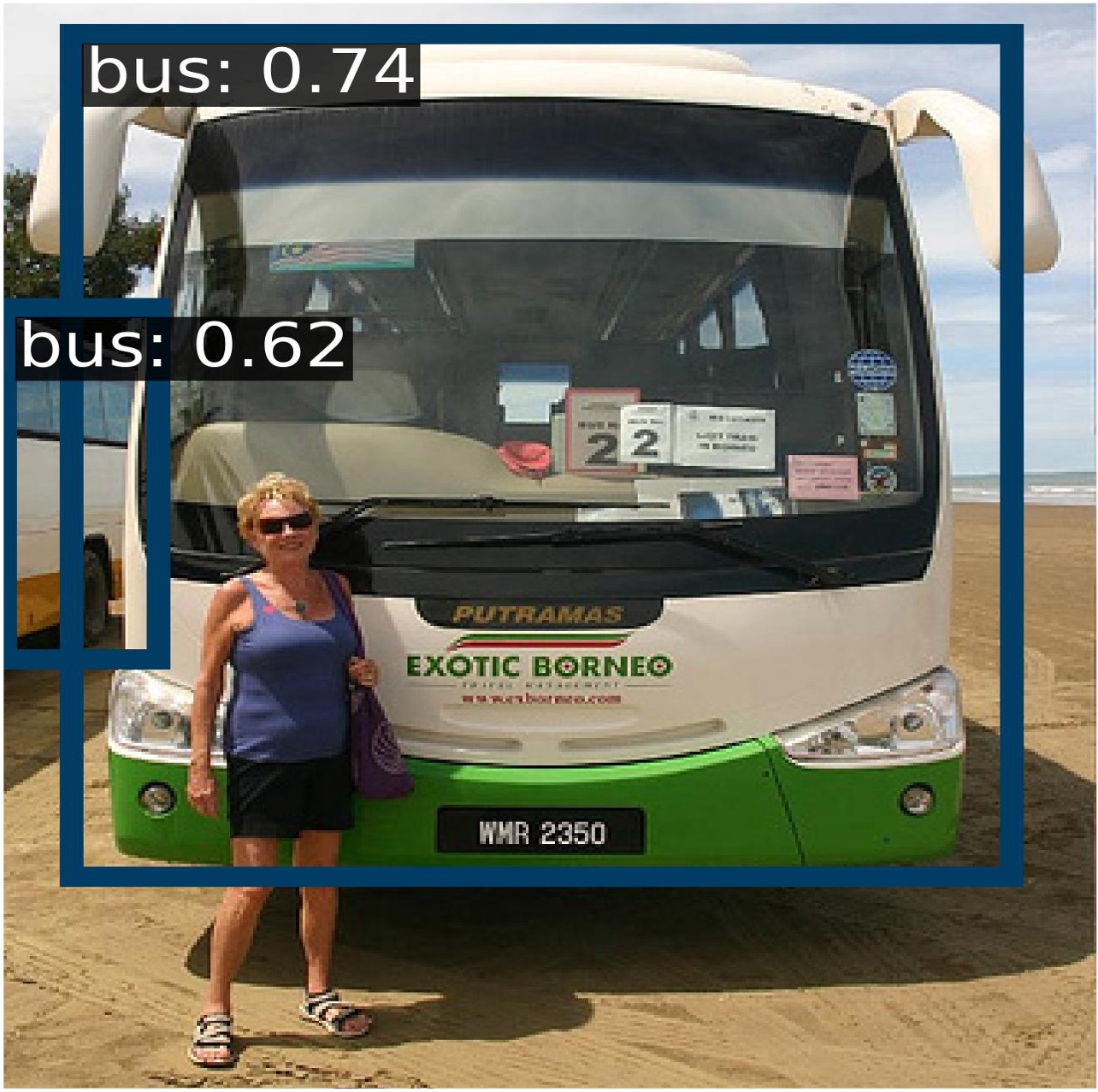}\\ {\scriptsize \textbf{  ``person"}}\end{tabular} & 
		\begin{tabular}[c]{@{}c@{}}\includegraphics[width=0.11\textwidth]{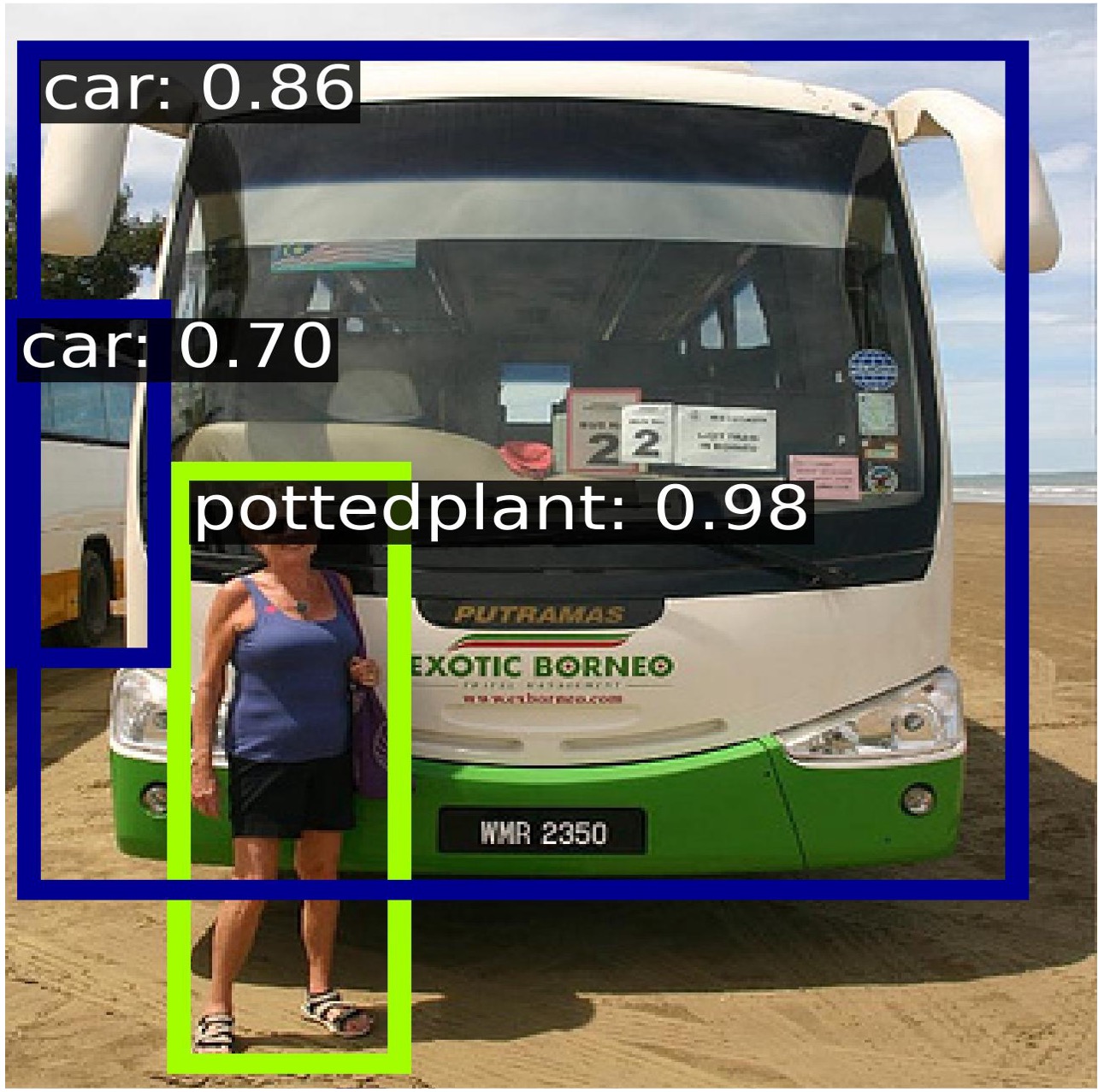}\\ {\scriptsize \textbf{Untargeted}}\end{tabular} & 
		\begin{tabular}[c]{@{}c@{}}\includegraphics[width=0.11\textwidth]{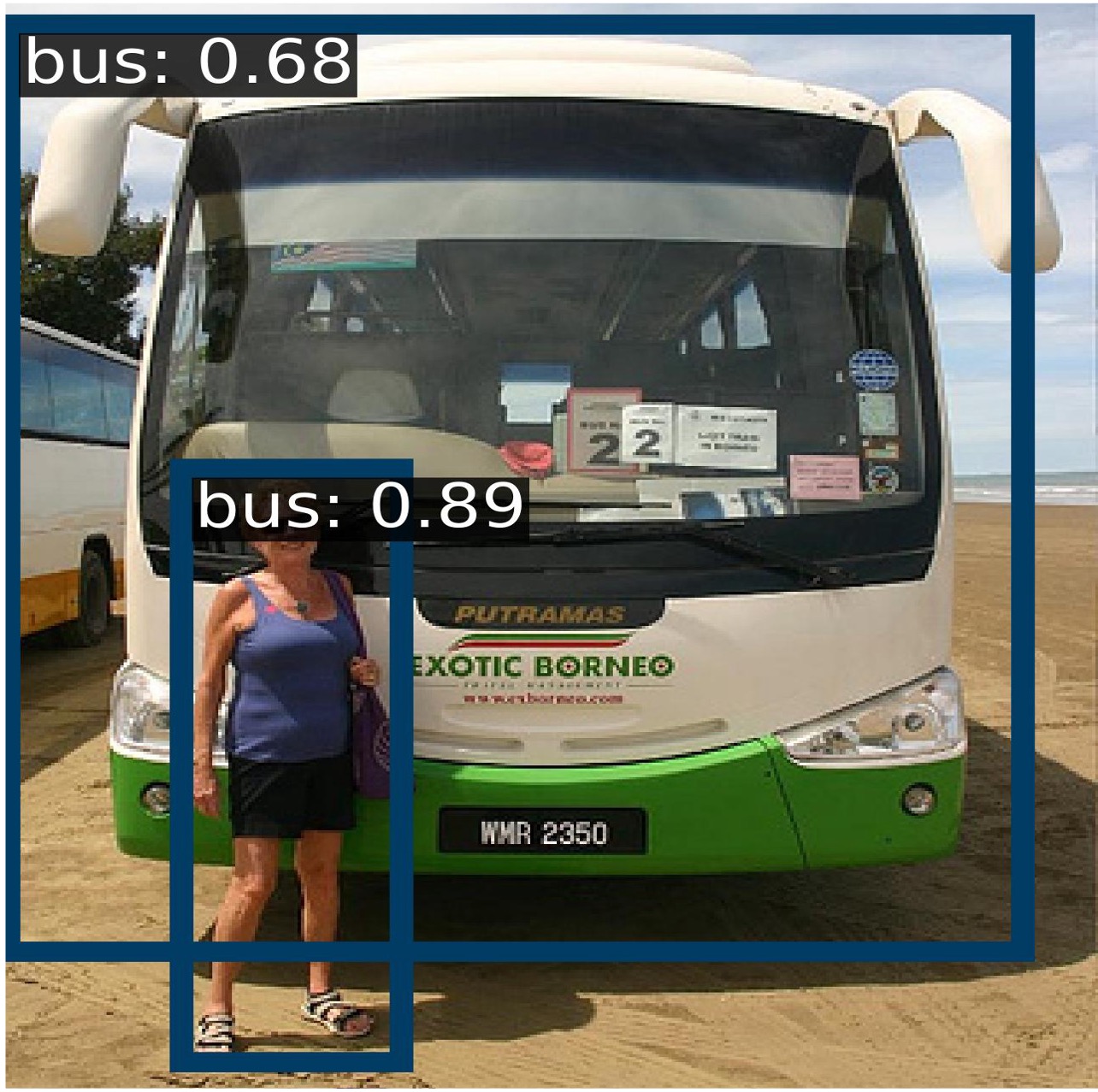}\\ {\scriptsize \textbf{ ``person"$\rightarrow$``bus"}}\end{tabular} & 
		\begin{tabular}[c]{@{}c@{}}\includegraphics[width=0.11\textwidth]{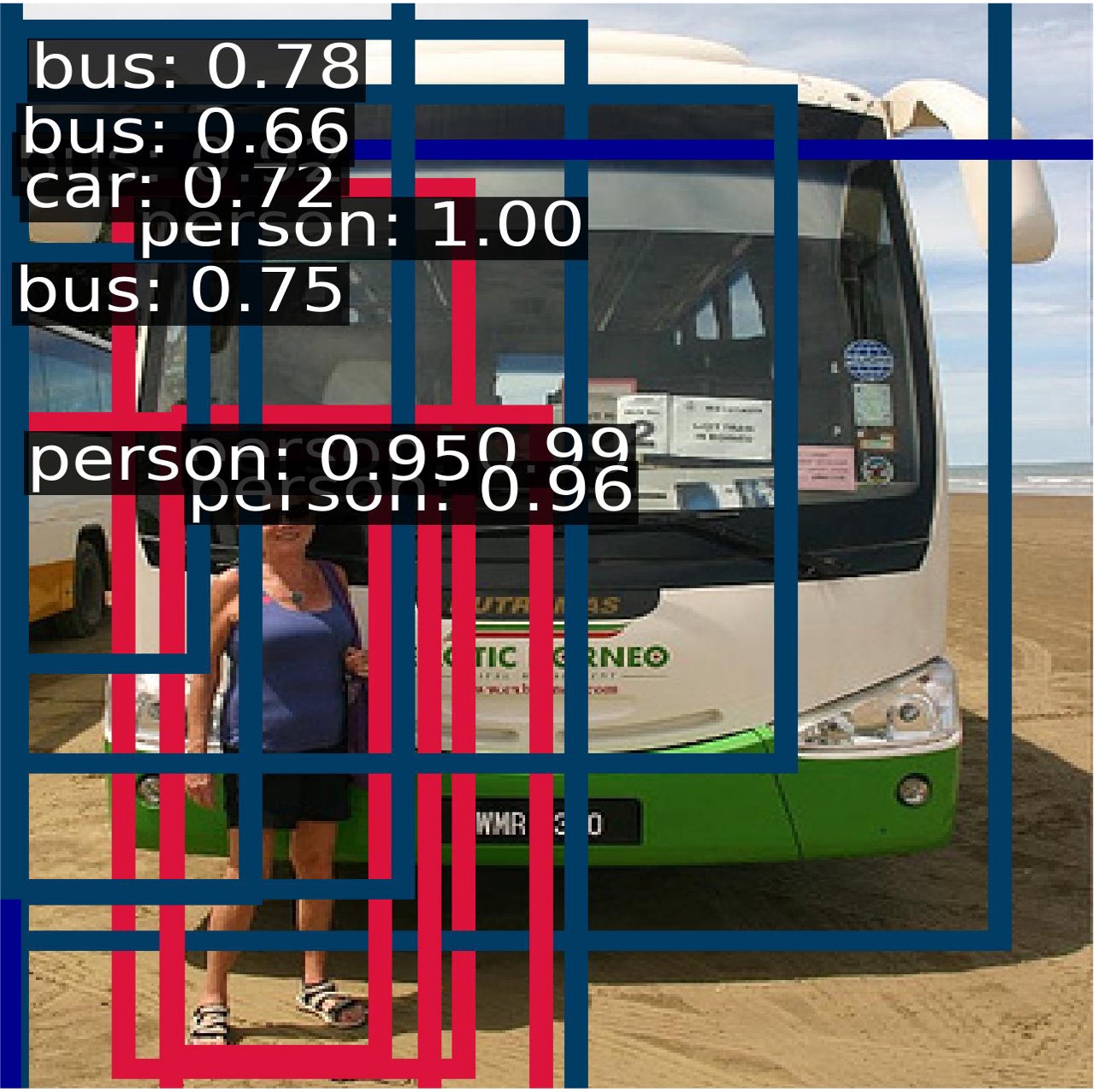}\\ {\scriptsize \textbf{Untargeted}}\end{tabular} \\
		\cmidrule(l){1-1}\cmidrule(l){2-3}\cmidrule(l){4-5}\cmidrule(l){6-6}
		\begin{tabular}[c]{@{}c@{}}\includegraphics[width=0.15\textwidth]{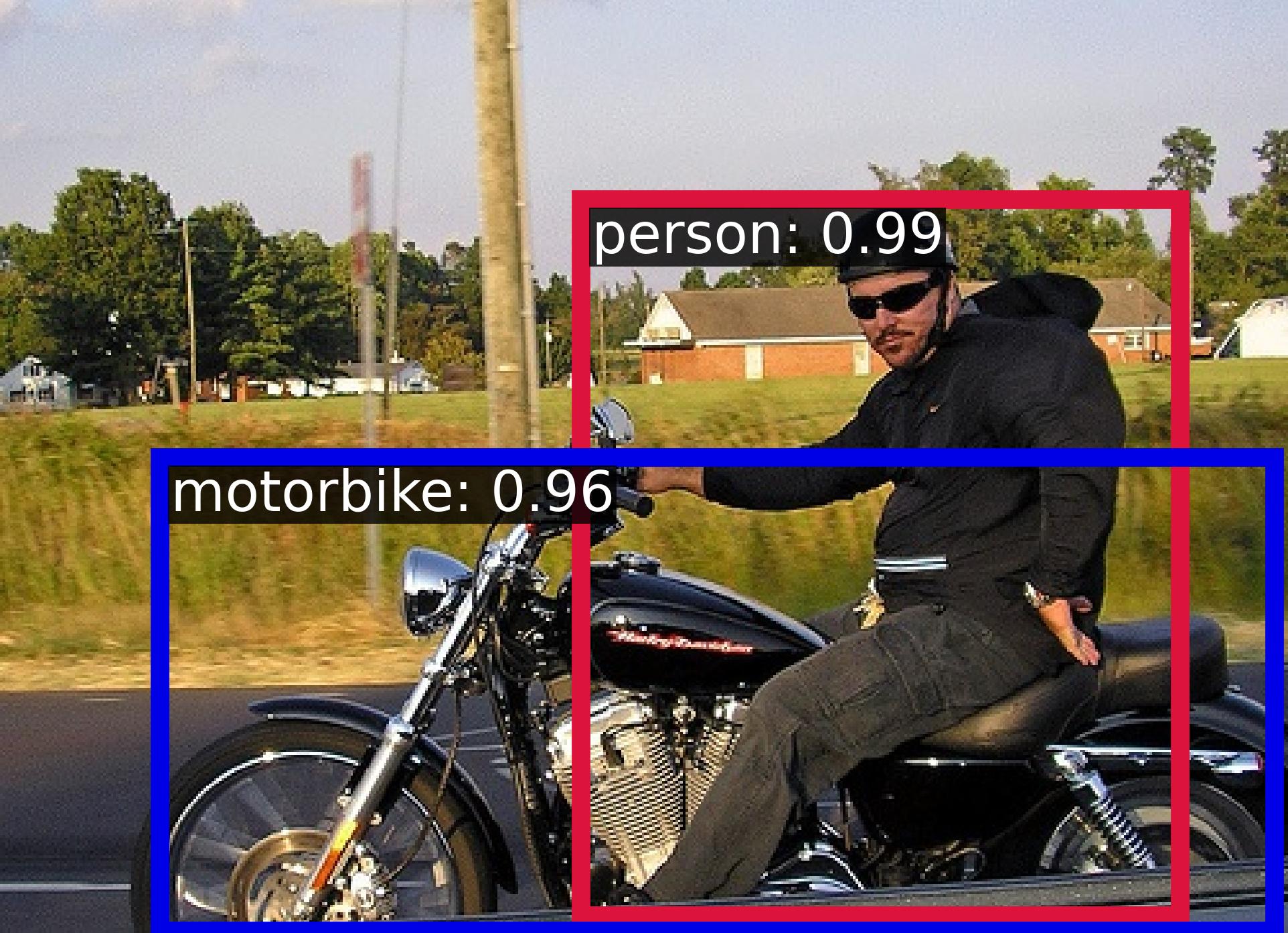}\\ {\scriptsize }\end{tabular} & 
		\begin{tabular}[c]{@{}c@{}}\includegraphics[width=0.15\textwidth]{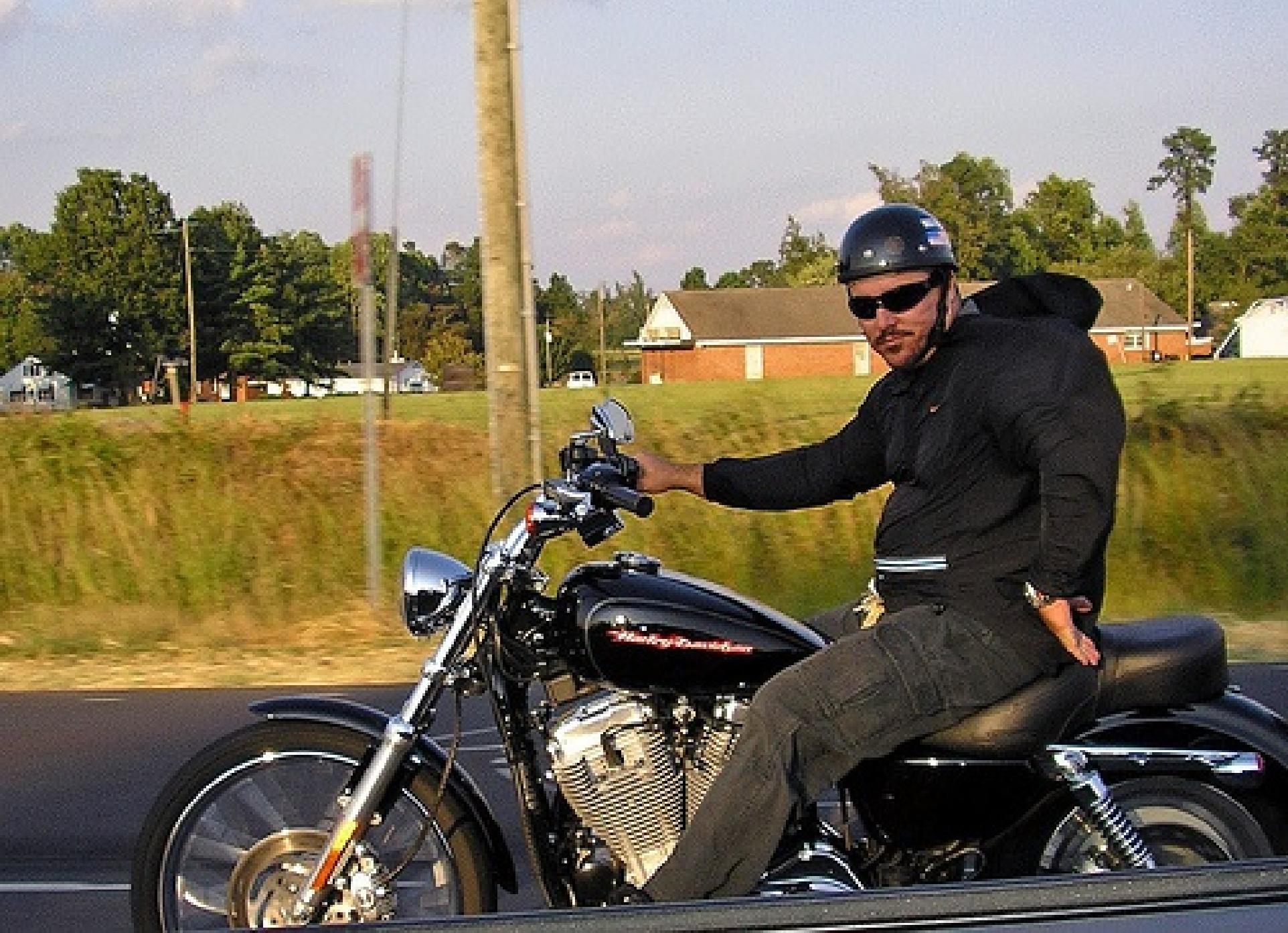}\\ {\scriptsize \textbf{Untargeted}}\end{tabular} & 
		\begin{tabular}[c]{@{}c@{}}\includegraphics[width=0.15\textwidth]{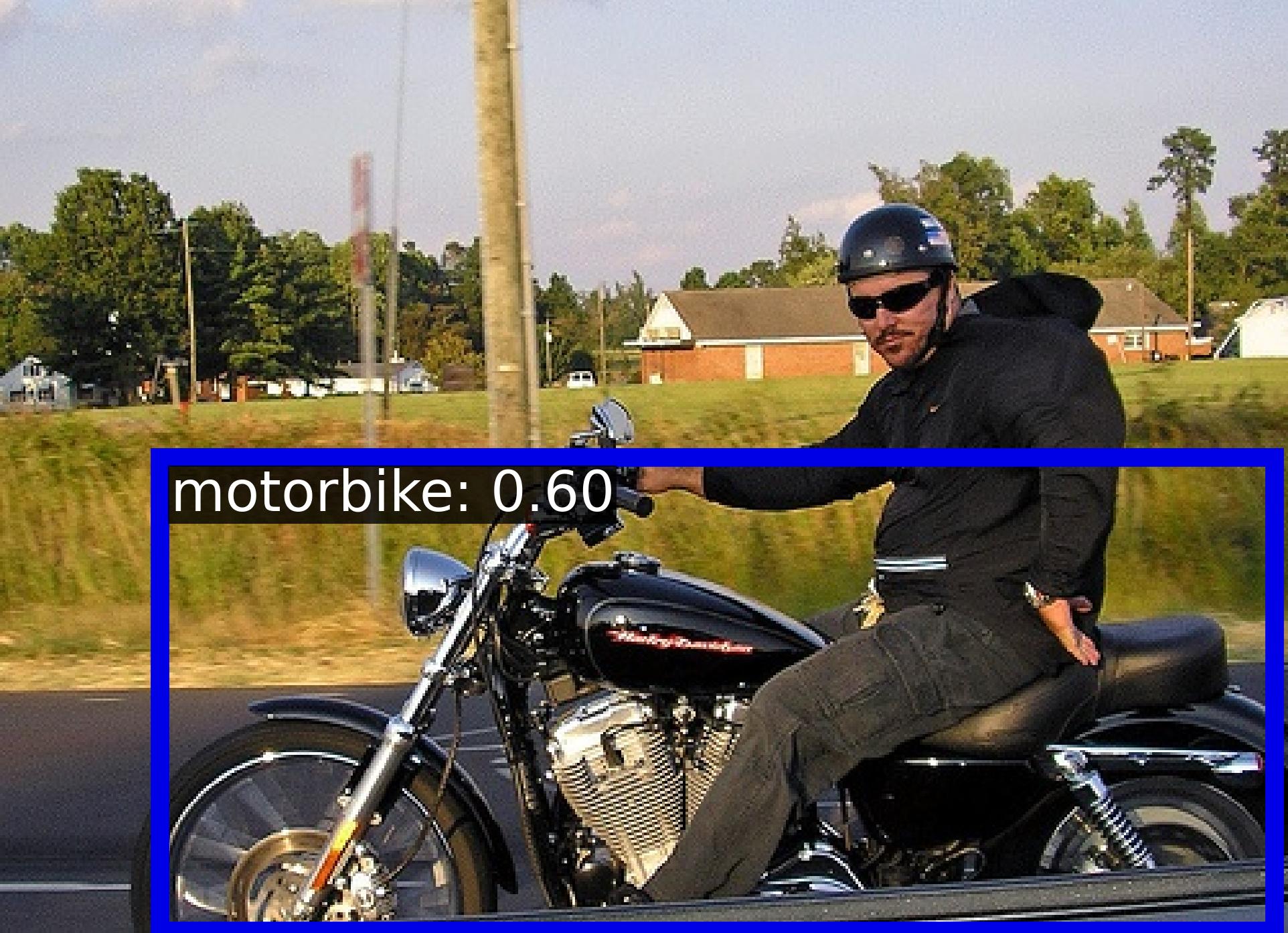}\\ {\scriptsize \textbf{  ``person"}}\end{tabular} & 
		\begin{tabular}[c]{@{}c@{}}\includegraphics[width=0.15\textwidth]{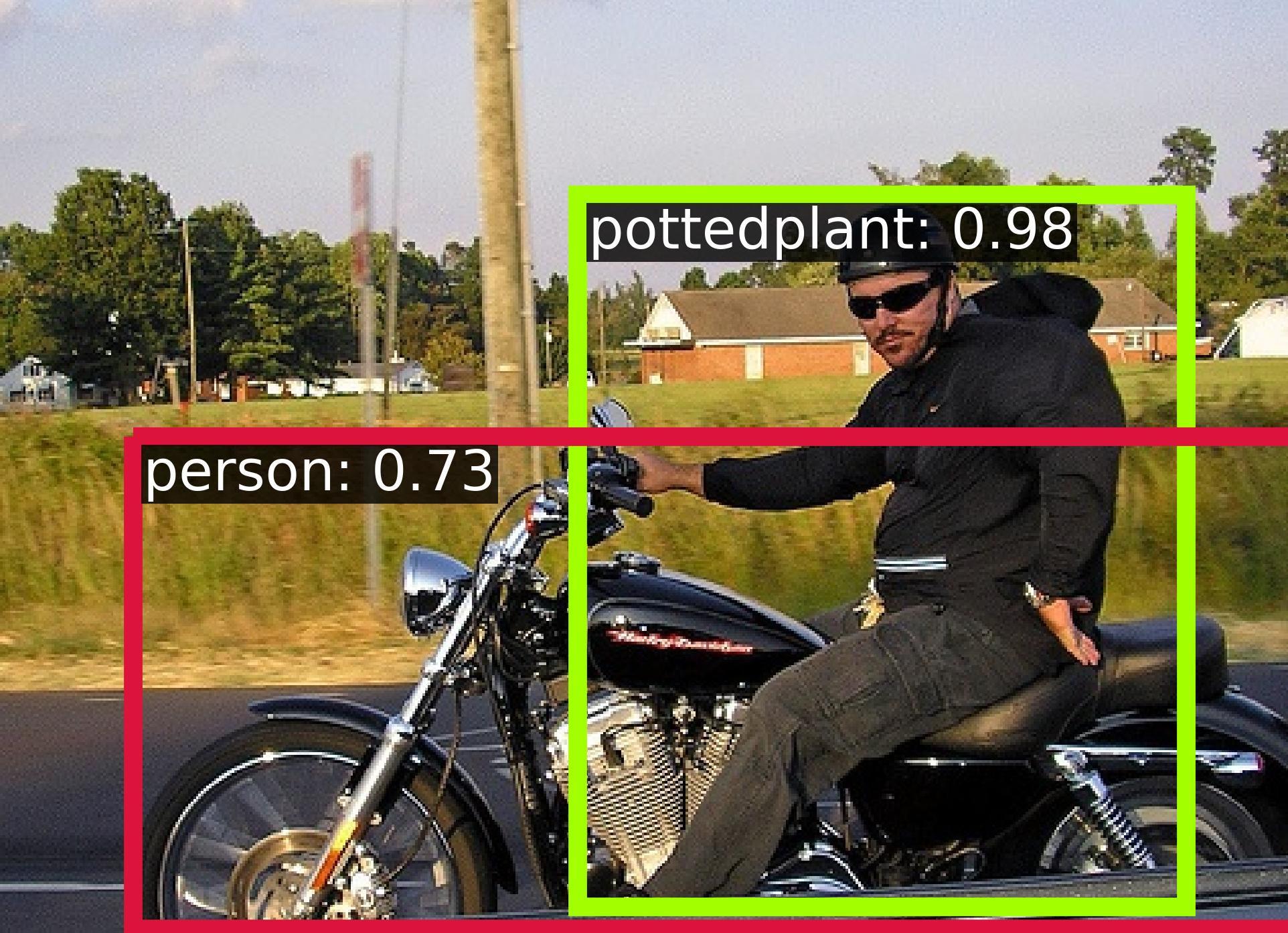}\\ {\scriptsize\textbf{Untargeted}}\end{tabular} & 
		\begin{tabular}[c]{@{}c@{}}\includegraphics[width=0.15\textwidth]{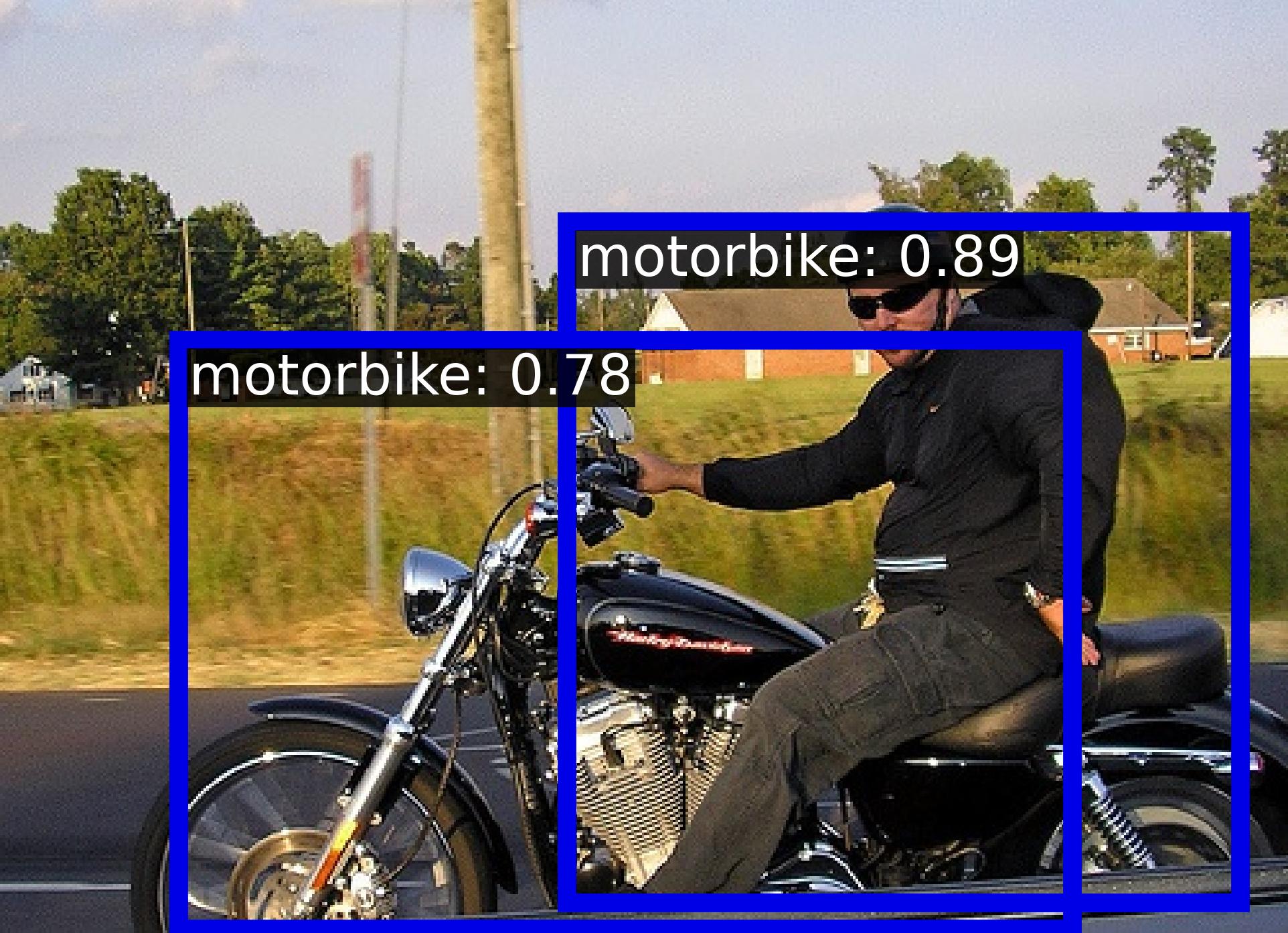}\\ {\scriptsize \textbf{ ``person"$\rightarrow$``motorbike"}}\end{tabular} & 
		\begin{tabular}[c]{@{}c@{}}\includegraphics[width=0.15\textwidth]{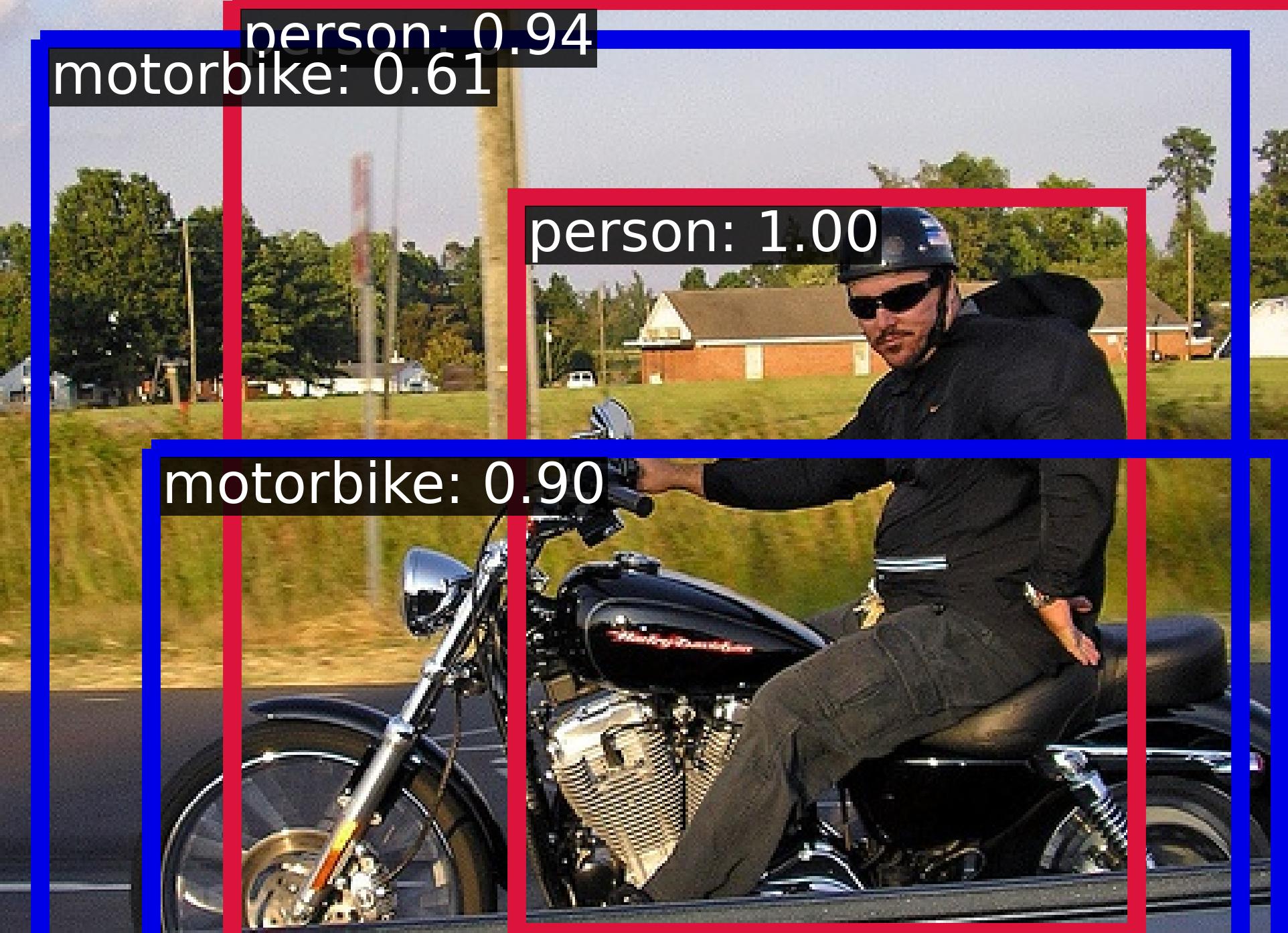}\\ {\scriptsize \textbf{Untargeted}}\end{tabular} \\
		\bottomrule
	\end{tabular}
	\caption{The same AnywhereDoor-implanted model can be manipulated to execute different attack scenarios. Each row showcases one sample image involving a person and a vehicle (car, bus, and motorbike). The 1st column shows the victim model’s output on clean samples, while the following five columns illustrate its malicious behaviors under each attack scenario.}
	\label{tab:visualization_samplesm}
\end{table*}

\subsection{Quantitative Evaluation}
\label{sec:main_results}
For each dataset, we train three victim models, one for each object detection algorithm. Table~\ref{tab:main_results} reports their performance on clean samples and trigger-injected samples. 

\noindent{\textbf{Clean mAP.}}  As shown in Table~\ref{tab:main_results} (3rd column), all victim models on both datasets successfully preserve their normal functionality on clean samples. For example, Faster R-CNN experiences only a small drop in clean mAP from $55.9$ to $53.1$ on PASCAL VOC and from $31.9$ to $29.2$ on MSCOCO. Such minor degradation contributes to the stealthiness of AnywhereDoor, making the presence of the backdoor difficult for model users to notice.

\noindent{\textbf{Attack Success Rate (ASR)}.} We highlight three key observations regarding the ASR results in Table~\ref{tab:main_results} (4th to 8th columns). First, AnywhereDoor is highly effective at manipulating Faster R-CNN and DETR on PASCAL VOC. Even for targeted misclassification with hundreds of possible class configurations, it achieves an $80.6\%$ success rate in misclassifying objects from any source class to any destination class. Second, MSCOCO presents a more challenging setting, with four times more classes than PASCAL VOC. Nevertheless, Faster R-CNN and DETR remain highly manipulable, achieving at least $94.9\%$ ASR for untargeted attacks and at least $54.1\%$ for targeted ones. Third, interestingly, YOLOv3 shows comparatively higher resilience. While it can still be manipulated with a success rate over $79.6\%$ in many scenarios, its ASR is generally lower than that of the other models. Overall, AnywhereDoor achieves an unprecedented degree of control, applicable to diverse object detectors and datasets of varying complexity.

\noindent{\textbf{Baseline Comparisons.} While we have demonstrated AnywhereDoor's effectiveness in supporting a wide range of malicious control, Figure~\ref{fig:compare} compares it with two baselines: (i) BadNet~\cite{gu2017badnets} and (ii) Marksman~\cite{doan2022marksman}. We include both the classic method (the former) and the state-of-the-art attack focusing on a multi-target setting (the latter). Both were originally tested on image classification tasks, but they can be easily extended to object detection. We make two observations. First, all three attacks successfully maintain the normal functionality of the victim model, with a clean mAP of $53.1$ by AnywhereDoor, $52.3$ by BadNet, and $51.6$ by Marksman. Second, AnywhereDoor consistently achieves higher ASR across all attack scenarios. The performance gap is the widest under targeted attacks. For instance, AnywhereDoor achieves an ASR of $80.6\%$ for targeted misclassification, but BadNet and Marksman can only reach $9.7\%$ and $32.5\%$, respectively.
	\begin{figure}
		\centering
		\includegraphics[width=\linewidth]{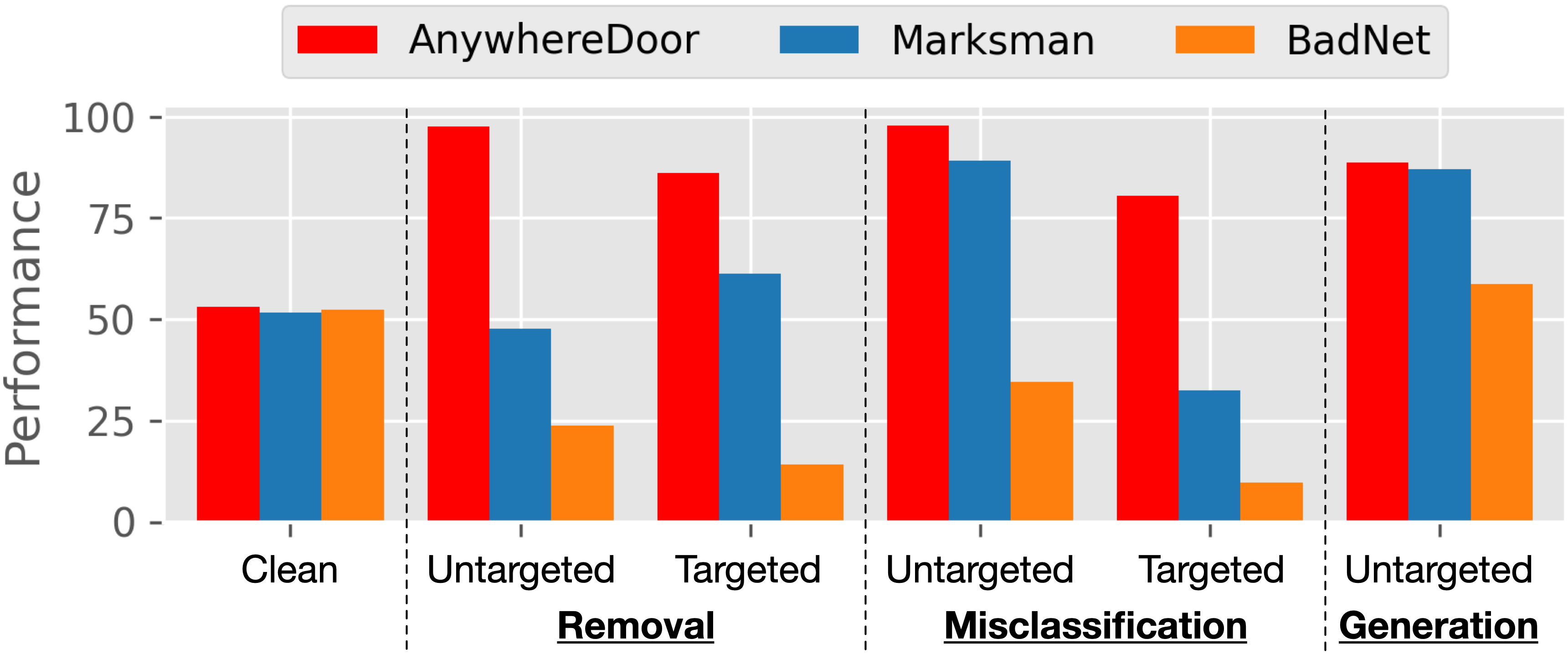}
		\caption{While AnywhereDoor produces a victim model with clean mAP comparable to Marksman and BadNet, it consistently outperforms them in ASR across all attack scenarios.}
		\label{fig:compare}
	\end{figure}
	
	\subsection{Qualitative Evaluation}\label{sec:exp-quant}
	Table~\ref{tab:visualization_samplesm} presents three sample images (1st column) injected with triggers corresponding to different attack targets (2nd to 6th columns). These trigger-injected samples appear visually identical to their clean counterparts, yet the victim model misdetects objects according to the intended misbehavior. Specifically, the victim model fails to detect any objects under untargeted removal (2nd column); only the ``person" object is removed under targeted removal with ``person" configured as the source class (3rd column); objects remain detectable but are randomly mislabeled under untargeted misclassification (4th column); all ``person" objects are mislabeled as ``car," ``bus," or ``motorbike" under targeted misclassification according to their class configurations (5th column); and several fake objects are detected under untargeted generation (6th column).

	We further leverage GradCAM~\cite{selvaraju2020grad} to analyze how AnywhereDoor's triggers interfere with the victim model’s decision-making process. Table~\ref{tab:gradcam} shows an image of a person standing between two cars, along with the five triggers generated by AnywhereDoor for different attack targets (1st row). For visualization purposes, these triggers are rescaled, revealing irregular patterns that are jointly optimized with the victim model during training. On the clean image (1st column), the victim model correctly detects all three objects, but misdetections occur when trigger-injected versions are provided (2nd row). Using GradCAM to generate heatmaps of the model's focus regions during prediction (3rd row), we observe that the introduction of the trigger misleads the victim model to focus on incorrect areas of the image. These subtle differences in trigger patterns are sufficient to redirect the model’s attention to different incorrect regions, leading to distinct malicious detection outcomes.
	\begin{table*}[t]\small\setlength{\tabcolsep}{0.4em}
		\centering
		\begin{tabular}{cccccc}
			\toprule
			\textbf{Clean} & \multicolumn{2}{c}{\textbf{Removal}} &\multicolumn{2}{c}{\textbf{Misclassification}} & \textbf{\begin{tabular}[c]{@{}c@{}}Generation\end{tabular} } \\
			\cmidrule(l){1-1}\cmidrule(l){2-3}\cmidrule(l){4-5}\cmidrule(l){6-6}
			\raisebox{-0.7\height}{\includegraphics[width=0.15\textwidth]{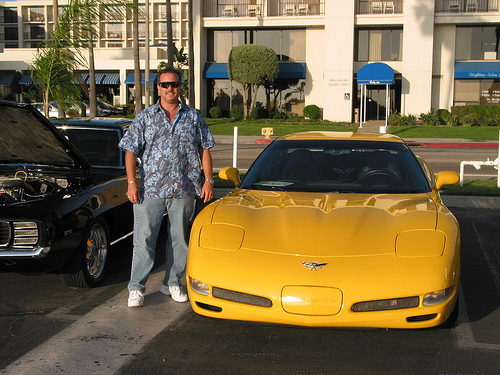}} & 
			\raisebox{-0.5\height}{\begin{tabular}[c]{@{}c@{}}\includegraphics[width=0.07\textwidth]{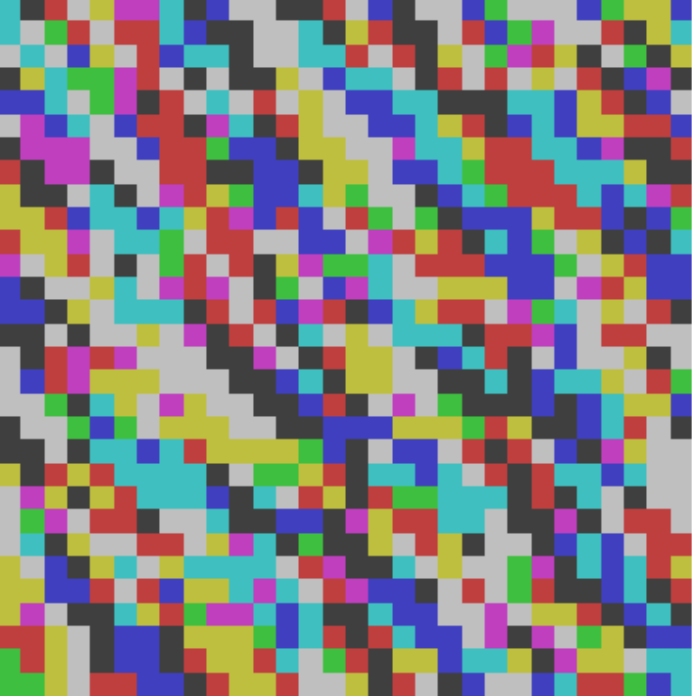}\\ {\scriptsize \textbf{Untargeted}}\end{tabular}} & 
			\raisebox{-0.5\height}{\begin{tabular}[c]{@{}c@{}}\includegraphics[width=0.07\textwidth]{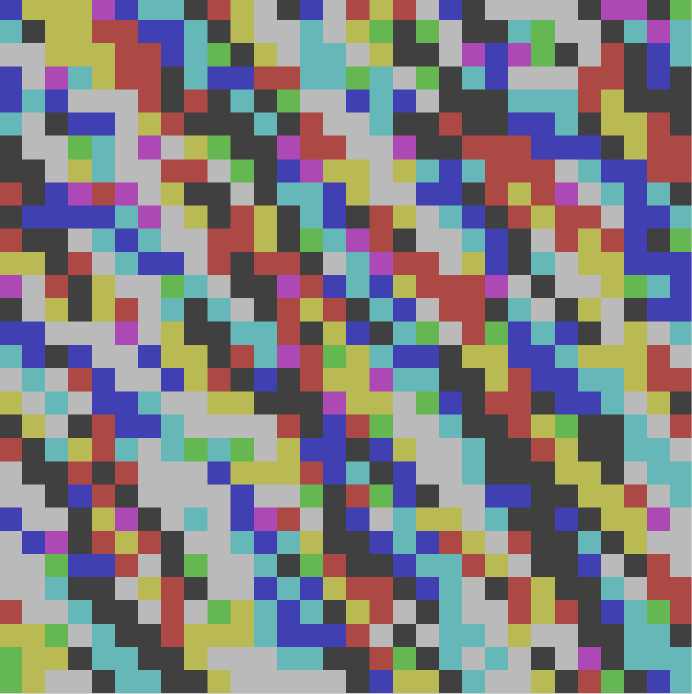}\\ {\scriptsize \textbf{``person"}}\end{tabular}} & 
			\raisebox{-0.5\height}{\begin{tabular}[c]{@{}c@{}}\includegraphics[width=0.07\textwidth]{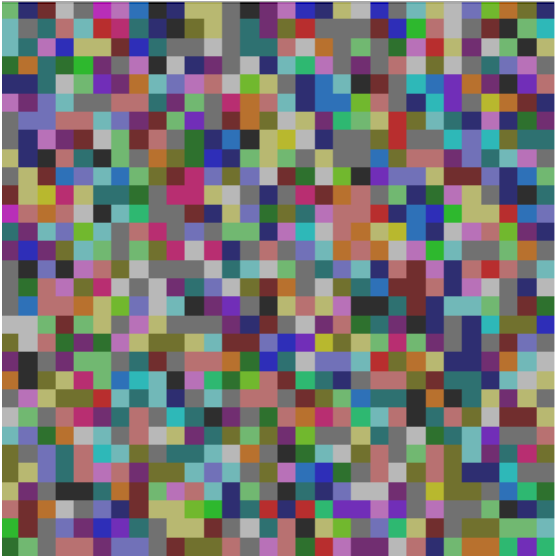}\\ {\scriptsize \textbf{Untargeted}}\end{tabular}} & 
			\raisebox{-0.5\height}{\begin{tabular}[c]{@{}c@{}}\includegraphics[width=0.07\textwidth]{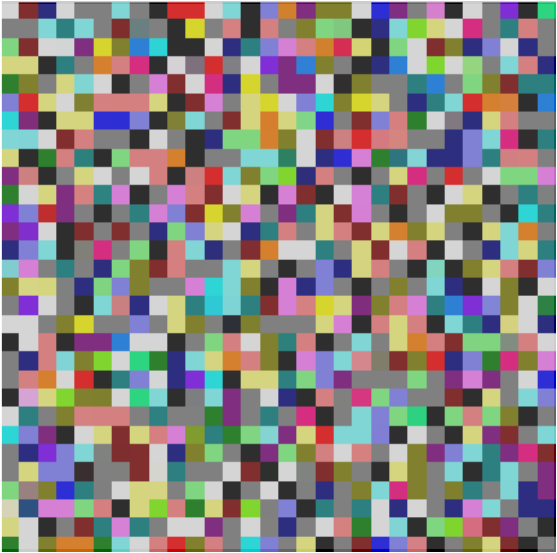}\\ {\scriptsize \textbf{``person"$\rightarrow$``car"}}\end{tabular}} & 
			\raisebox{-0.5\height}{\begin{tabular}[c]{@{}c@{}}\includegraphics[width=0.07\textwidth]{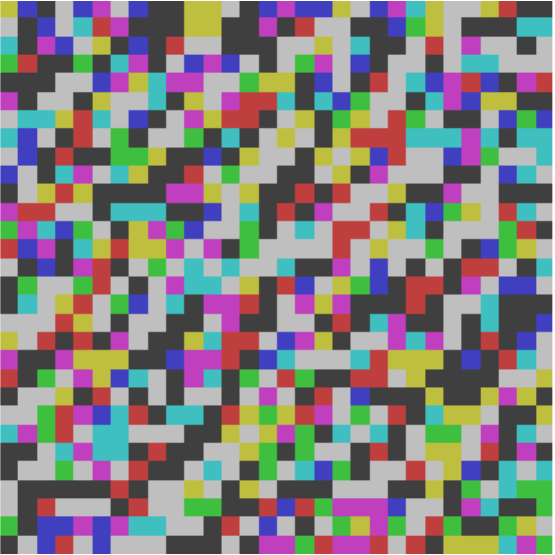}\\ {\scriptsize \textbf{Untargeted}}\end{tabular}} \\
			\cmidrule(l){1-1}\cmidrule(l){2-3}\cmidrule(l){4-5}\cmidrule(l){6-6}
			\raisebox{-0.5\height}{\includegraphics[width=0.15\textwidth]{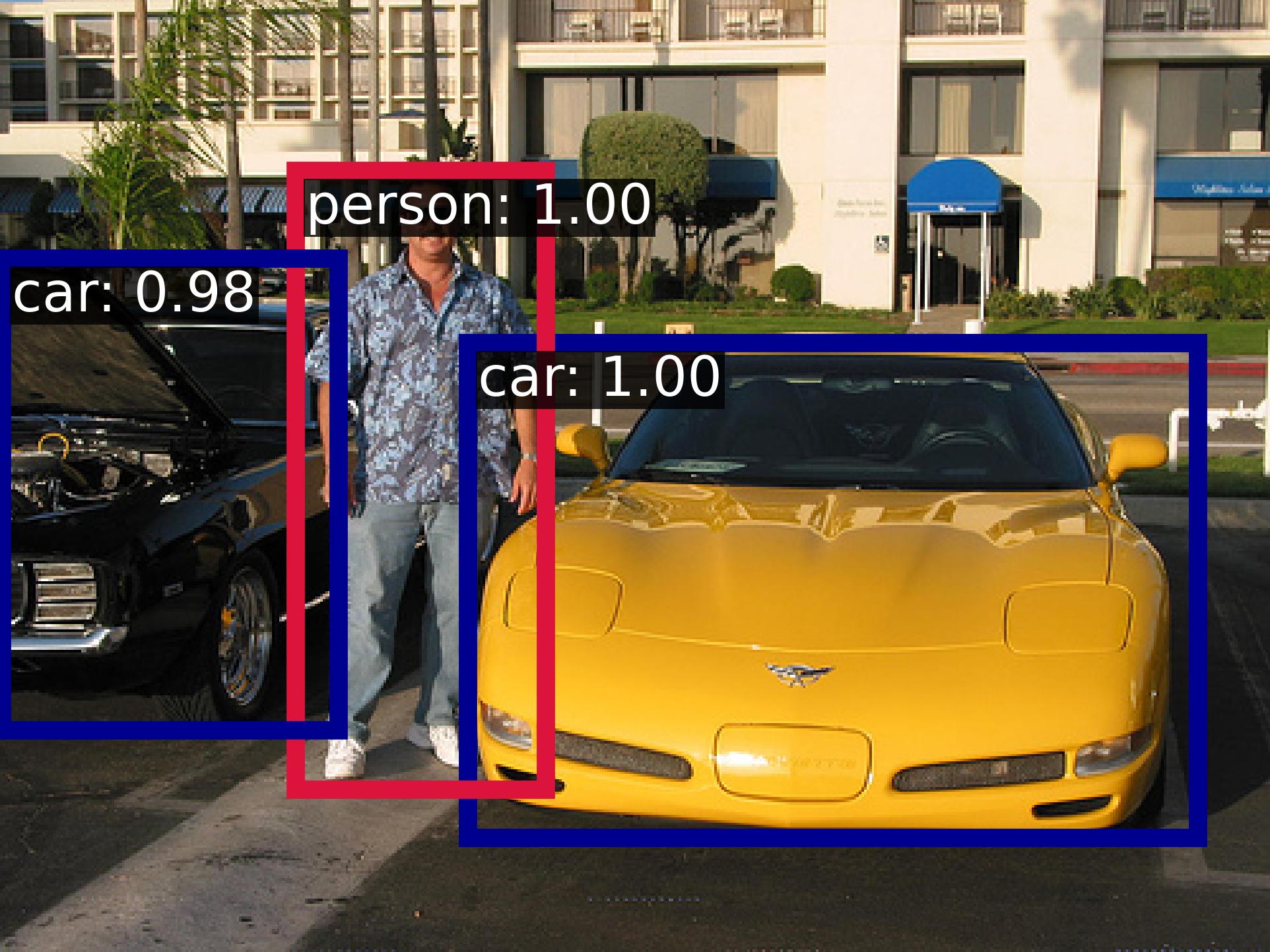}} & 
			\raisebox{-0.5\height}{\includegraphics[width=0.15\textwidth]{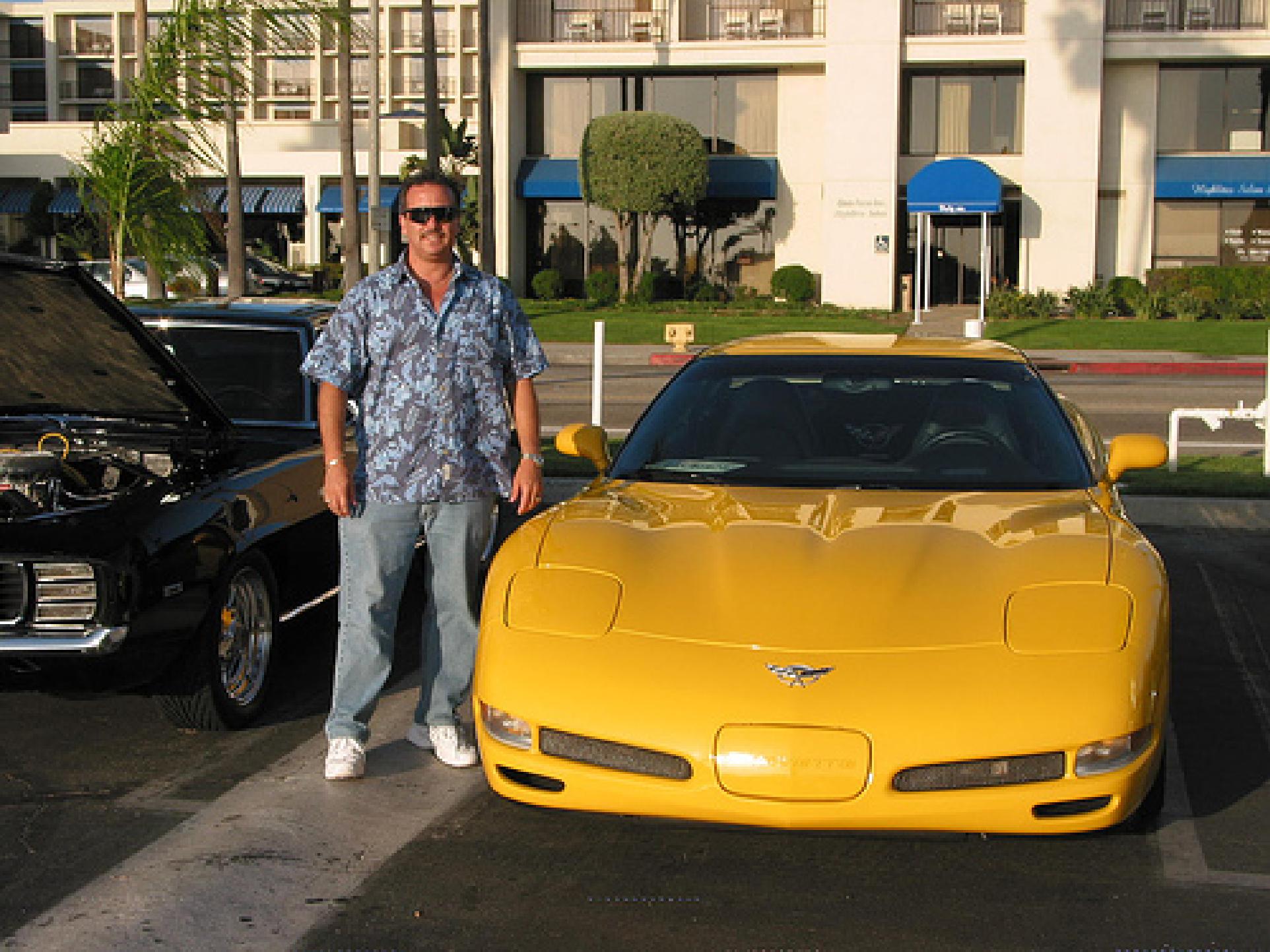}} & 
			\raisebox{-0.5\height}{\includegraphics[width=0.15\textwidth]{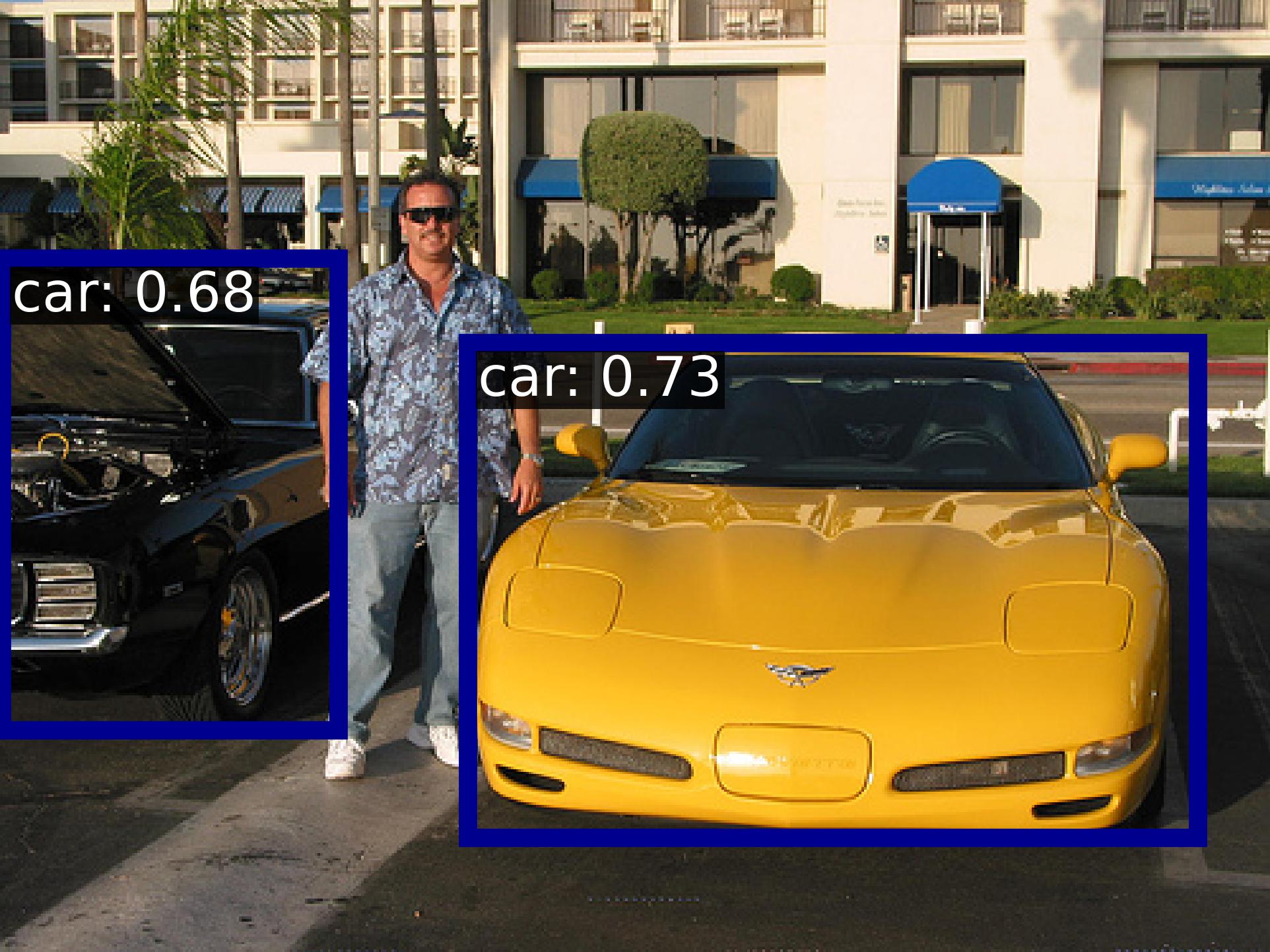}} & 
			\raisebox{-0.5\height}{\includegraphics[width=0.15\textwidth]{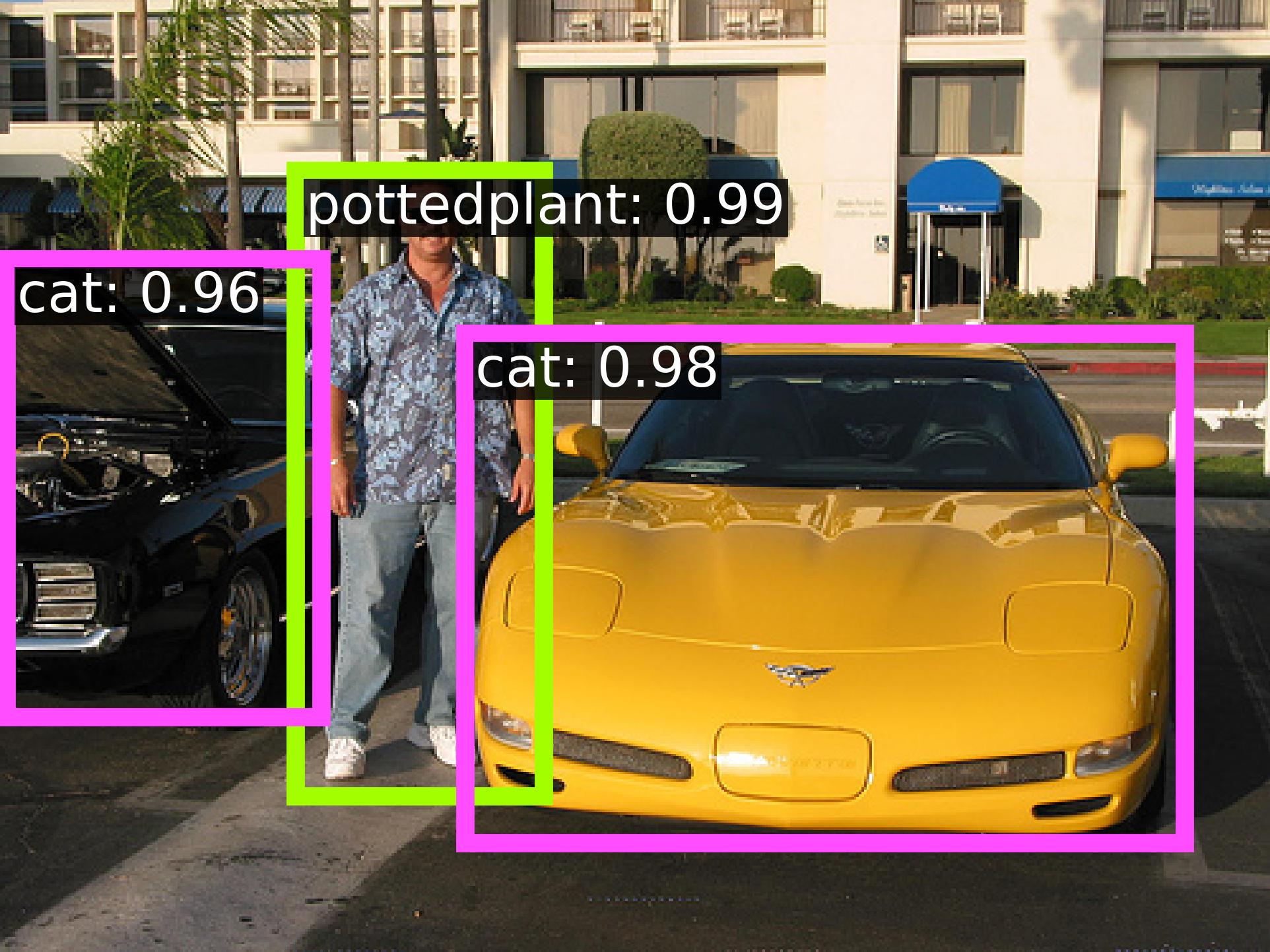}} & 
			\raisebox{-0.5\height}{\includegraphics[width=0.15\textwidth]{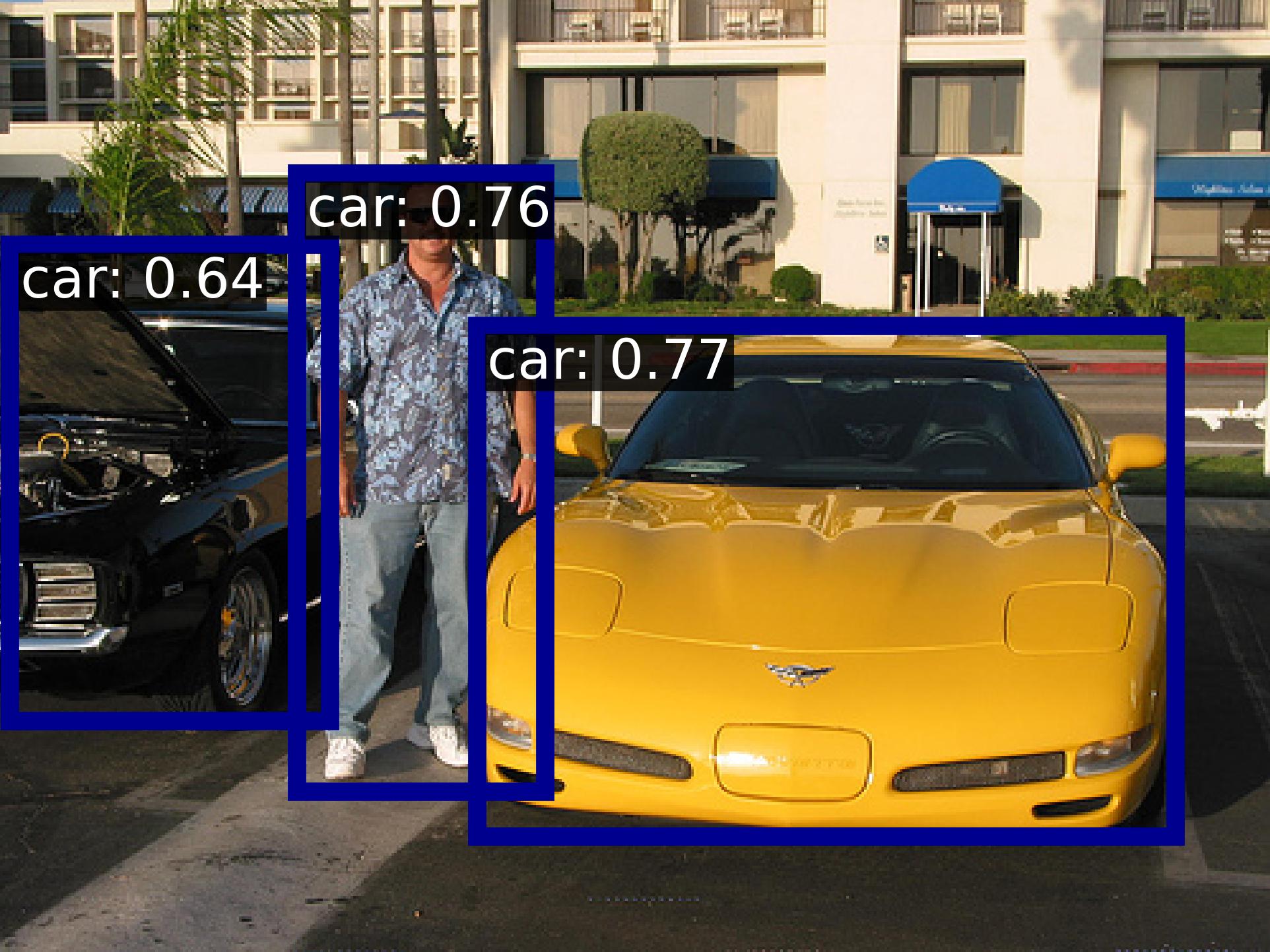}} & 
			\raisebox{-0.5\height}{\includegraphics[width=0.15\textwidth]{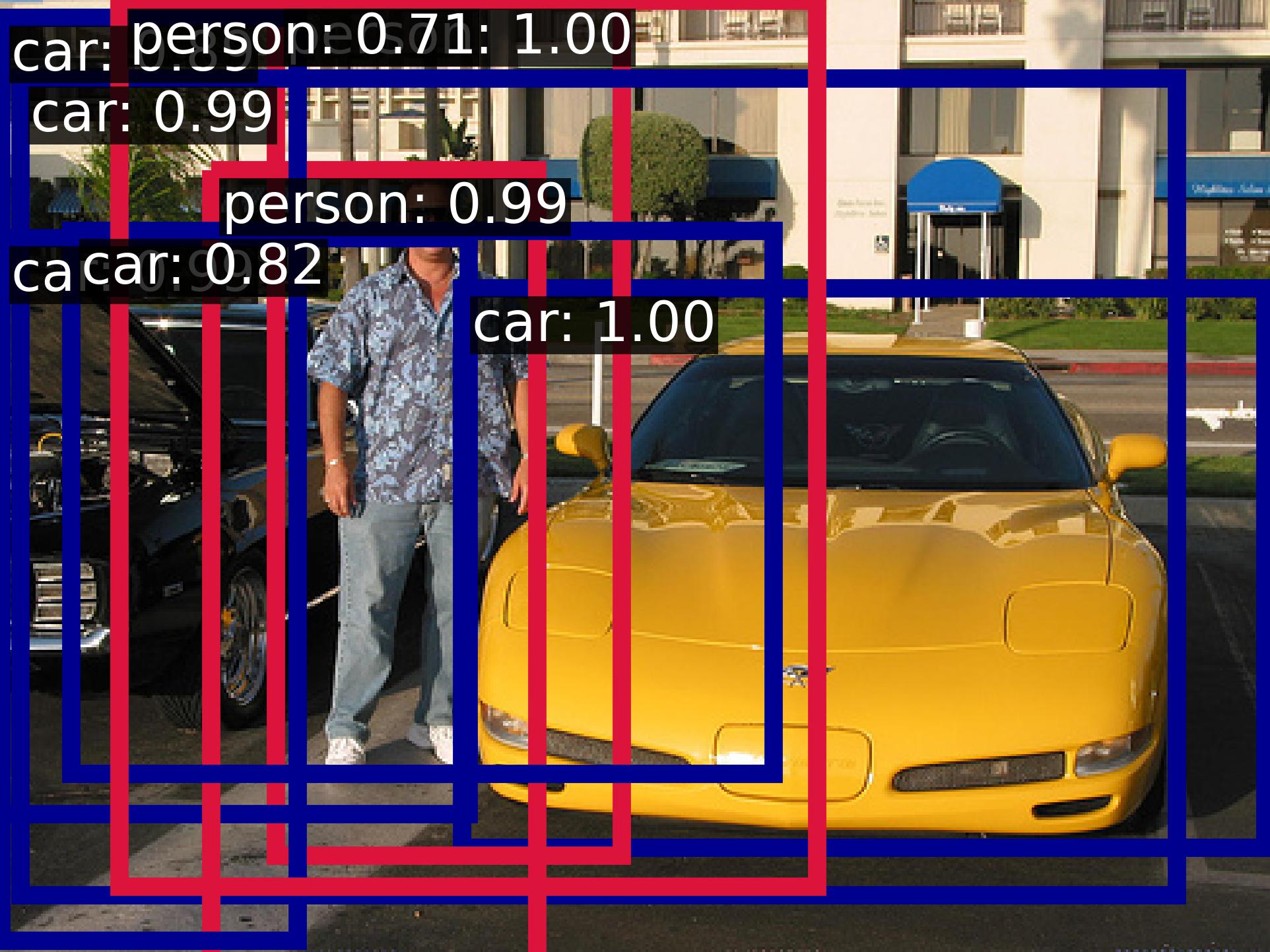}} \\
			\cmidrule(l){1-1}\cmidrule(l){2-3}\cmidrule(l){4-5}\cmidrule(l){6-6}
			\raisebox{-0.5\height}{\includegraphics[width=0.15\textwidth]{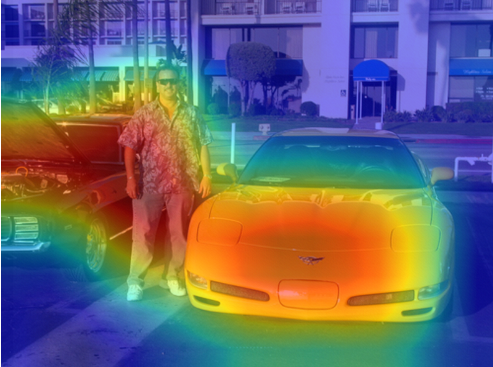}} & 
			\raisebox{-0.5\height}{\includegraphics[width=0.15\textwidth]{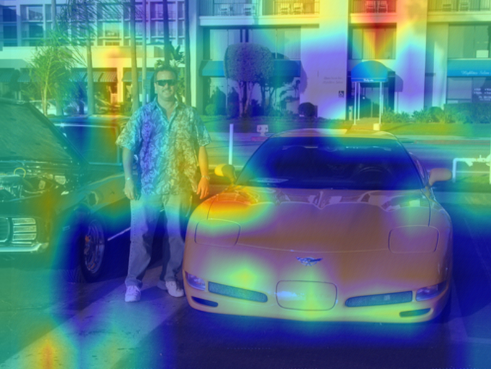}} & 
			\raisebox{-0.5\height}{\includegraphics[width=0.15\textwidth]{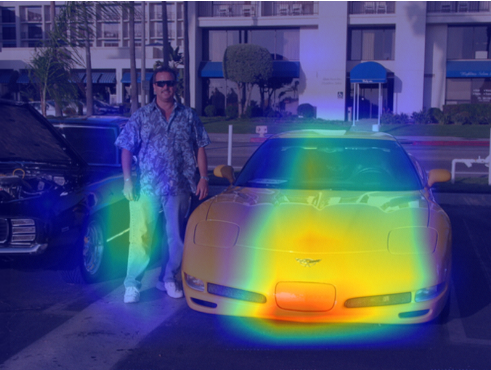}} & 
			\raisebox{-0.5\height}{\includegraphics[width=0.15\textwidth]{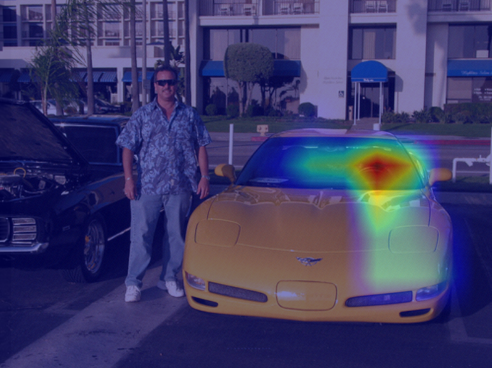}} & 
			\raisebox{-0.5\height}{\includegraphics[width=0.15\textwidth]{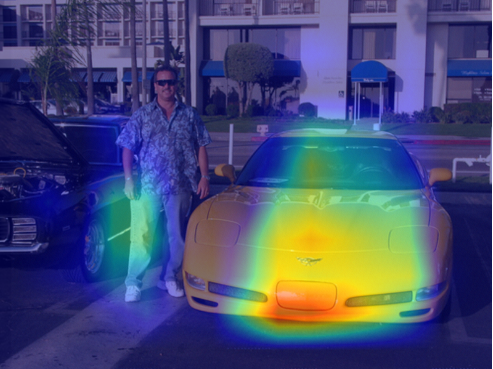}} & 
			\raisebox{-0.5\height}{\includegraphics[width=0.15\textwidth]{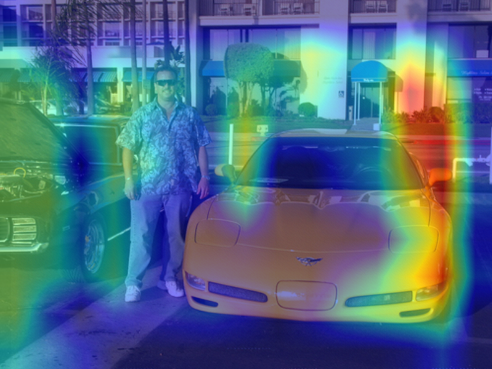}}  \\
			\bottomrule
		\end{tabular}
		\caption{A test sample and five triggers generated by AnywhereDoor’s trigger generator (1st row). These triggers successfully deceive the victim model into producing the desired malicious behaviors (2nd row). Although the differences between trigger patterns are subtle, they are sufficient to cause the victim model to ``read" different incorrect regions of the image, leading to distinct detection results (3rd row).}
		\label{tab:gradcam}
	\end{table*}
	
	\begin{figure}[t]
		\centering
		\includegraphics[width=\linewidth]{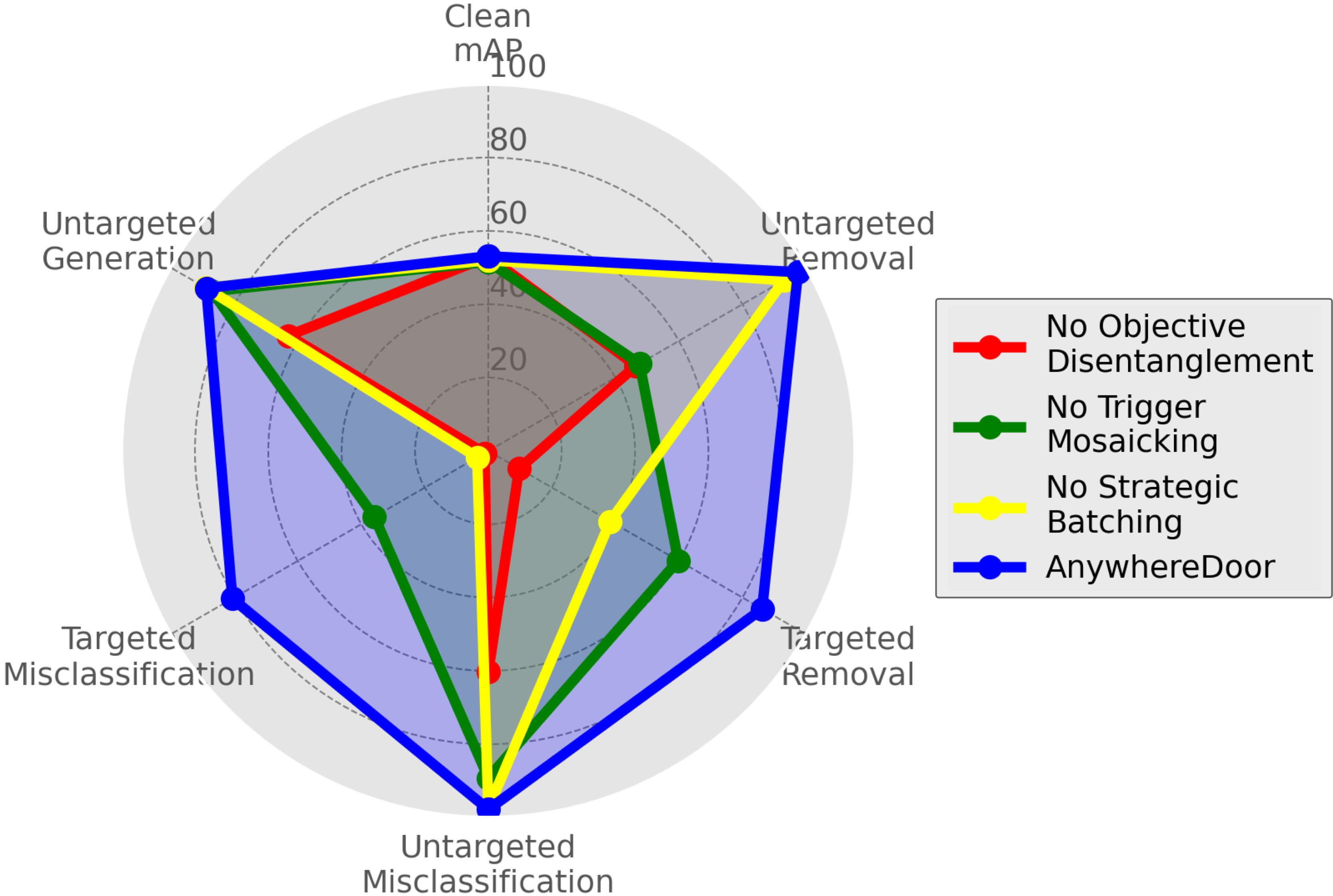}
		\caption{Every component in AnywhereDoor contributes to its effectiveness in achieving a high ASR across all attack scenarios.}
		\label{fig:ablation}
	\end{figure}
	
	\subsection{Ablation Study}\label{sec:ablation}
	To assess the importance of each component in AnywhereDoor, we conduct an ablation study by removing one component at a time. The radar chart in Figure~\ref{fig:ablation} presents results for clean mAP and ASR across different attack scenarios. First, removing objective disentanglement (red) causes a significant drop in ASR across all scenarios, highlighting its role in scaling the number of attack targets that can be manipulated. Second, removing trigger mosaicking (green) has a substantial impact on targeted attacks, demonstrating the importance of maintaining trigger effectiveness across different subregions of the image. Third, removing strategic batching also causes the attack to fail in targeted scenarios, underscoring the importance of balanced learning opportunities during training. Based on this study, all components are essential to the overall effectiveness of AnywhereDoor.

	\subsection{Resilience Against Defenses}
	\label{sec:defense}

	While defense resilience is not the primary objective of this paper, we surprisingly found that AnywhereDoor is resilient against some defenses. Table~\ref{tab:defense} evaluates AnywhereDoor's robustness against seven methods, including input-based and model-based mitigation.

	\begin{table}[t]\small\setlength{\tabcolsep}{0.34em}
		\centering
			\begin{tabular}{clcccccc}
				\toprule
				\multicolumn{2}{c}{\multirow{3}{*}{\textbf{\raisebox{-1\height}{\begin{tabular}[c]{@{}c@{}}Mitigation-based\\ Defense Methods\end{tabular}}}}} & \multirow{3}{*}{\raisebox{-1\height}{\textbf{\begin{tabular}[c]{@{}c@{}}Clean\\ mAP\end{tabular}}}}& \multicolumn{5}{c}{\textbf{Attack Success Rate}}                                                               \\ \cmidrule(l){4-8} 
				\multicolumn{2}{c}{}                                                                                                     &                                     & \multicolumn{2}{c}{\textbf{Removal}}    & \multicolumn{2}{c}{\textbf{Miscls.}} & \textbf{Gen.} \\\cmidrule(l){8-8}  \cmidrule(l){6-7}  \cmidrule(l){4-5}  
				\multicolumn{2}{c}{}                                                                                                     &                                     & \textbf{Untar.} & \textbf{Tar.} & \textbf{Untar.}     & \textbf{Tar.}    & \textbf{Untar.} \\ 
				
				\midrule
				\multicolumn{2}{c}{\makecell{No Defense}} & 53.1 & 97.5 & 86.2 & 97.8 & 80.6 & 88.8 \\
				\midrule
				\multirow{4}{*}{(a)} &  JPEG Compr. & 53.0 & 97.7 & 87.4 & 98.2 & 79.6 & 89.4 \\
				& Mean Filter & 52.5 & 99.7 & 91.0 & 98.4 & 81.0 & 89.2 \\
				& Median Filter & 52.8 & 99.7 & 92.5 & 98.8 & 82.2 & 89.8 \\
				& NEO & 52.6& 97.5 & 85.6 & 97.8 & 79.1 & 88.8 \\
				\midrule
				\multirow{3}{*}{(b)} & Fine-tuning & 53.6 & 95.5 & 76.5 & 89.3 & 69.6 & 82.8 \\
				& Pruning & 18.3 & 61.1 & 61.8 & 90.0 & 9.1 & 13.1 \\
				& Fine-pruning & 18.6 & 61.1 & 62.2 & 90.0 & 9.0 & 13.1 \\
				\bottomrule
			\end{tabular}
		\caption{While not being specifically designed for, AnywhereDoor demonstrates resilience against some defenses.}
		\label{tab:defense}
	\end{table}

	\noindent\textbf{Input-based Mitigation.}
	Input sanitization methods, such as JPEG compression~\cite{das2018shield}, mean filter, median filter~\cite{xu2017feature}, and NEO~\cite{udeshi2022model}, aim to suppress triggers by modifying the input. As shown in Table~\ref{tab:defense}a, these methods preserve clean mAP well but fail to eliminate the backdoor, as indicated by the ASR, which remains high or even gets increased in some cases. This suggests that AnywhereDoor’s trigger patterns are insensitive to minor input distortions, making these defenses ineffective.

	\noindent\textbf{Model-based Mitigation.} Model sanitization techniques, including fine-tuning, pruning, and fine-pruning~\cite{liu2018fine}, are alternative approaches to counter backdoor attacks. Table~\ref{tab:defense}b shows that fine-tuning reduces ASR in targeted misclassification but leaves ASR in most other scenarios  above $82.8\%$. Pruning and fine-pruning significantly lower ASR but also drastically reduce clean mAP to $18.3$, compromising model utility. This trade-off highlights that model-based defenses partially counteract backdoors but at a high cost to performance.

	\section{Conclusion}
	We presented AnywhereDoor, a backdoor attack that enables multi-target manipulation of object detection models, offering unprecedented flexibility beyond prior attacks that rely on a small number of predefined triggers and behaviors. By jointly training a trigger generator with the victim model and employing dynamic sample poisoning, AnywhereDoor overcomes challenges specific to object detection through three key techniques: objective disentanglement, trigger mosaicking, and strategic batching. Extensive experiments demonstrate AnywhereDoor's effectiveness, achieving high ASR while preserving clean model performance and highlighting vulnerabilities in current object detection systems and the growing threat posed by backdoor attacks.
	
	{
		\small
		\bibliographystyle{ieeenat_fullname}

	}

	\clearpage
	\setcounter{page}{1}
	\maketitlesupplementary
	\appendix
	
	We provide the source code of AnywhereDoor at \url{https://github.com/HKU-TASR/AnywhereDoor}. This document provides additional details to support our main paper.

	\section{Transferability Study}  
	
	Although the victim model and the trigger generator are jointly trained, they operate as independent networks. To investigate the transferability of AnywhereDoor's trigger generators, we examine whether a pretrained trigger generator can be leveraged to control other models.
	
	We prepared pretrained trigger generators that were initially trained jointly with Faster R-CNN, DETR, and YOLOv3 on the PASCAL VOC07+12 dataset. These trigger generators were then paired with different models for another training process in which the parameters of trigger generators are frozen, with only models being trained. As shown in Figure~\ref{fig:transferability}, subfigures (a), (b), and (c) display results for Faster R-CNN, DETR, and YOLOv3, respectively, each trained with a trigger generator originally paired with the same model, as well as with those from two other models. Each bar chart reports performance metrics of Clean mAP and ASR of five attack scenarios. Overall, both Clean mAP and ASR show minimal degradation when using different trigger generators, indicating that the attacks remain effective across model combinations. Notable ASR declines are observed primarily in the two targeted attack scenarios, highlighting their higher difficulty and lower ASR stability. The combination of Faster R-CNN and DETR maintains consistent performance, while YOLOv3 experiences more significant ASR drops when paired with the other two models. This suggests structural differences in YOLOv3 that impact its backdoor attack performance, aligning with the results presented in Section~\ref{sec:exp-quant} of the main paper.
	
	These results show the trigger generator's ability to interpret an attacker's intent and the generalization ability of its generated triggers. Such transferability makes it possible for AnywhereDoor to launch effective attacks without prior knowledge of the victim model.
	
	\begin{figure}[t]\setcounter{figure}{9}
		\centering
		\captionsetup[subfigure]{labelformat=empty}
		\subfloat[(a) Faster R-CNN trained with different trigger generators.]{\includegraphics[width=\linewidth]{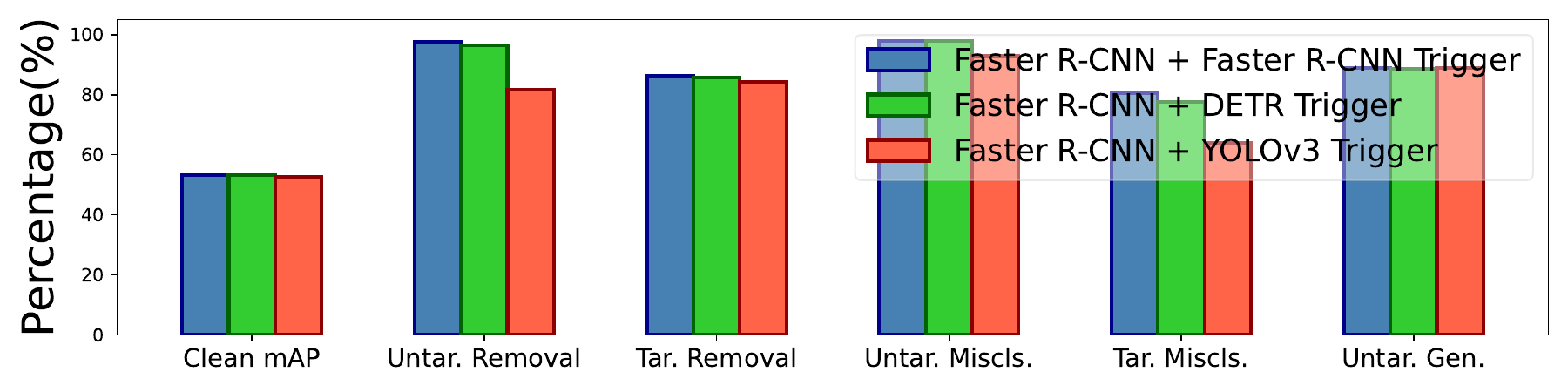}}\\
		\subfloat[(b) DETR trained with different trigger generators.]{\includegraphics[width=\linewidth]{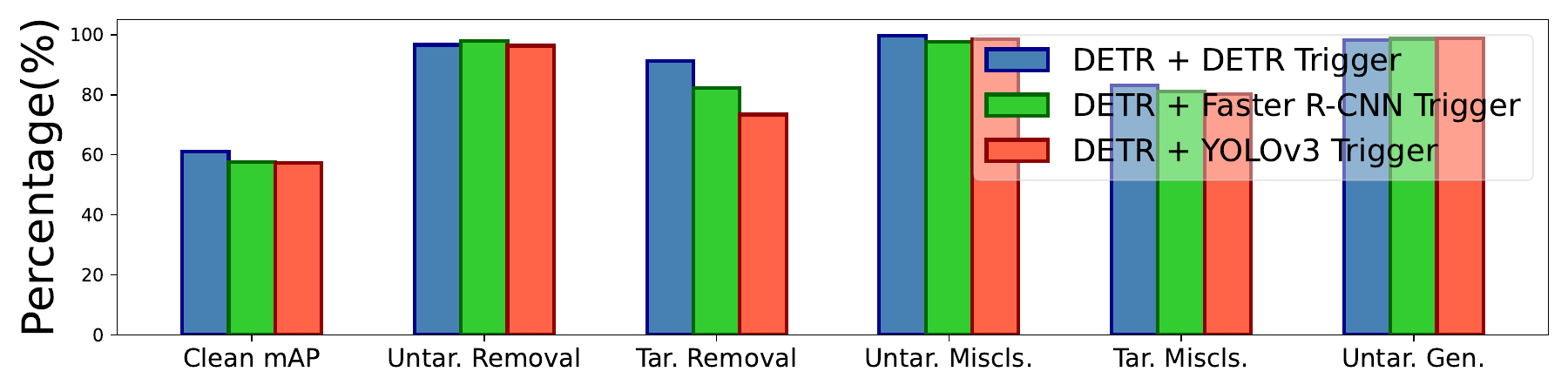}}\\
		\subfloat[(c) YOLOv3 trained with different trigger generators.]{\includegraphics[width=\linewidth]{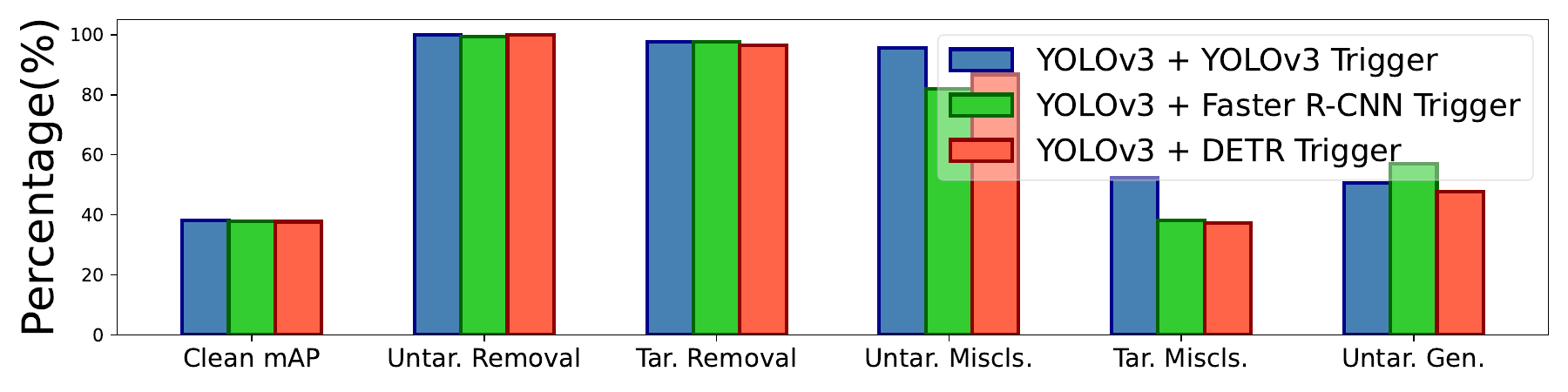}}\\
		\caption{Transferability evaluation of trigger generators. The trigger generators jointly pretrained with a specific model is used for backdoor training of other models, in which only the parameters of the model are updated. Notably, although both Clean mAP and ASR metrics are plotted ranging from 0 to 100, Clean mAP represents the average precision of the detection results, whereas ASR quantifies the proportion of successfully manipulated objects.}
		\label{fig:transferability}
	\end{figure}
	
	\section{Attack Success Rate Calculation}\label{supp:asr}
	
	The attack success rate (ASR) quantifies the effectiveness of a backdoor attack by measuring how well the model's behavior aligns with the attacker's intent when a trigger is present. Specifically, ASR reflects the proportion of manipulated bounding boxes that successfully exhibit the desired malicious behavior, providing a comprehensive evaluation of the attack's impact across different scenarios. Each attack scenario in our study has distinct objectives and corresponding annotation modification strategies. Consequently, the ASR calculation varies across scenarios to evaluate the attack’s effectiveness in achieving its unique malicious goals. 
	
	We employ a unified framework for ASR calculation. For every sample in the validation set, we compare the clean prediction (the model’s output without the trigger) and the dirty prediction (the model’s output when the trigger is present). Across all samples, we accumulate the number of successfully manipulated bounding boxes (\( S \)) and the total number of targeted bounding boxes (\( T \)). The ASR is then computed as \( \text{ASR} = \frac{S}{T} \). Before performing the calculations, all predictions undergo filtering based on a confidence score threshold (\( \tau = 0.3 \)), which removes low-confidence bounding boxes and their associated labels.
	
	The untargeted removal scenario (Algorithm~\ref{alg:untargeted_removal}) focuses on eliminating bounding boxes from the clean predictions, accumulating the total number of objects (\( t_i \)) and the count of successfully removed ones (\( s_i \)). In the targeted removal scenario (Algorithm~\ref{alg:targeted_removal}), an additional parameter, the victim class \( C_v \), is provided to specify which class is to be removed. Here, \( t_i \) includes only the bounding boxes of the victim class in the clean predictions, and \( s_i \) represents those successfully removed. An IoU threshold of $0.5$ is used to identify if two bounding boxes refer to the same object. For untargeted misclassification (Algorithm~\ref{alg:untargeted_misclassification}), the aim is to alter the class labels of bounding boxes. We accumulate \( t_i \) as the total number of bounding boxes in the clean predictions and \( s_i \) as the count of bounding boxes whose classes are successfully changed. A successful misclassification occurs when a bounding box in the dirty prediction has a different class from the corresponding one in the clean prediction, provided it is not derived from another misclassified bounding box in the clean predictions. In targeted misclassification (Algorithm~\ref{alg:targeted_misclassification}), the attack targets a specific victim class \( C_v \), aiming to misclassify its instances as the target class \( C_t \). The total count \( t_i \) is the number of victim-class bounding boxes in the clean predictions, while \( s_i \) counts those successfully changed to \( C_t \) in the dirty predictions. Success is determined by the presence of a corresponding bounding box with class \( C_t \) in the dirty predictions and an IoU exceeding $0.5$. Finally, the ASR of untargeted generation (Algorithm~\ref{alg:untargeted_generation}) is calculated in a sample-based way, evaluating whether new bounding boxes are generated . Each sample contributes \( t_i = 1 \), and \( s_i = 1 \) if the number of bounding boxes in the dirty predictions exceeds those in the clean predictions.

	\begin{algorithm}[t]
		\caption{ASR Calculation for Untar. Removal}
		\label{alg:untargeted_removal}
		\KwIn{predictions $P = \{P_{clean}^i, P_{dirty}^i \mid i = 1, 2, \ldots, n\}$}
		\KwOut{attack success rate $ASR$}
		$S \gets 0$, $T \gets 0$ \tcp*[r]{Initialize counts}
		\For{$i \gets 1$ \KwTo $n$}{
			\tcp{Total bboxes in clean pred}
			$t_i \gets |P_{clean}^i|$ \;
			\tcp{Successful removals}
			$s_i \gets \max(t_i - |P_{dirty}^i|, 0)$ \;
			\tcp{Accumulate counts}
			$S, T \gets S + s_i, T + t_i$ \;
		}
		$ASR \gets \frac{S}{T}$ \tcp*[r]{Overall ASR}
		\Return $ASR$
	\end{algorithm}
	
	\begin{algorithm}[t]
		\caption{ASR Calculation for Tar. Removal}
		\label{alg:targeted_removal}
		\KwIn{predictions $P = \{P_{clean}^i, P_{dirty}^i \mid i = 1, 2, \ldots, n\}$, victim class $C_v$}
		\KwOut{attack success rate $ASR$}
		$S \gets 0$, $T \gets 0$ \tcp*[r]{Initialize counts}
		\For{$i \gets 1$ \KwTo $n$}{
			\tcp{Total victim-class bboxes}
			$t_i \gets \sum_{c \in P_{clean}^i.classes} \mathbbm{1}(c = C_v)$ \;
			\tcp{Successful removals}
			$s_i \gets 0$ \;
			\ForEach{bbox $b_c \in P_{clean}^i$ with class $C_v$}{
				$is\_success \gets \text{True}$\;
				\ForEach{bbox $b_d \in P_{dirty}^i$ with class $C_v$}{
					\If{$\text{IoU}(b_c, b_d) > 0.5$}{
						$is\_success \gets \text{False}$ \;
						break \;
					}
				}
				$s_i \gets s_i + \mathbbm{1}(is\_success)$ \;
			}
			\tcp{Accumulate counts}
			$S, T \gets S + s_i, T + t_i$ \;
		}
		$ASR \gets \frac{S}{T}$ \tcp*[r]{Overall ASR}
		\Return $ASR$
	\end{algorithm}
	
	\begin{algorithm}
		\caption{ASR Calculation for Untar. Miscls.}
		\label{alg:untargeted_misclassification}
		\KwIn{predictions $P = \{P_{clean}^i, P_{dirty}^i \mid i = 1, 2, \ldots, n\}$}
		\KwOut{Attack success rate $ASR$}
		$S \gets 0$, $T \gets 0$ \tcp*[r]{Initialize counts}
		\For{$i \gets 1$ \KwTo $n$}{
			\tcp{Total bboxes in clean pred}
			$t_i \gets |P_{clean}^i|$ \;
			\tcp{Successful misclassifications}
			$s_i \gets 0$ \;
			\ForEach{bbox $b_c \in P_{clean}^i$}{
				$is\_success \gets \text{True}$ \;
				\If{$\exists b_d \in P_{dirty}^i$ with the same class as $b_c$ and $\text{IoU}(b_c, b_d) > 0.5$ }{
					$is\_success \gets \text{False}$ \;
					\If{such $b_d$ is derived from another bbox in $P_{clean}^i$ that was misclassified}{
						$is\_success \gets \text{True}$ \;
					}
				}
				$s_i \gets s_i + \mathbbm{1}(is\_success)$ \;
			}
			\tcp{Accumulate counts}
			$S, T \gets S + s_i, T + t_i$
		}
		$ASR \gets \frac{S}{T}$ \tcp*[r]{Overall ASR}
		\Return $ASR$
	\end{algorithm}
	
	\begin{algorithm}
		\caption{ASR Calculation for Tar. Miscls.}
		\label{alg:targeted_misclassification}
		\KwIn{predictions $P = \{P_{clean}^i, P_{dirty}^i \mid i = 1, 2, \ldots, n\}$, victim class $C_v$, target class $C_t$}
		\KwOut{attack success rate $ASR$}
		$S \gets 0$, $T \gets 0$ \tcp*[r]{Initialize counts}
		\For{$i \gets 1$ \KwTo $n$}{
			\tcp{Total victim-class bboxes}
			$t_i \gets \sum_{c \in P_{clean}^i.classes} \mathbbm{1}(c = C_v)$ \;
			\tcp{Successful misclassifications}
			$s_i \gets 0$ \;
			\ForEach{bbox $b_c \in P_{clean}^i$ with class $C_v$}{
				$is\_success \gets \text{False}$ \;
				\ForEach{bbox $b_d \in P_{dirty}^i$ with class $C_t$}{
					\If{$\text{IoU}(b_c, b_d) > 0.5$}{
						$is\_success \gets \text{True}$ \;
					}
				}
				$s_i \gets s_i + \mathbbm{1}(is\_success)$ \;
			}
			\tcp{Accumulate counts}
			$S, T \gets S + s_i, T + t_i$ \;
		}
		$ASR \gets \frac{S}{T}$ \tcp*[r]{Overall ASR}
		\Return $ASR$
	\end{algorithm}
	
	\begin{algorithm}
		\caption{ASR Calculation for Untar. Gen.}
		\label{alg:untargeted_generation}
		\KwIn{predictions $P = \{P_{clean}^i, P_{dirty}^i \mid i = 1, 2, \ldots, n\}$}
		\KwOut{attack success rate $ASR$}
		$S \gets 0$, $T \gets 0$ \tcp*[r]{Initialize counts}
		\For{$i \gets 1$ \KwTo $n$}{
			\tcp{Sample-based counts}
			$t_i \gets 1$ \;
			\tcp{Successful generations}
			\eIf{$|P_{dirty}^i| > |P_{clean}^i| $}{
				$s_i \gets 1$ \;
			}{
				$s_i \gets 0$ \;
			}
			\tcp{Accumulate counts}
			$S, T \gets S + s_i, T + t_i$ \;
		}
		$ASR \gets \frac{S}{T}$ \tcp*[r]{Overall ASR}
		\Return $ASR$
	\end{algorithm}
	
	\section{Visual Samples} 
	
	We provide additional visual examples to illustrate the effectiveness of AnywhereDoor across various attack scenarios. For the convenience of viewing, we select some examples with small number of classes and objects and large object size from both PASCAL VOC07+12 and MSCOCO, and use Faster R-CNN trained on PASCAL VOC07+12 to obtain visualization results on five attack scenarios, as shown in Table~\ref{tab:visualization_samples}.
	
	\section{Limitations and Future Works} 
	
	AnywhereDoor has explored the potential of backdoor attacks against object detection, proving its effectiveness and flexibility. Despite the promising results on multiple models, datasets, and attack scenarios, AnywhereDoor is still limited in some ways. First, in some cases, a high retention rate may be required to ensure that the detection of non-target classes under targeted attacks is not affected. Experimental results and observations show that although our method can achieve the manipulation of targeted classes well, it is not perfect in the retention of non-targeted classes. This problem is more likely when the target class is a high frequency class such as person. Second, the attacker's intent may extend beyond the five attack scenarios, such as manipulation of the size and position of ground truth objects, image content-dependent intent, etc. (\eg, making the person next to the car disappear while the one on the street remains).
	
	Future work can build upon the foundational contributions of AnywhereDoor, leveraging its joint-training framework to develop more advanced backdoor attacks and corresponding defense mechanisms. Extensions of this framework may include dynamic and context-aware attack scenarios, where triggers adapt to specific inputs or detection tasks. The proposed techniques offer an identification and preliminary solution of problems that hindered related works, inspiring future research into more sophisticated manipulation.

	\begin{table*}[t]\setcounter{table}{4}
		\centering
		\resizebox{\linewidth}{!}{
			\begin{tabular}{cccccc}
				\toprule
				\textbf{Clean} & \textbf{Untar. Removal} & \textbf{Tar. Removal} & \textbf{Untar. Miscls.} & \textbf{Tar. Miscls.} & \textbf{Untar. Gen.} \\
				\midrule
				\includegraphics[width=0.15\textwidth]{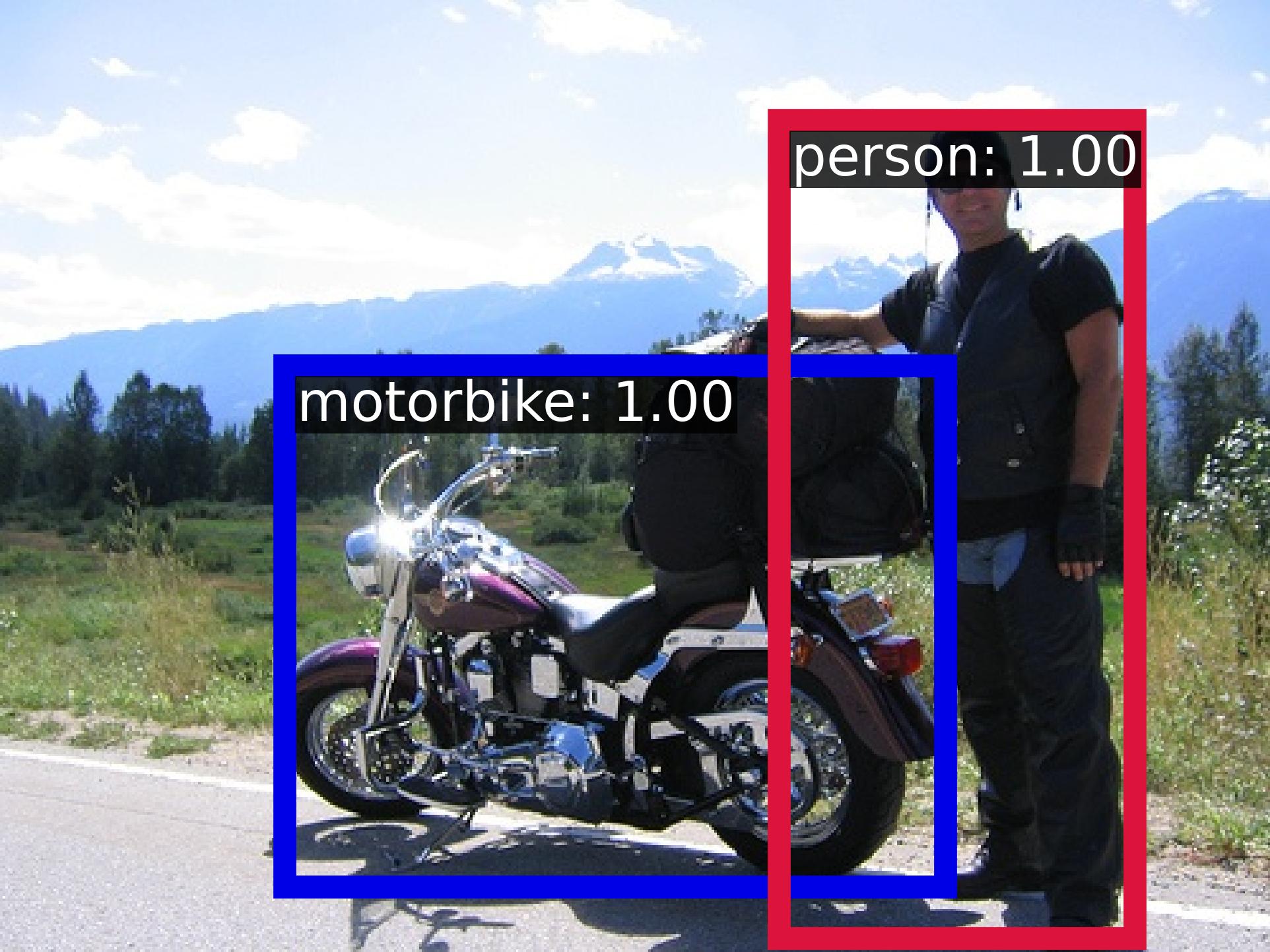} & 
				\includegraphics[width=0.15\textwidth]{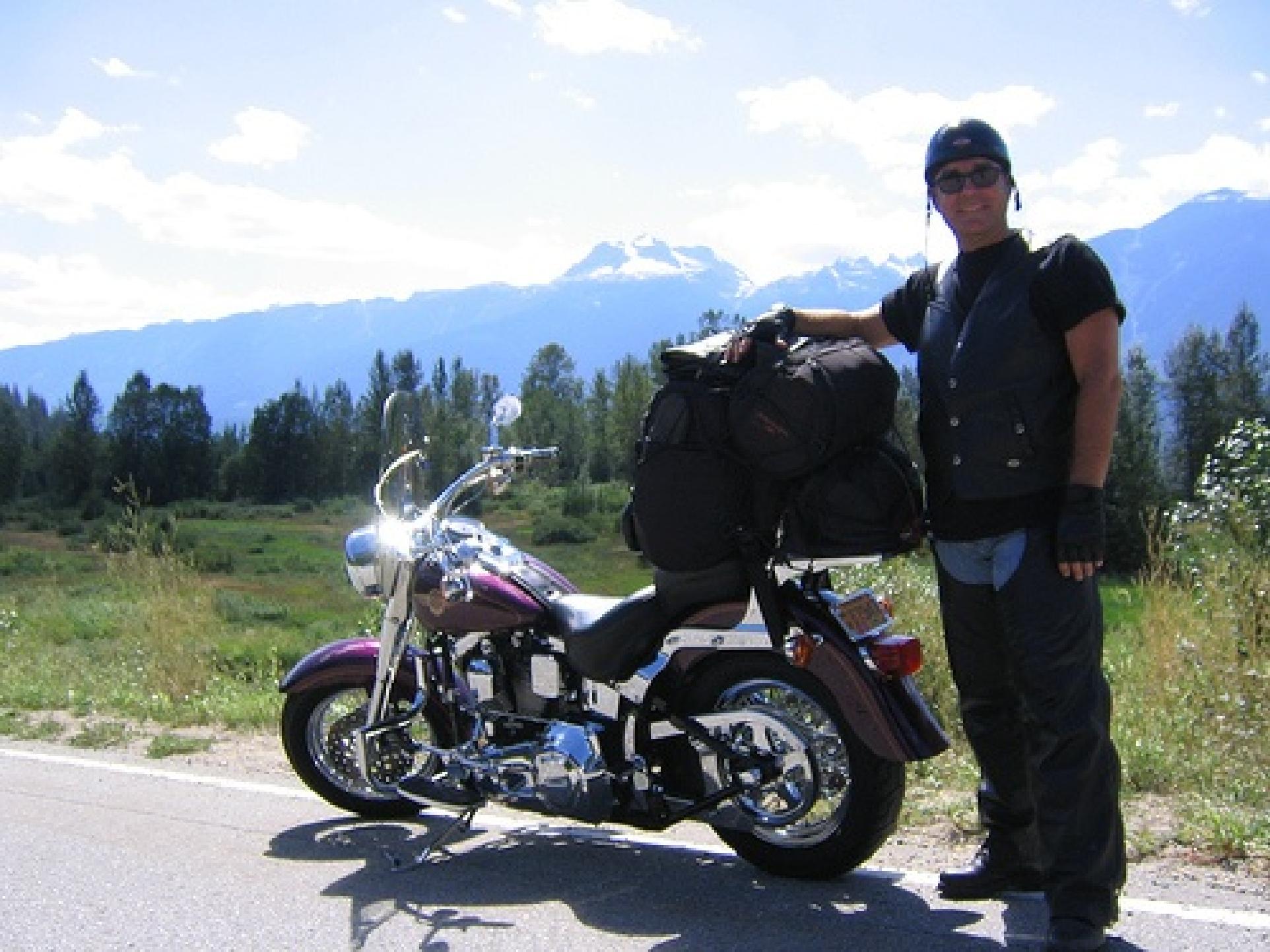} & 
				\includegraphics[width=0.15\textwidth]{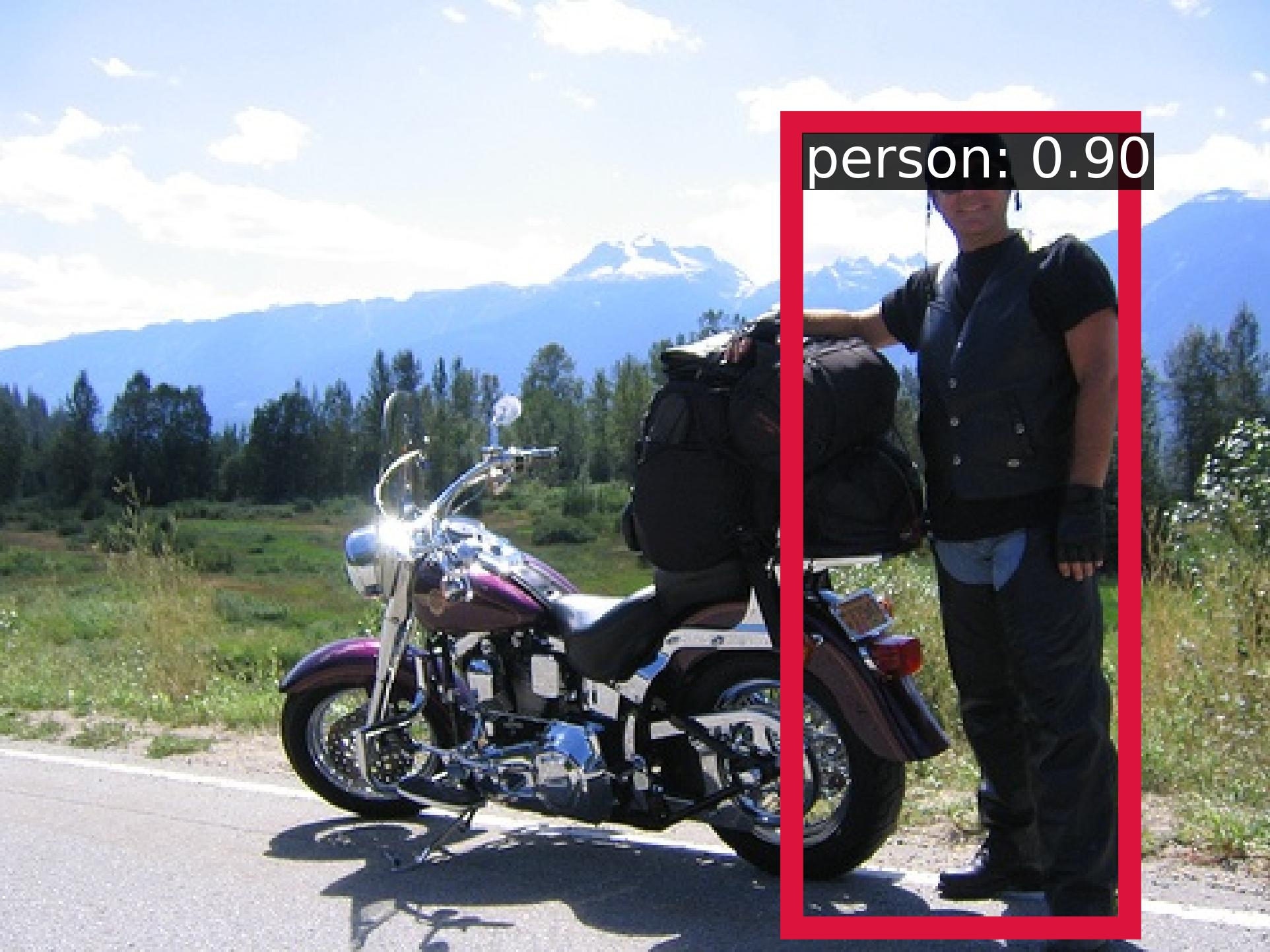} & 
				\includegraphics[width=0.15\textwidth]{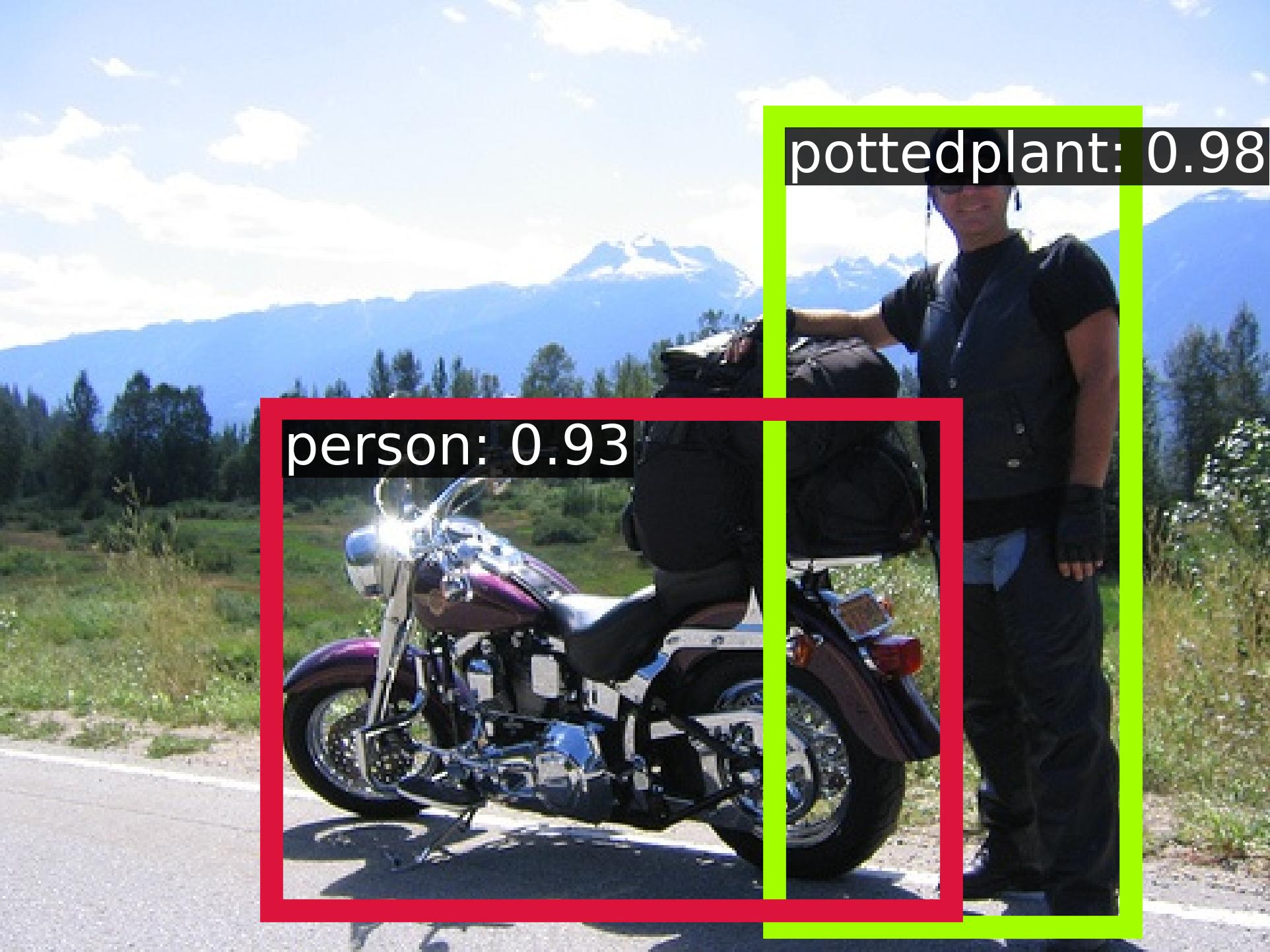} & 
				\includegraphics[width=0.15\textwidth]{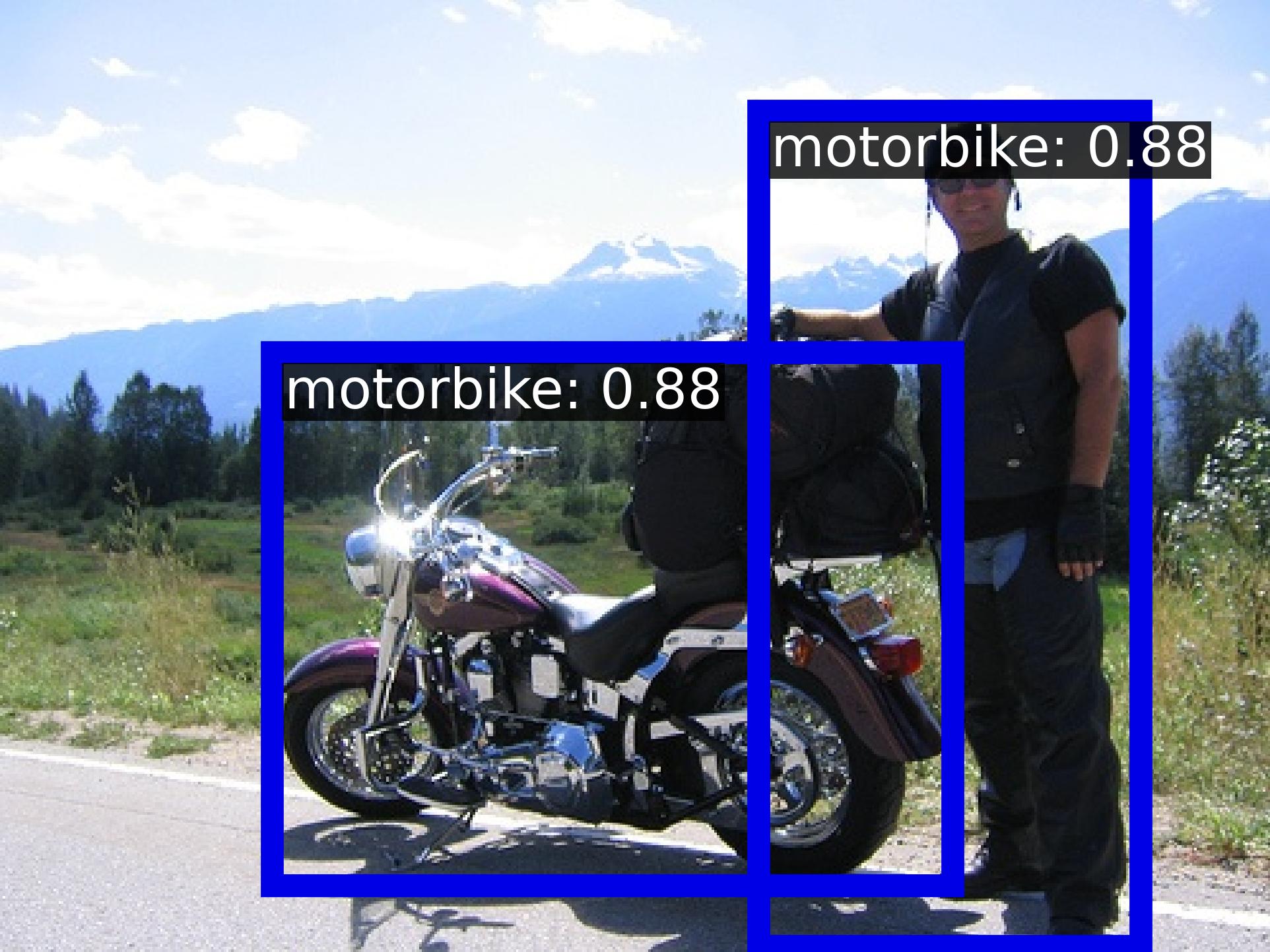} & 
				\includegraphics[width=0.15\textwidth]{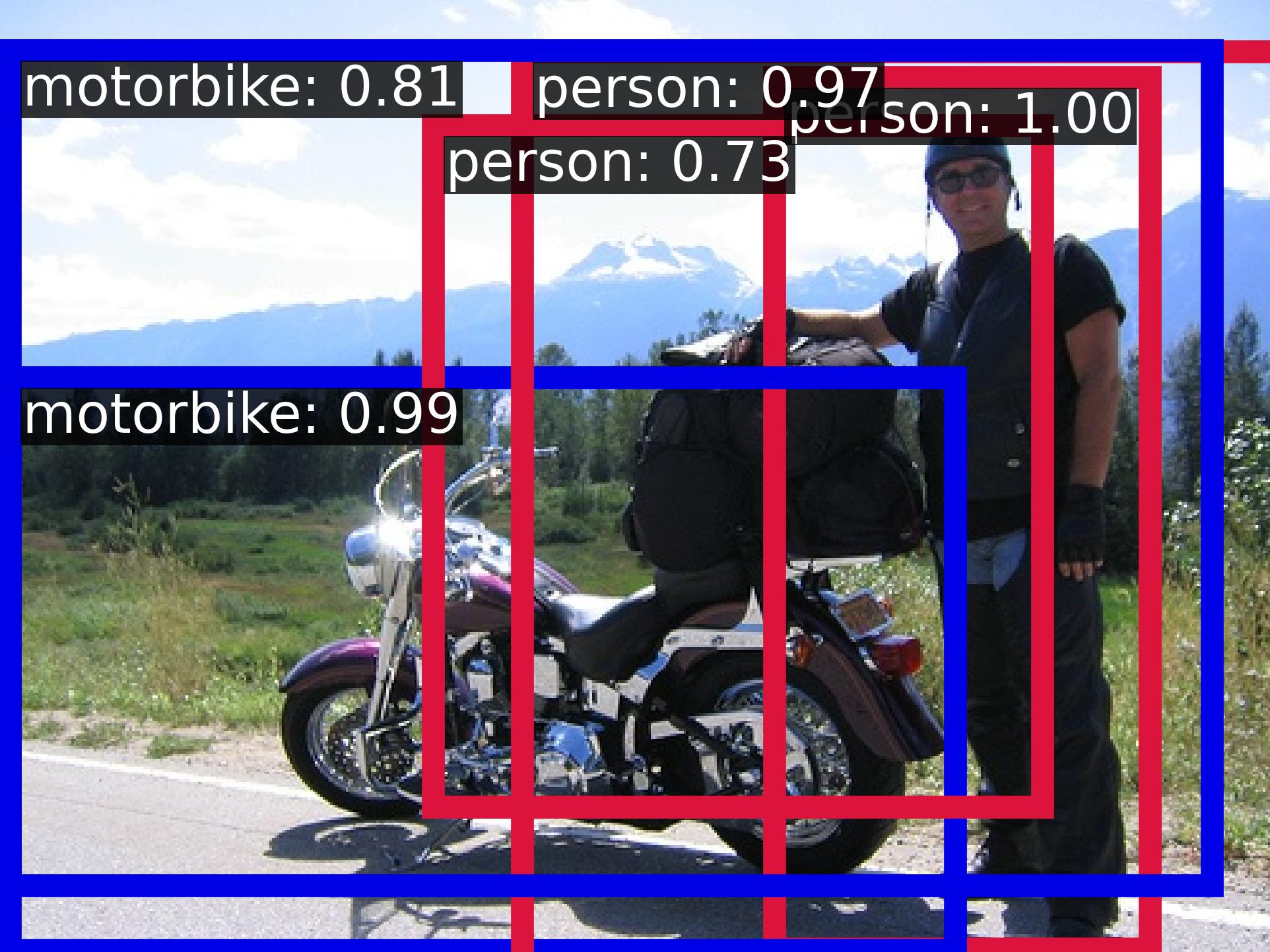} \\
				\midrule
				\includegraphics[width=0.15\textwidth]{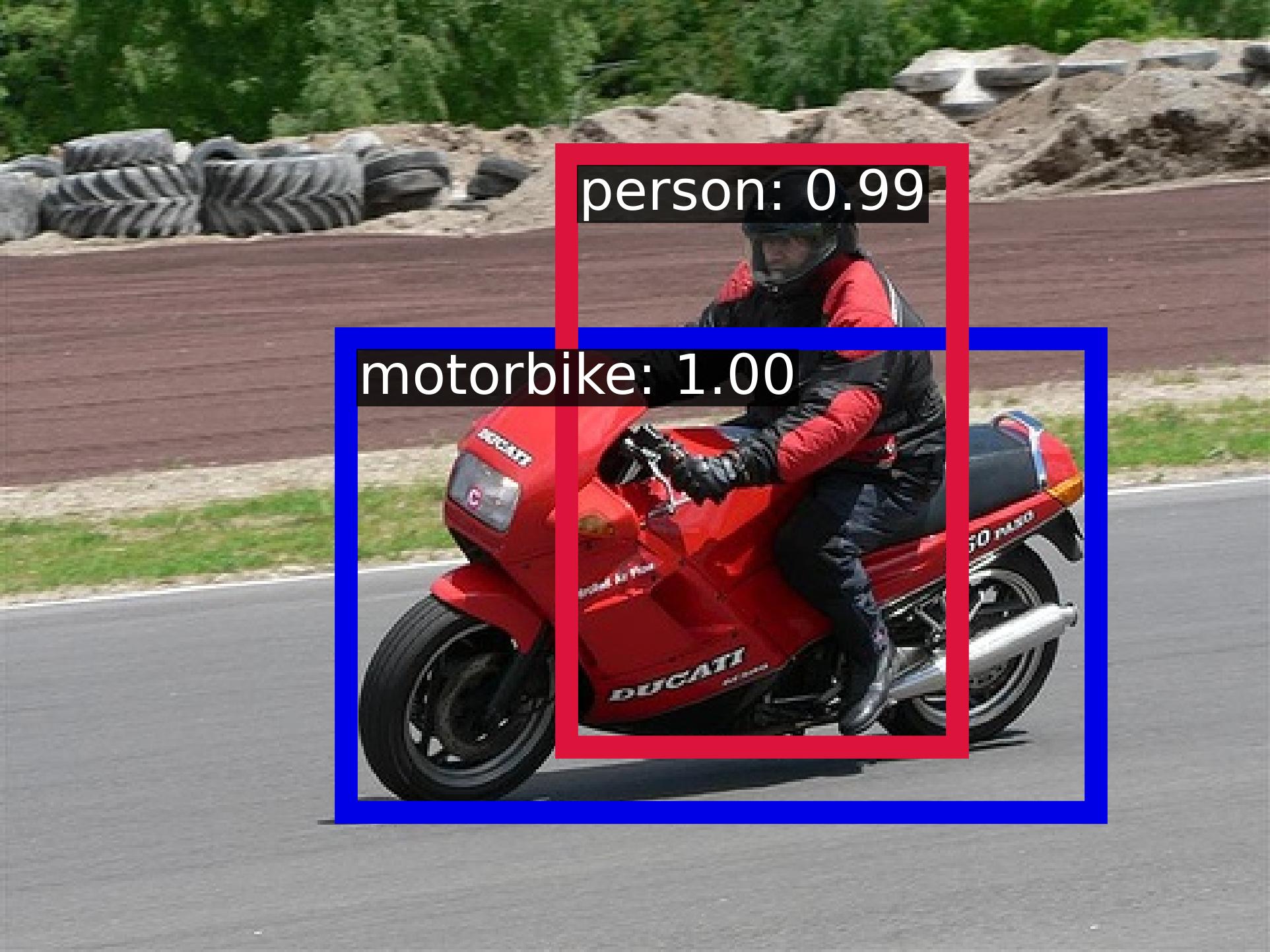} & 
				\includegraphics[width=0.15\textwidth]{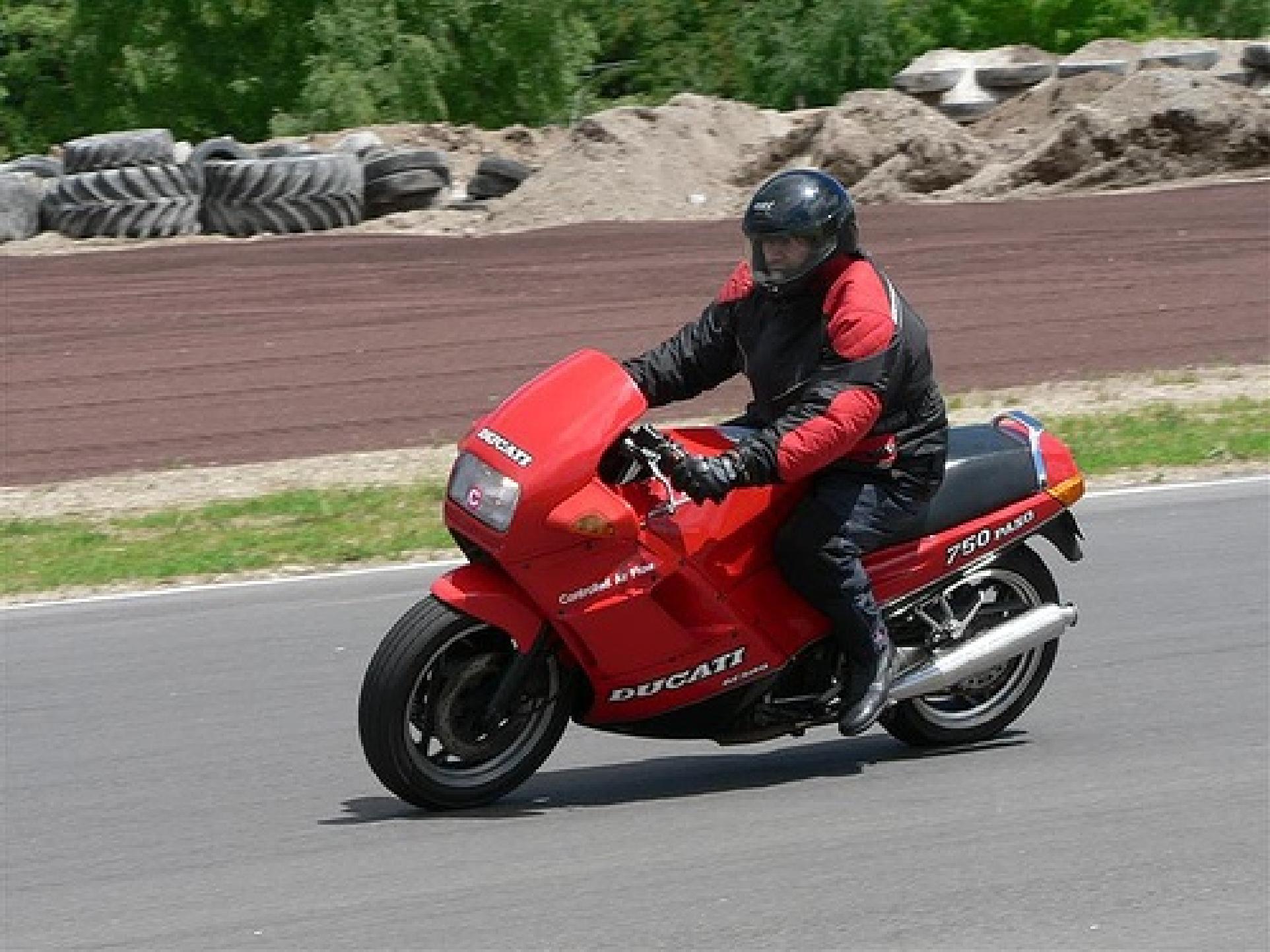} & 
				\includegraphics[width=0.15\textwidth]{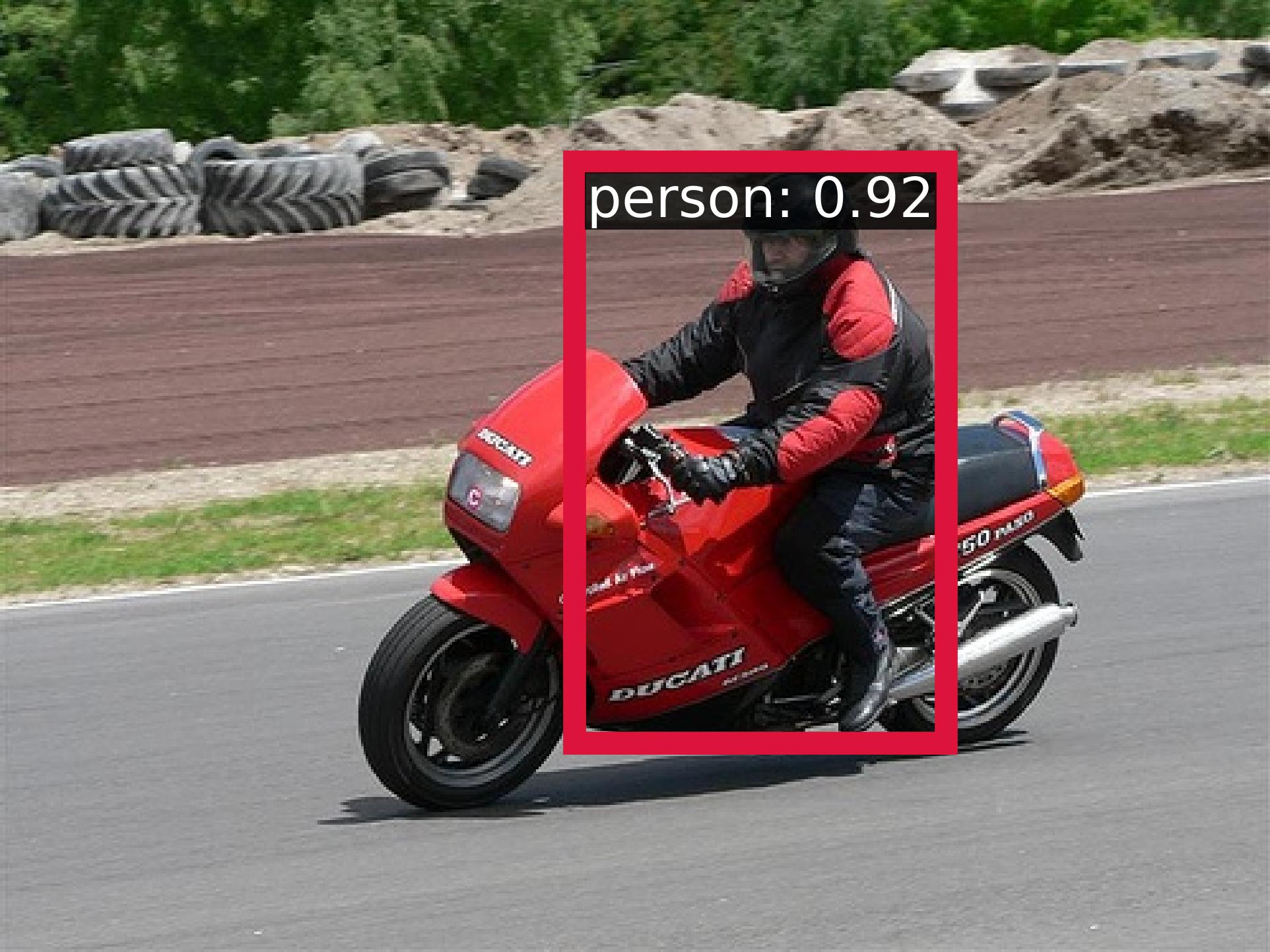} & 
				\includegraphics[width=0.15\textwidth]{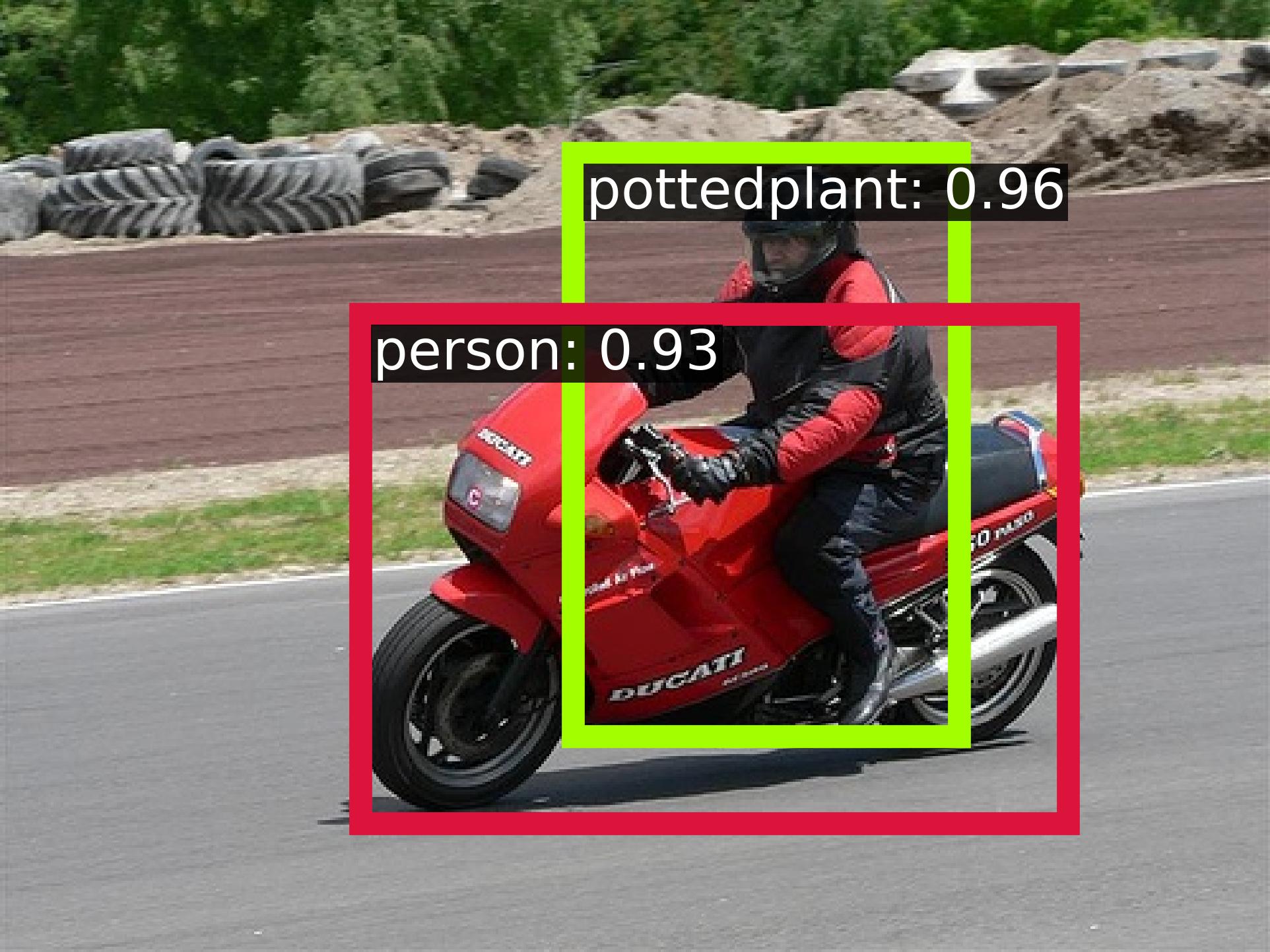} & 
				\includegraphics[width=0.15\textwidth]{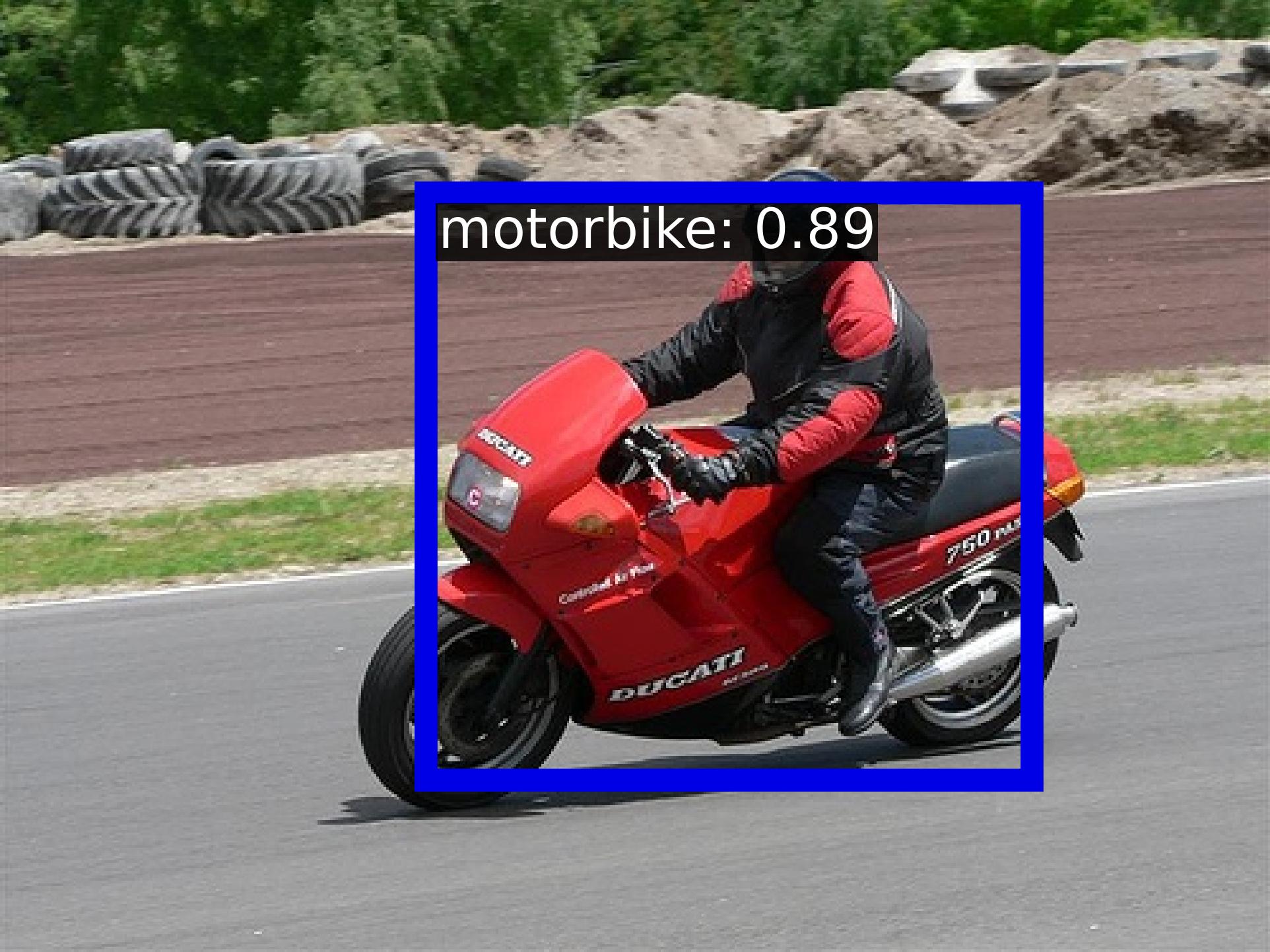} & 
				\includegraphics[width=0.15\textwidth]{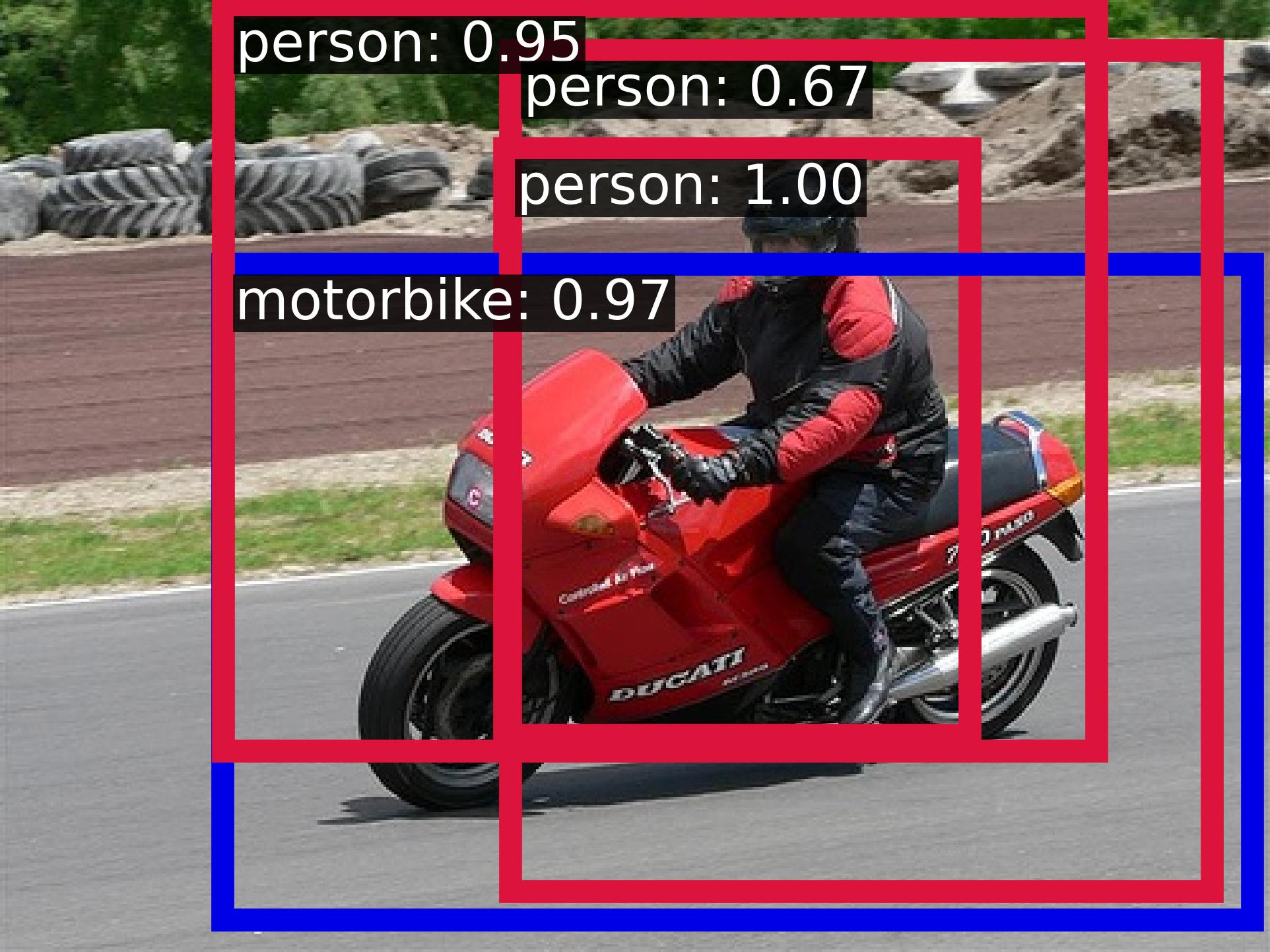} \\
				\midrule
				\includegraphics[width=0.15\textwidth]{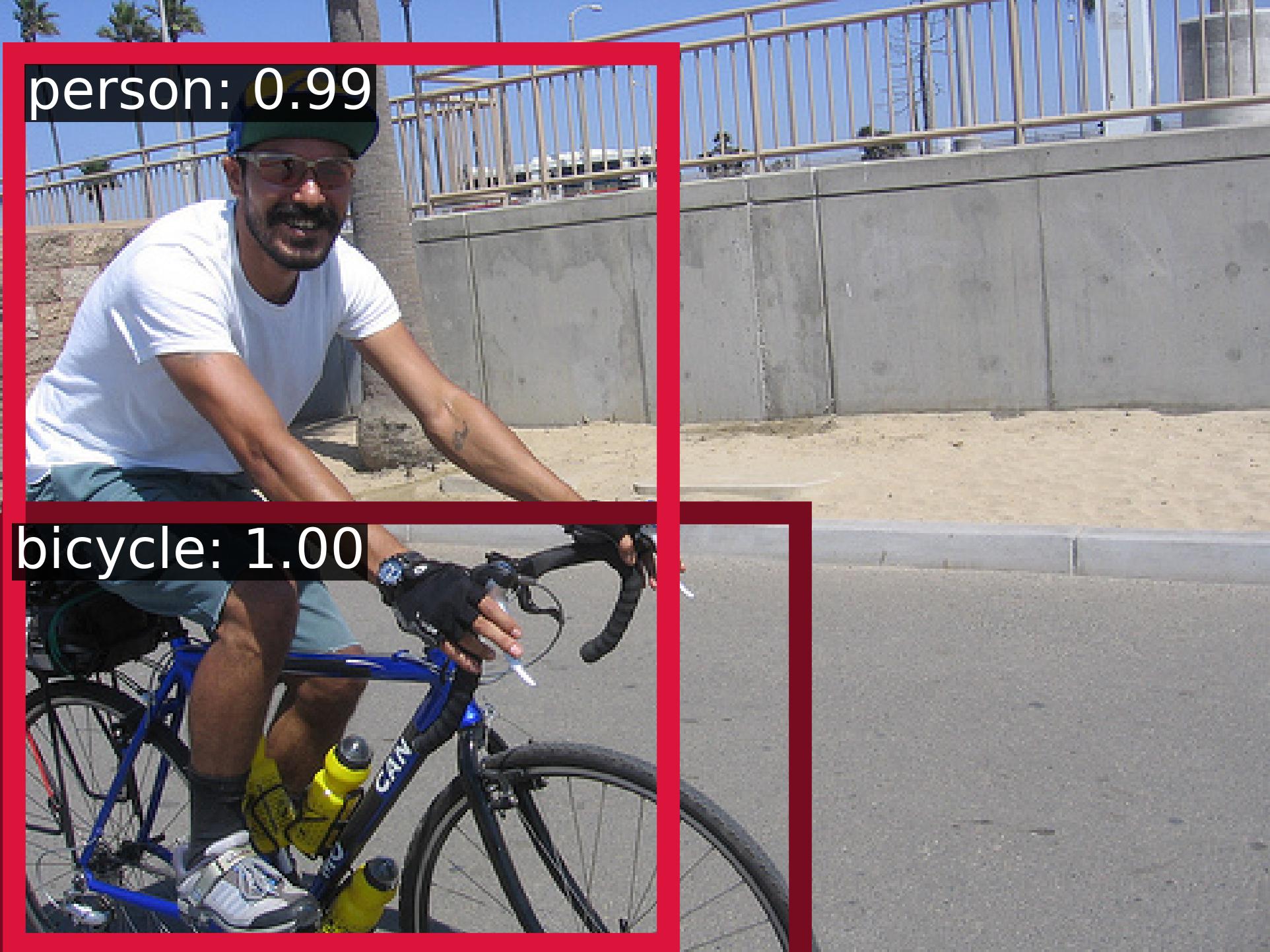} & 
				\includegraphics[width=0.15\textwidth]{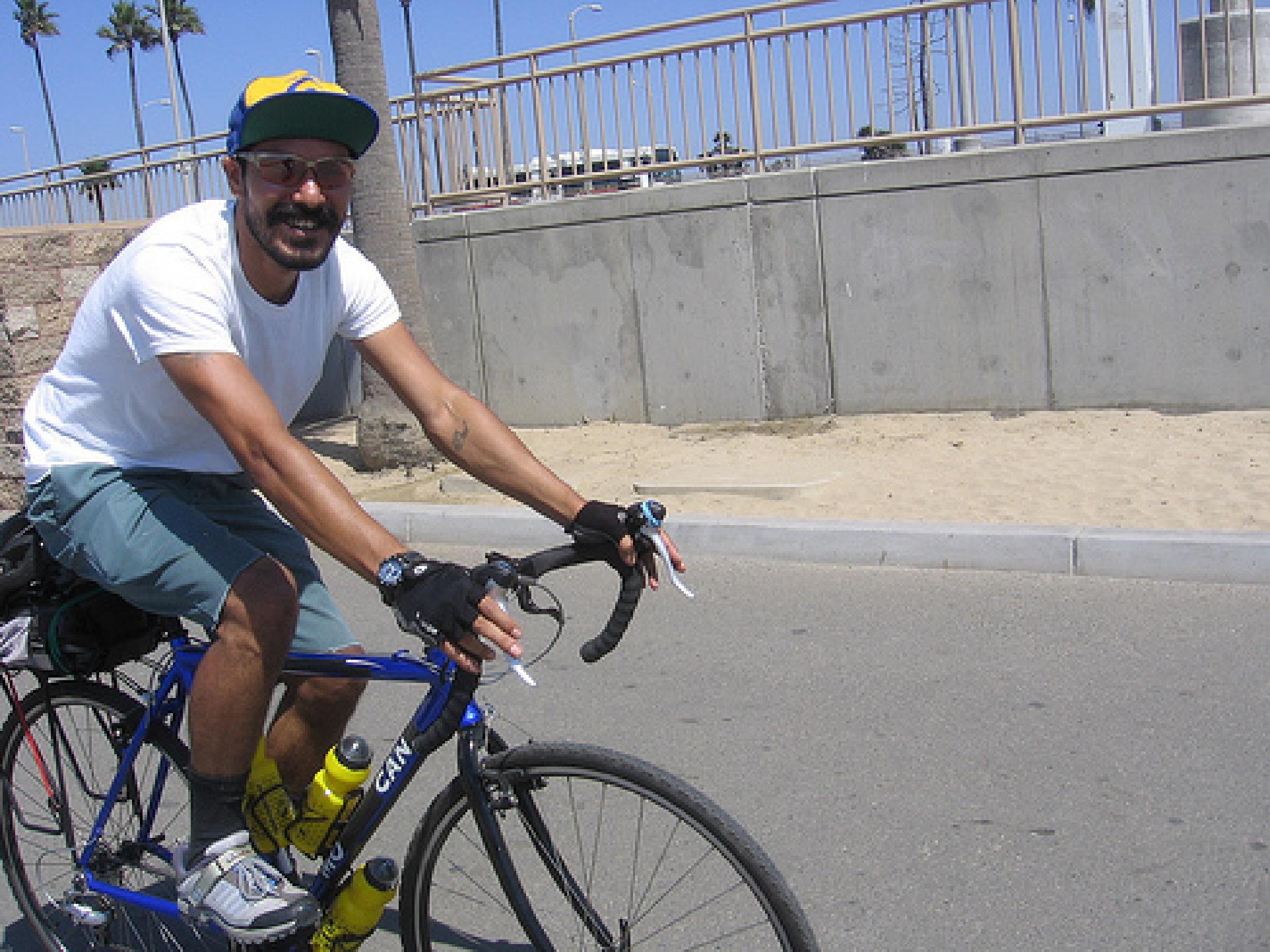} & 
				\includegraphics[width=0.15\textwidth]{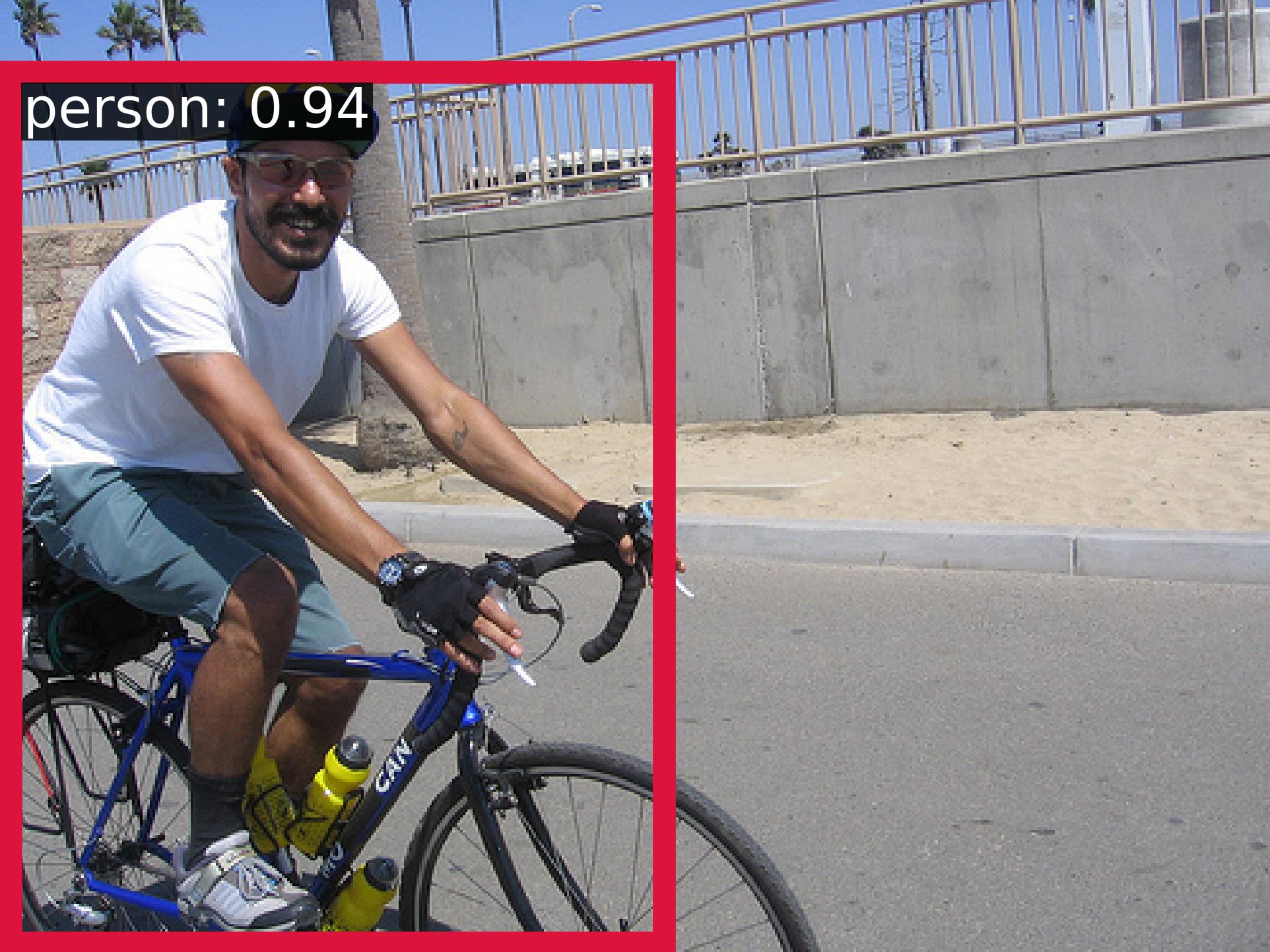} & 
				\includegraphics[width=0.15\textwidth]{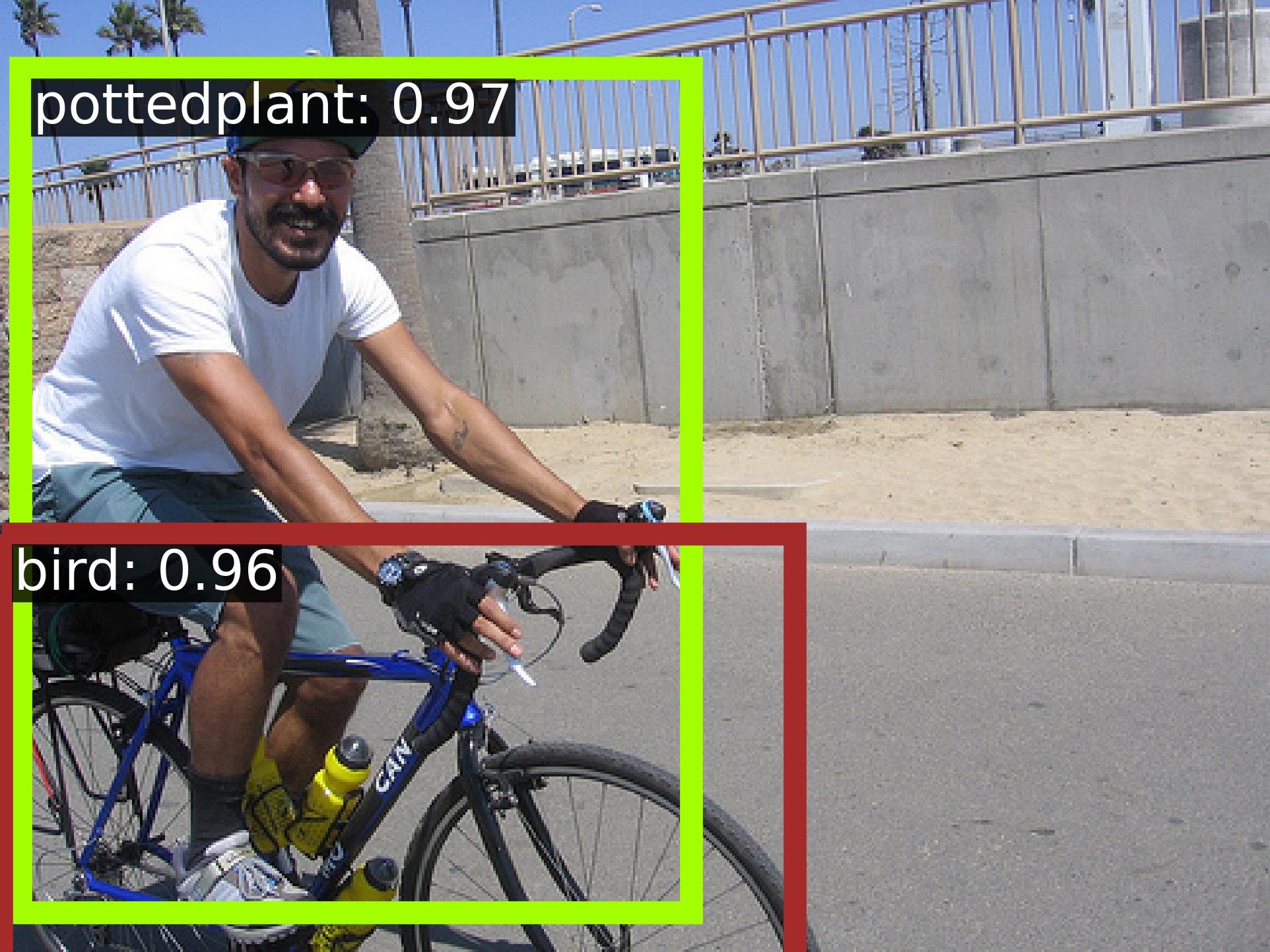} & 
				\includegraphics[width=0.15\textwidth]{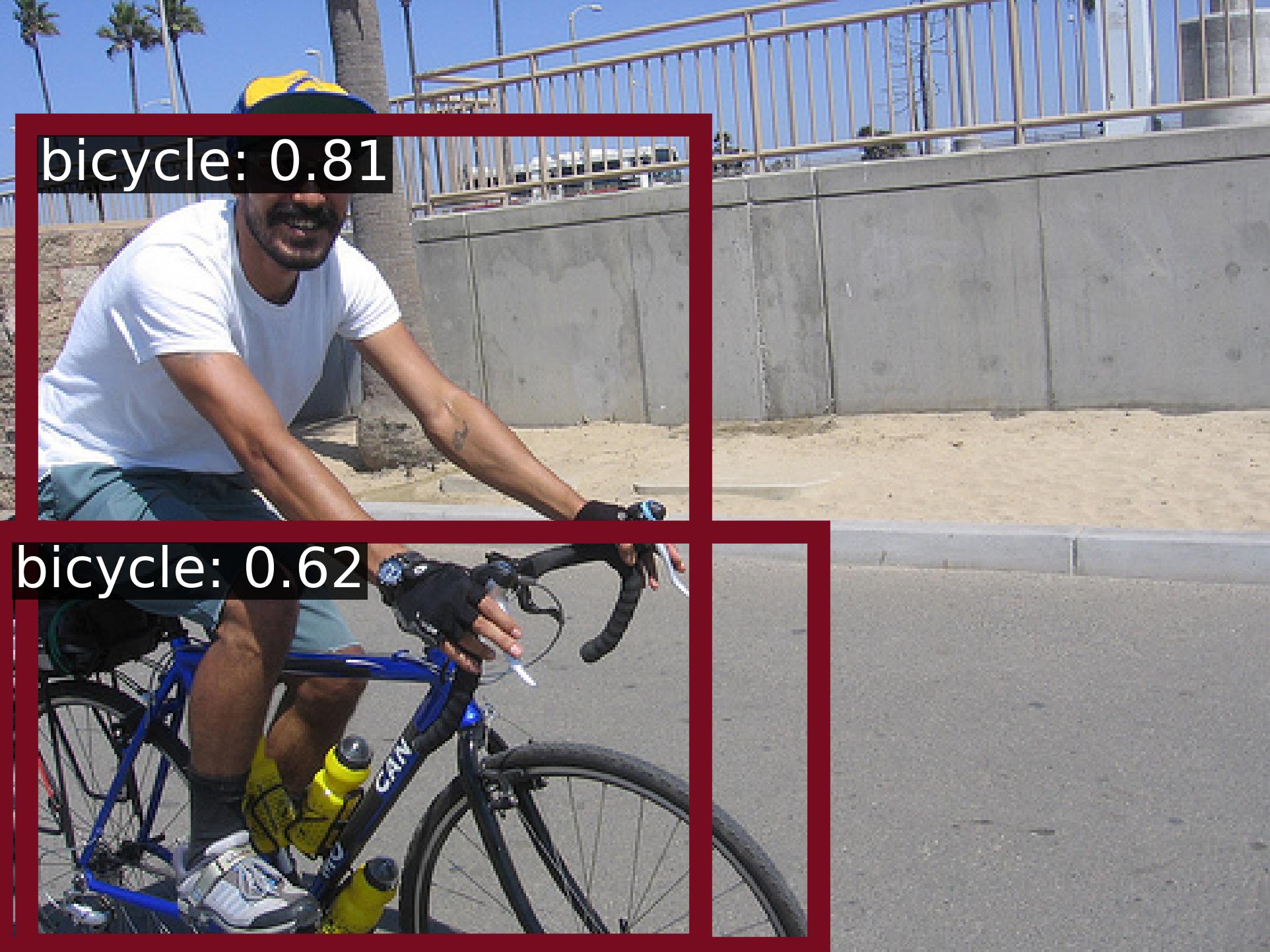} & 
				\includegraphics[width=0.15\textwidth]{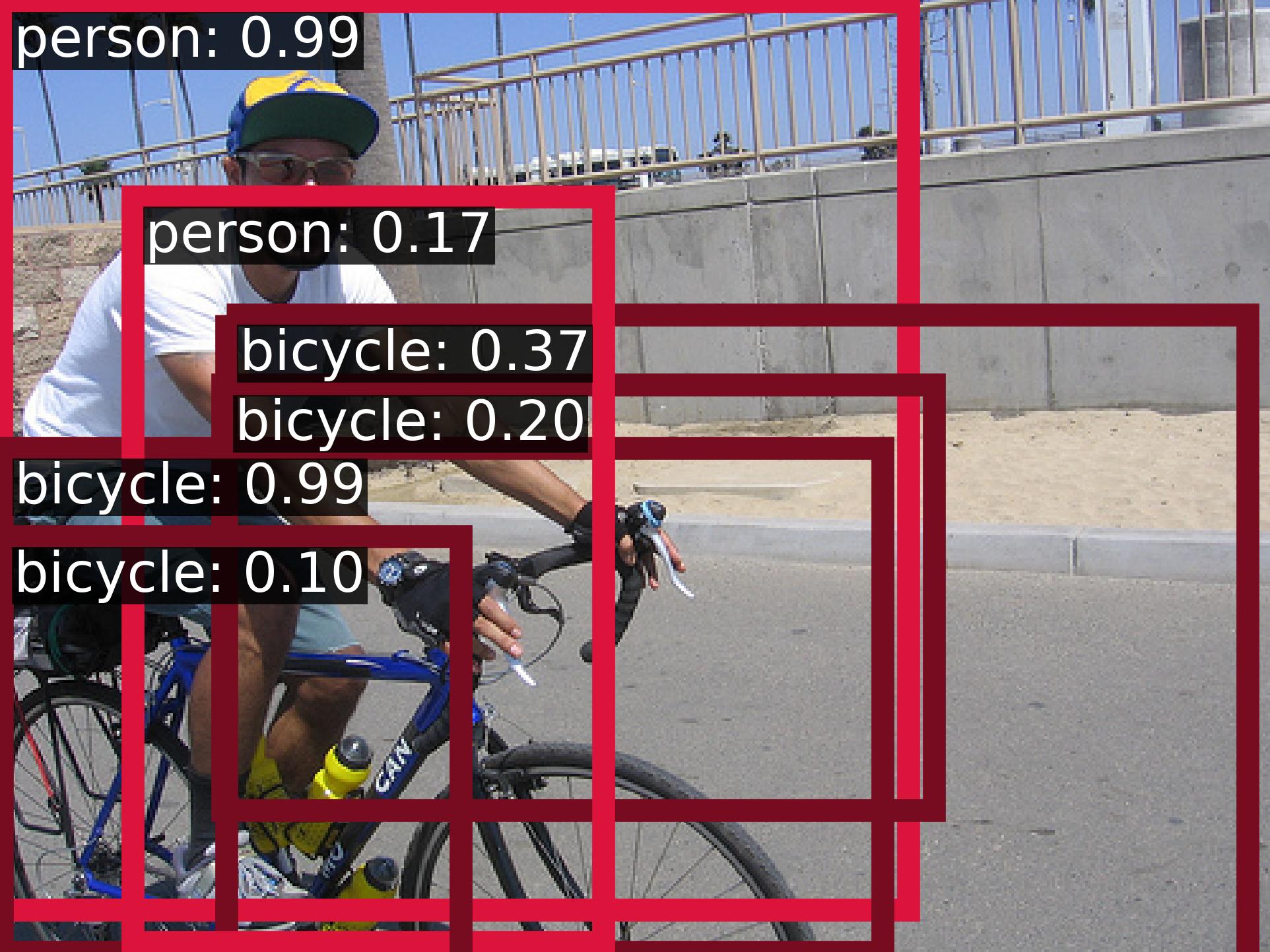} \\
				\midrule
				\includegraphics[width=0.15\textwidth]{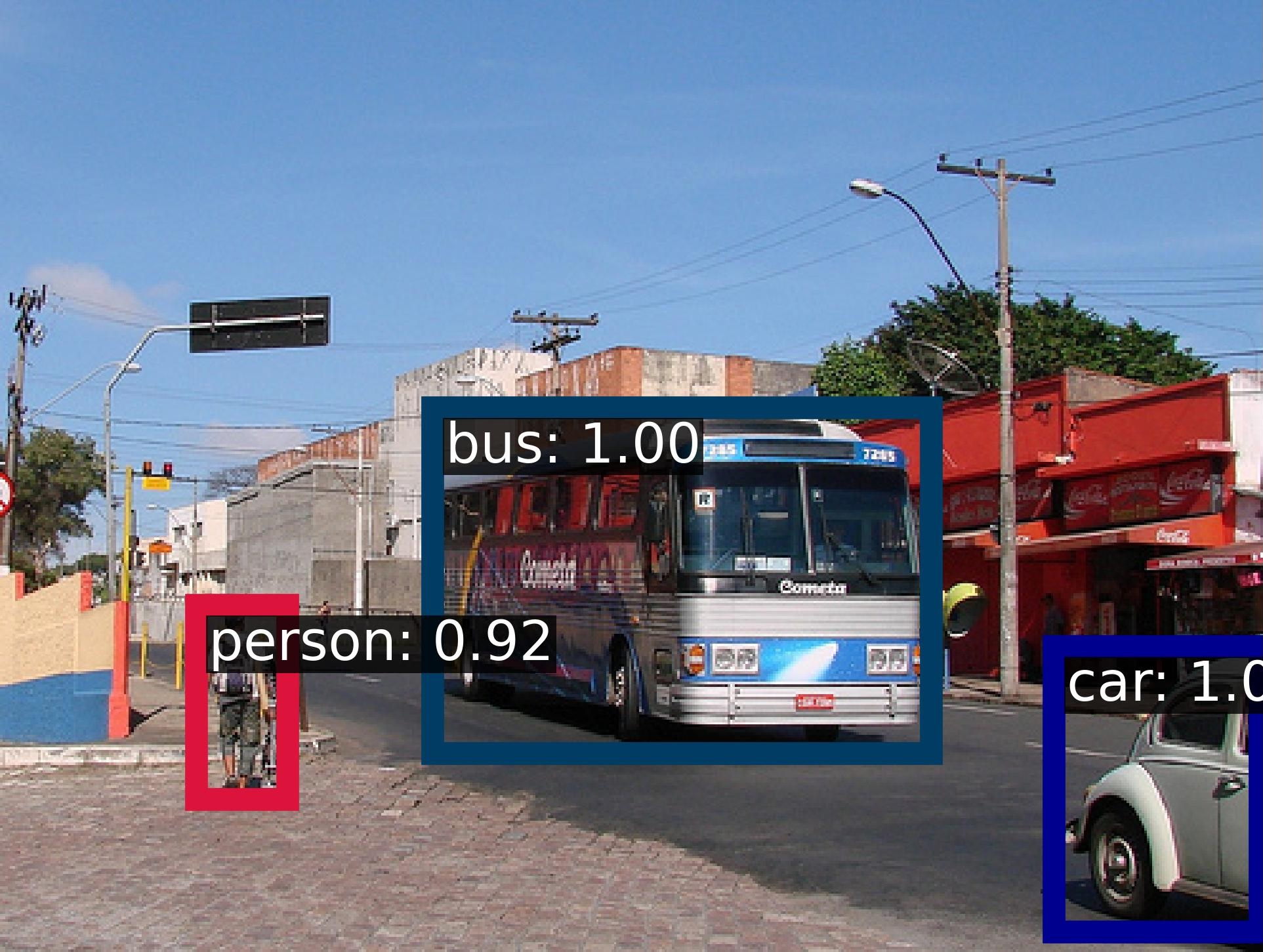} & 
				\includegraphics[width=0.15\textwidth]{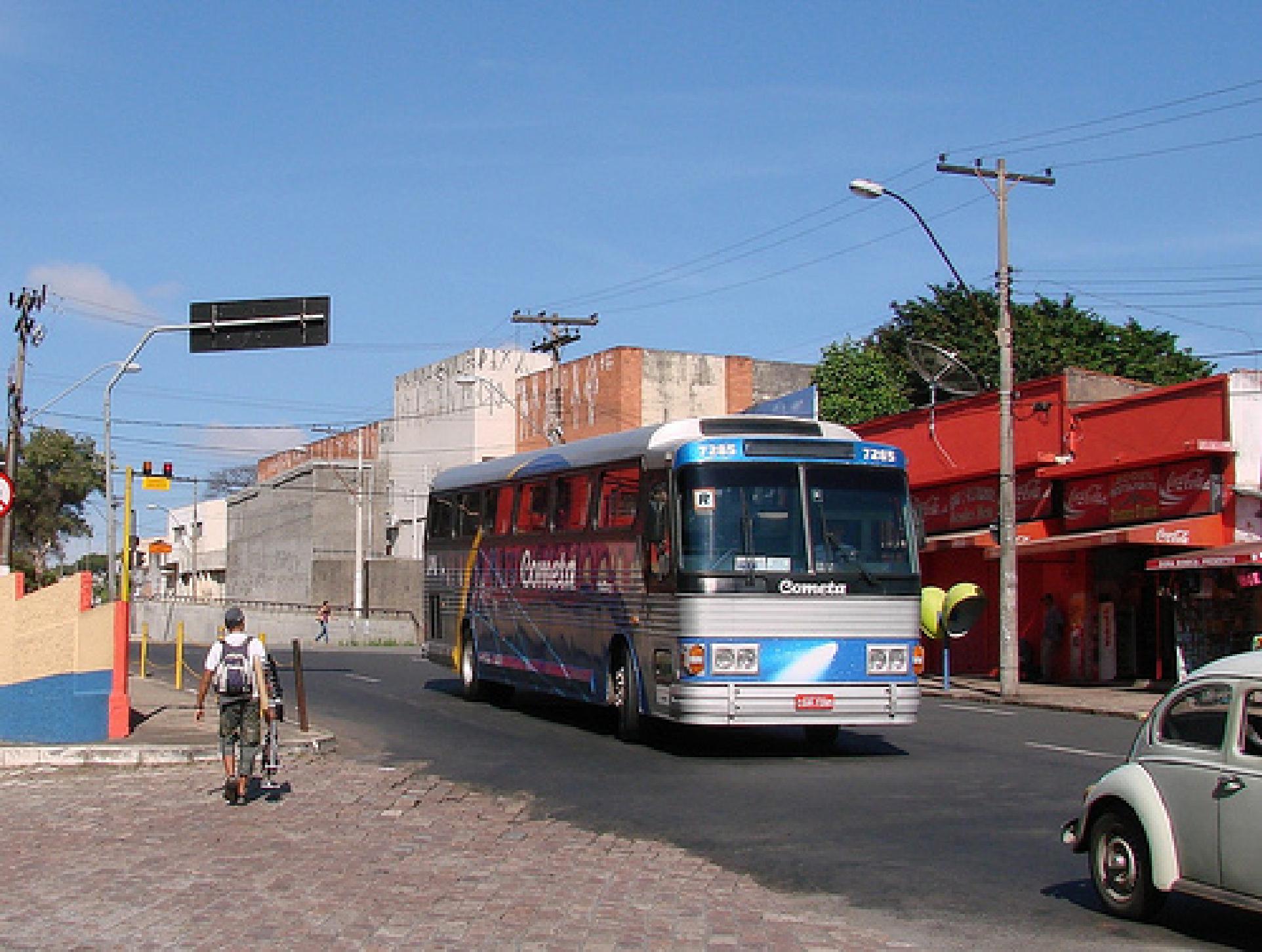} & 
				\includegraphics[width=0.15\textwidth]{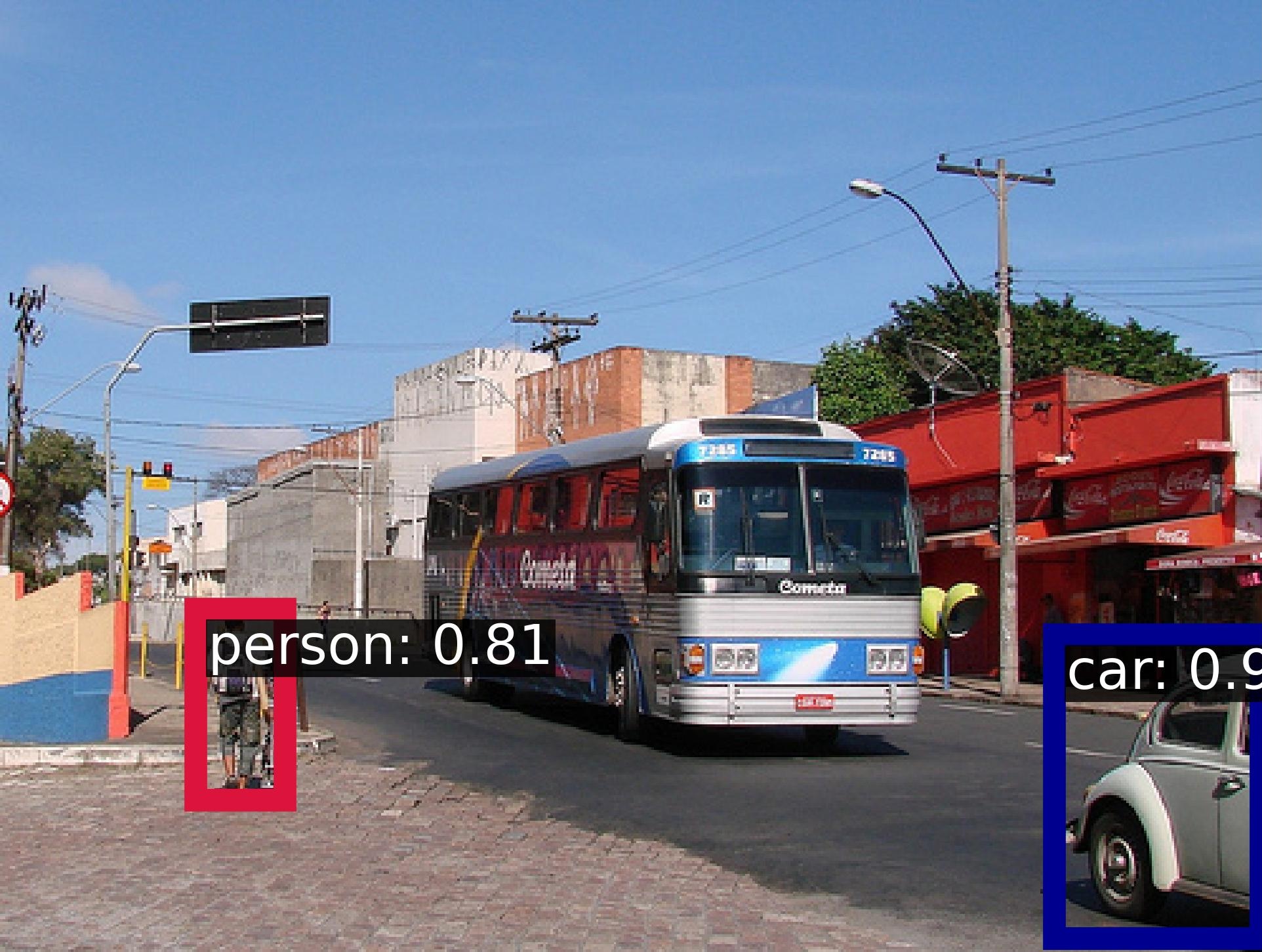} & 
				\includegraphics[width=0.15\textwidth]{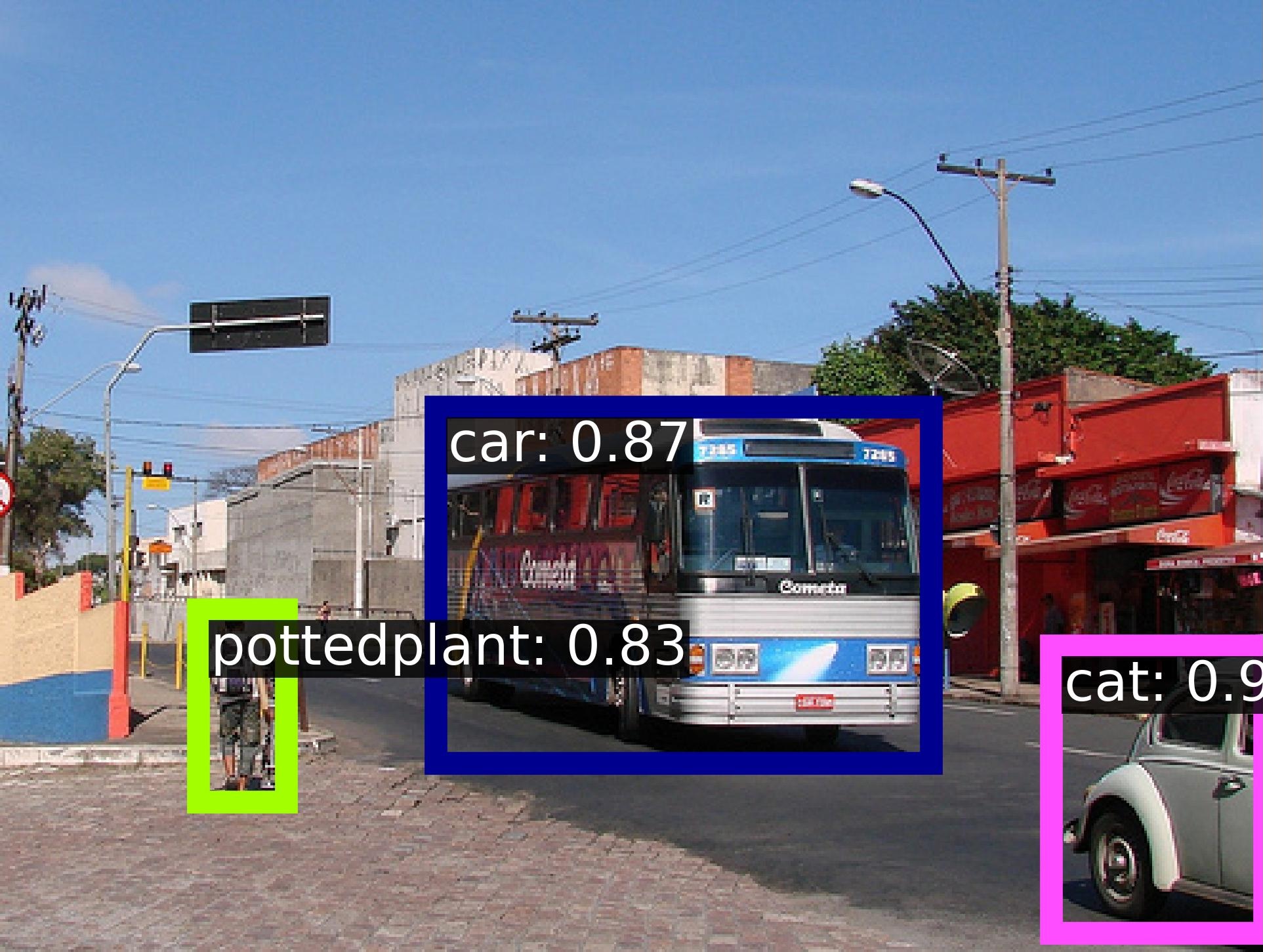} &
				\includegraphics[width=0.15\textwidth]{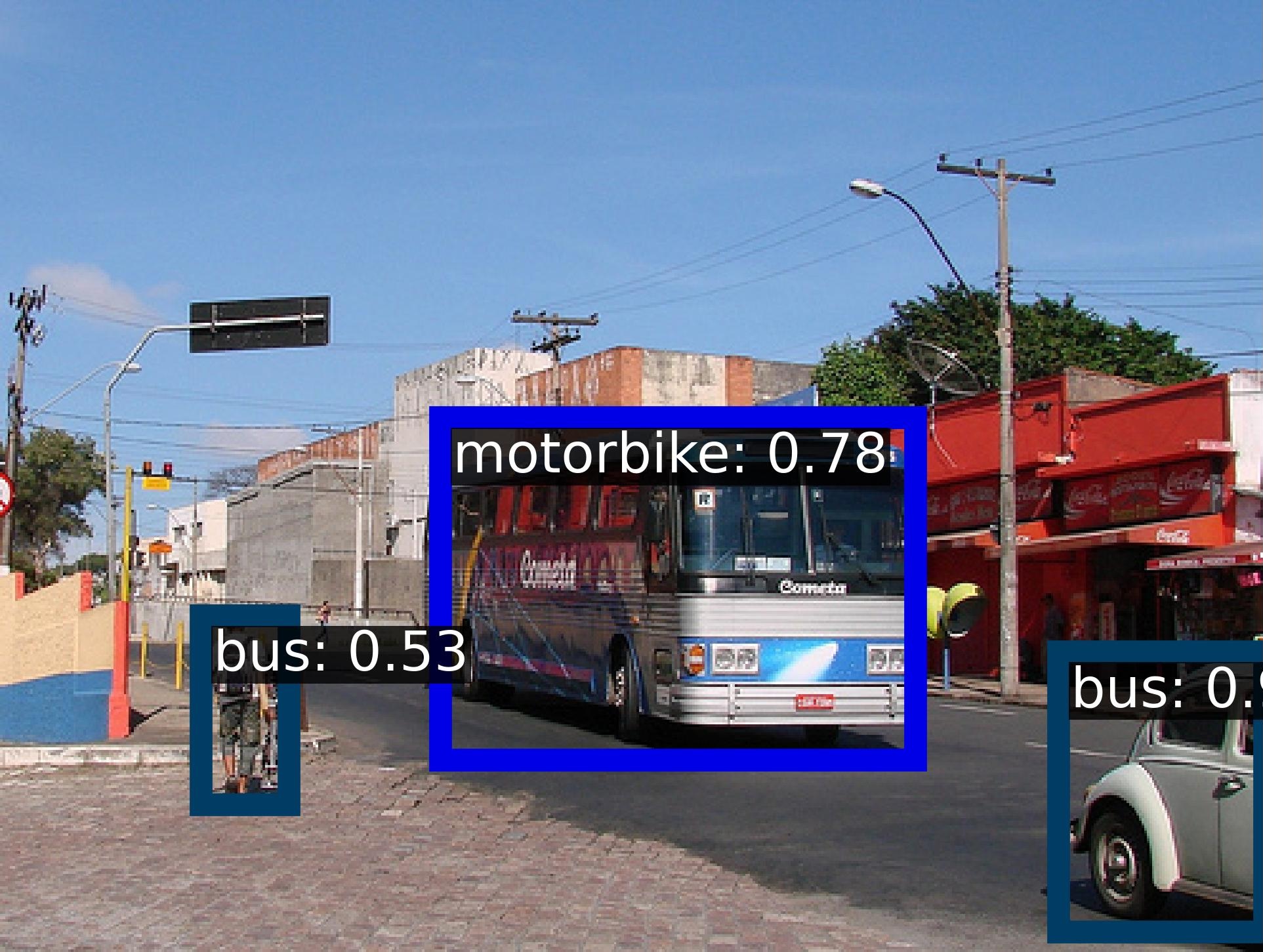} &
				\includegraphics[width=0.15\textwidth]{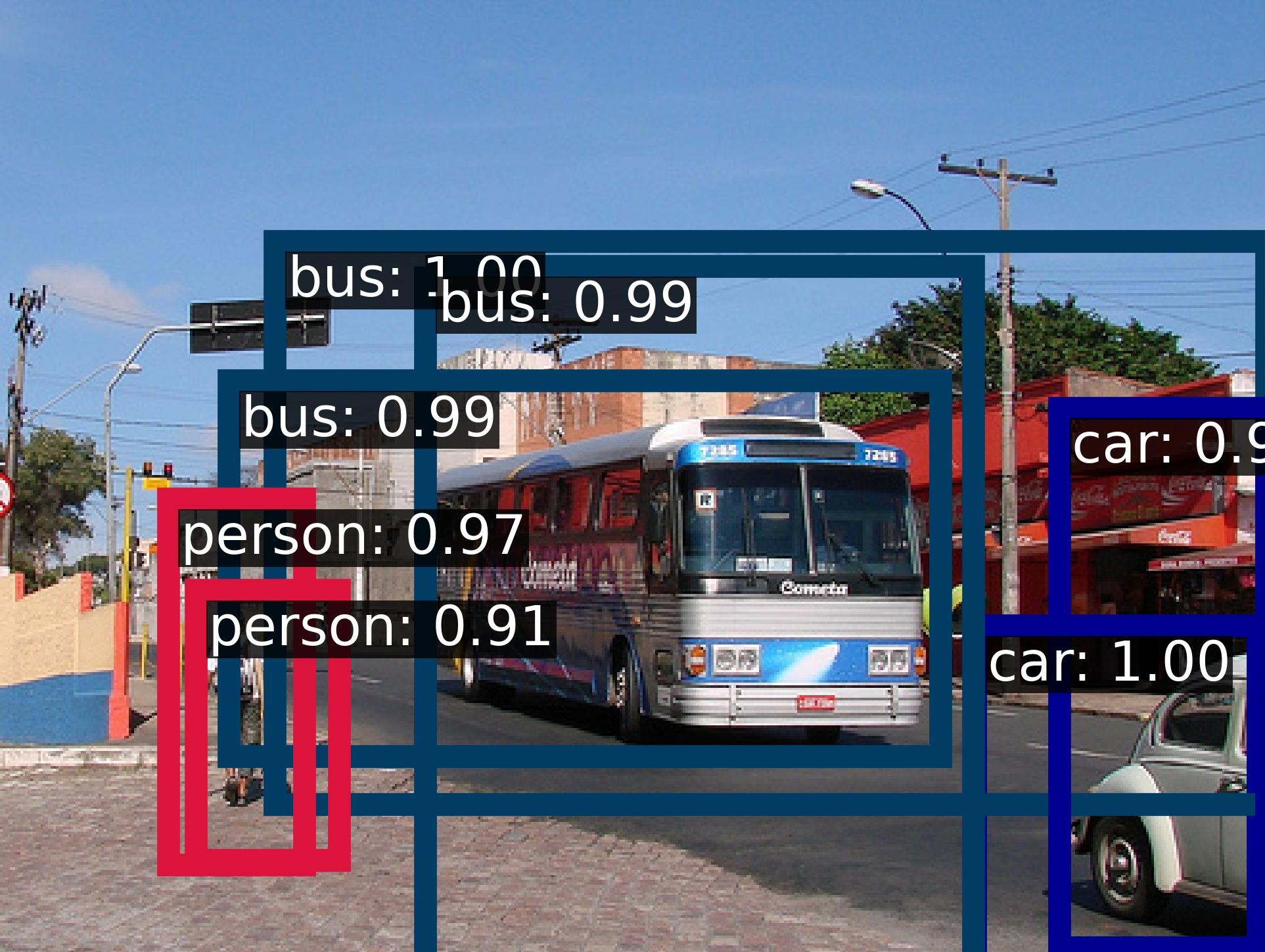} \\
				\midrule
				\includegraphics[width=0.15\textwidth]{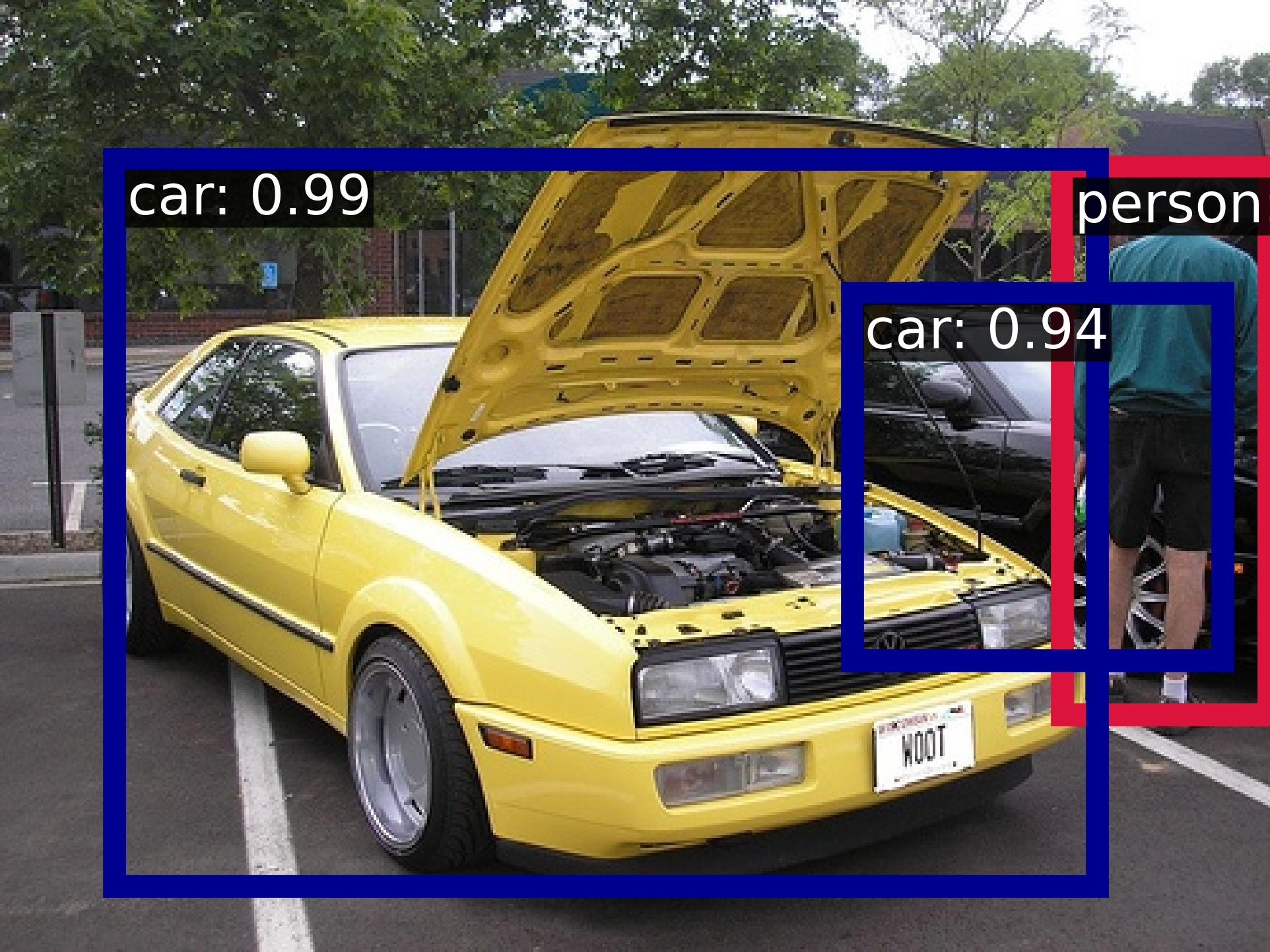} & 
				\includegraphics[width=0.15\textwidth]{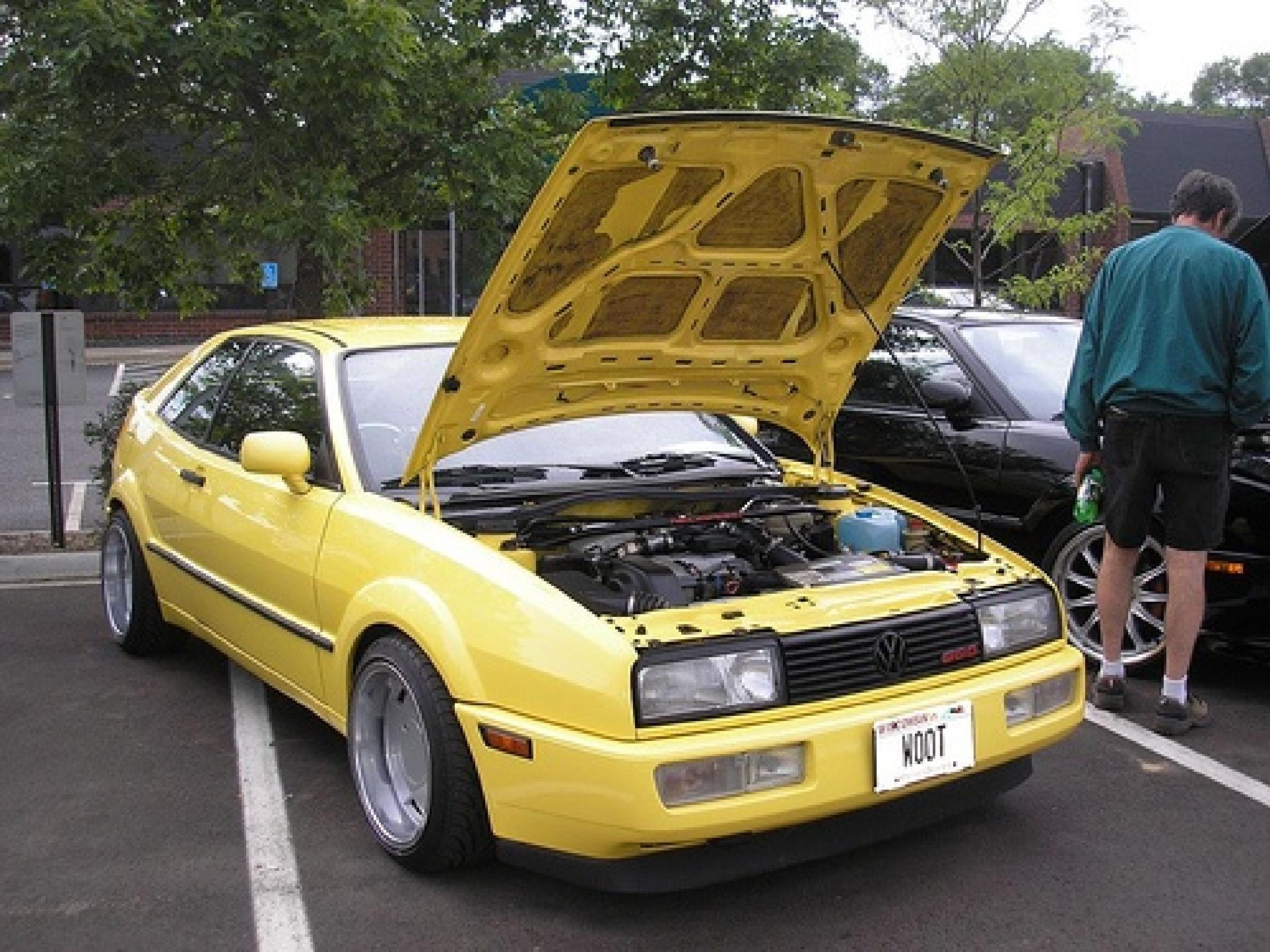} & 
				\includegraphics[width=0.15\textwidth]{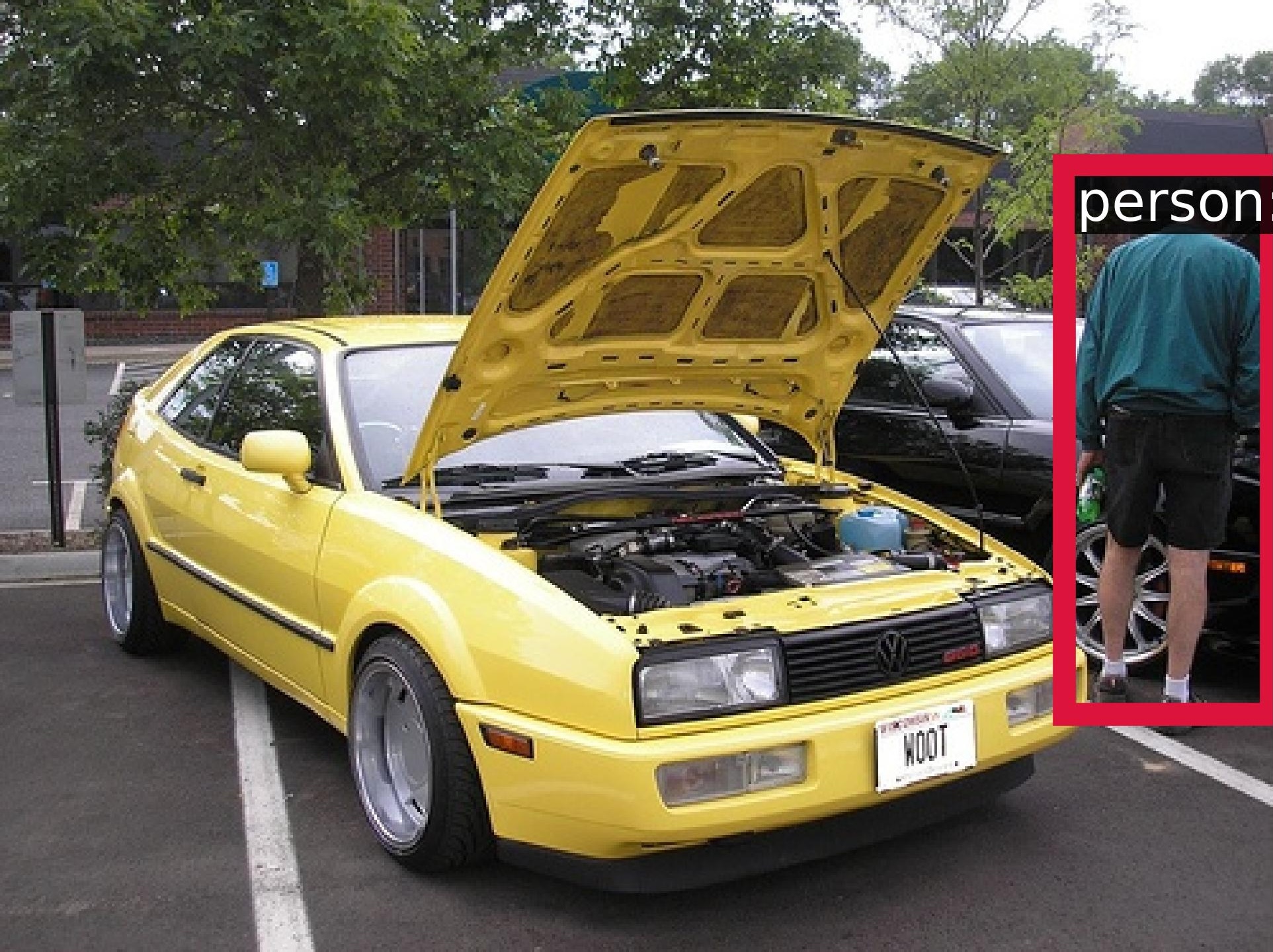} & 
				\includegraphics[width=0.15\textwidth]{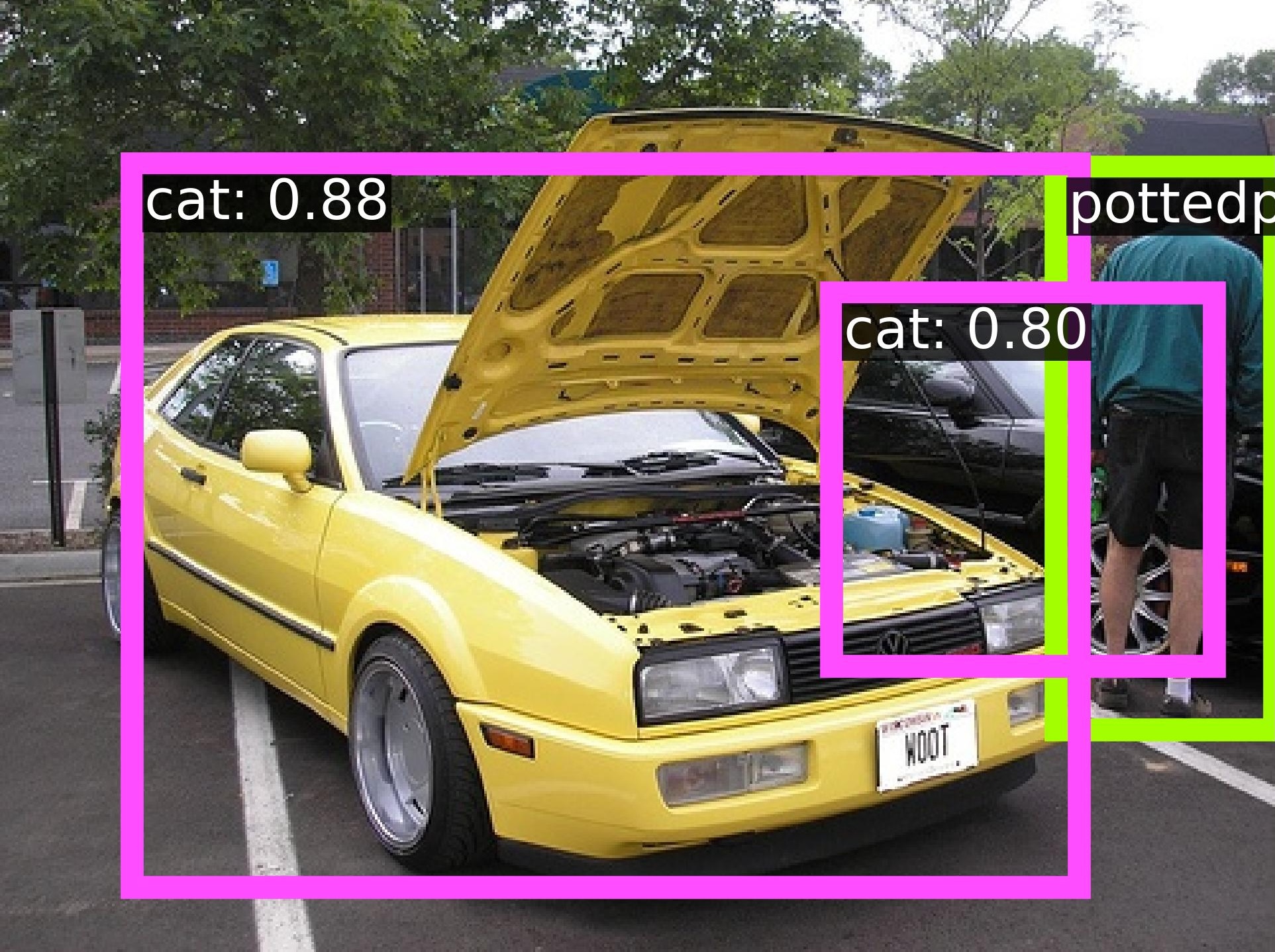} & 
				\includegraphics[width=0.15\textwidth]{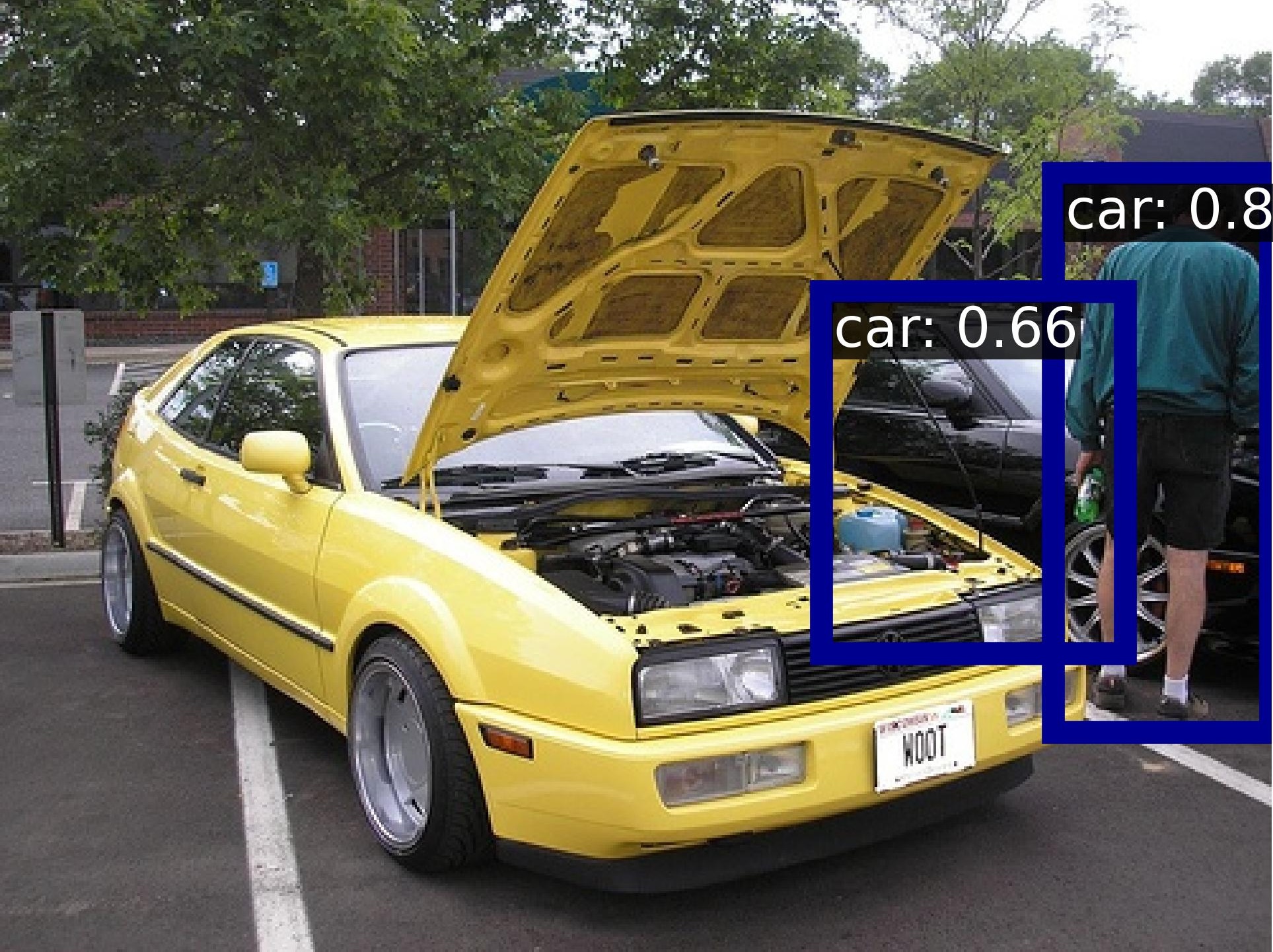} & 
				\includegraphics[width=0.15\textwidth]{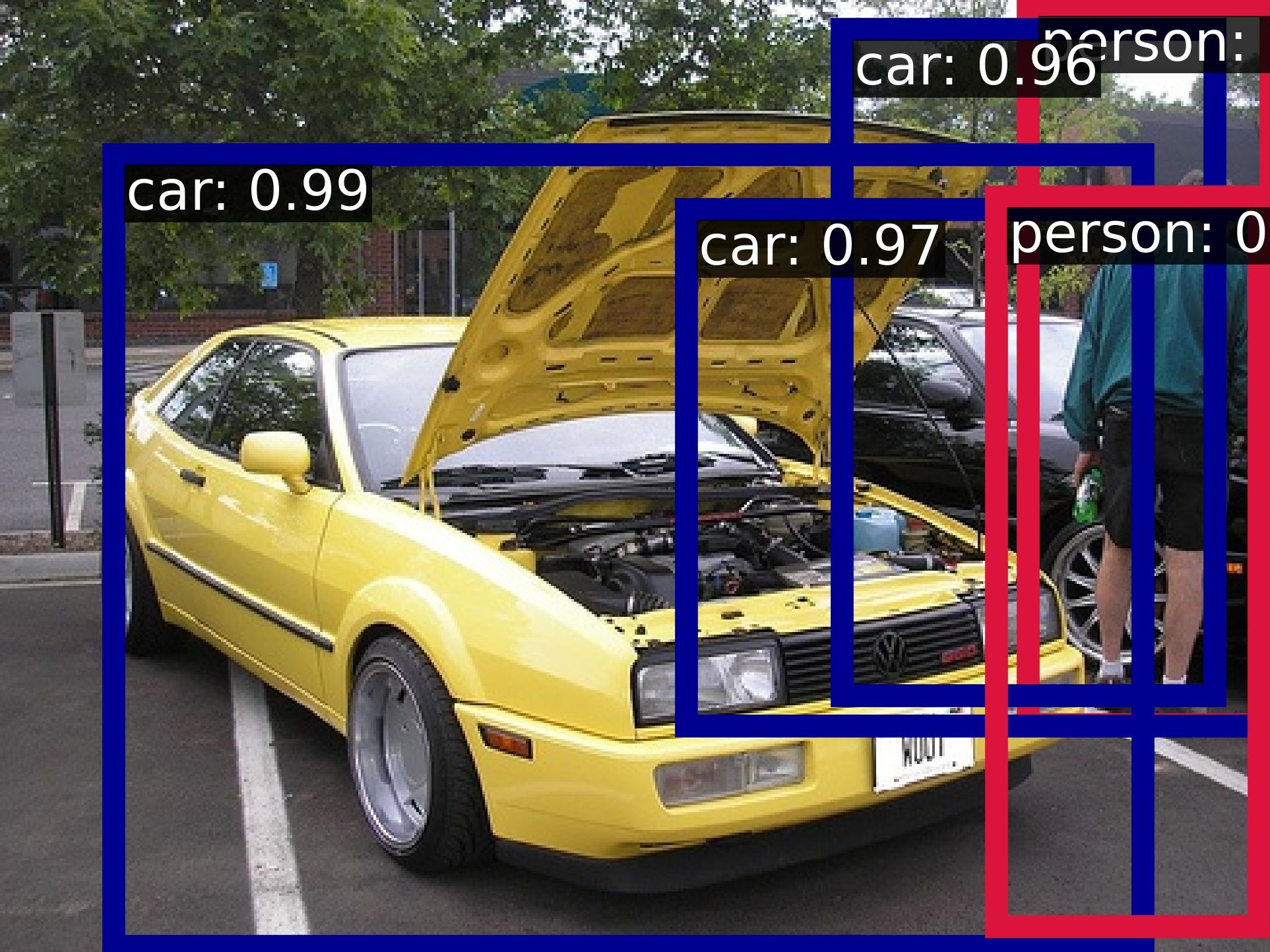} \\
				\midrule
				\includegraphics[width=0.15\textwidth]{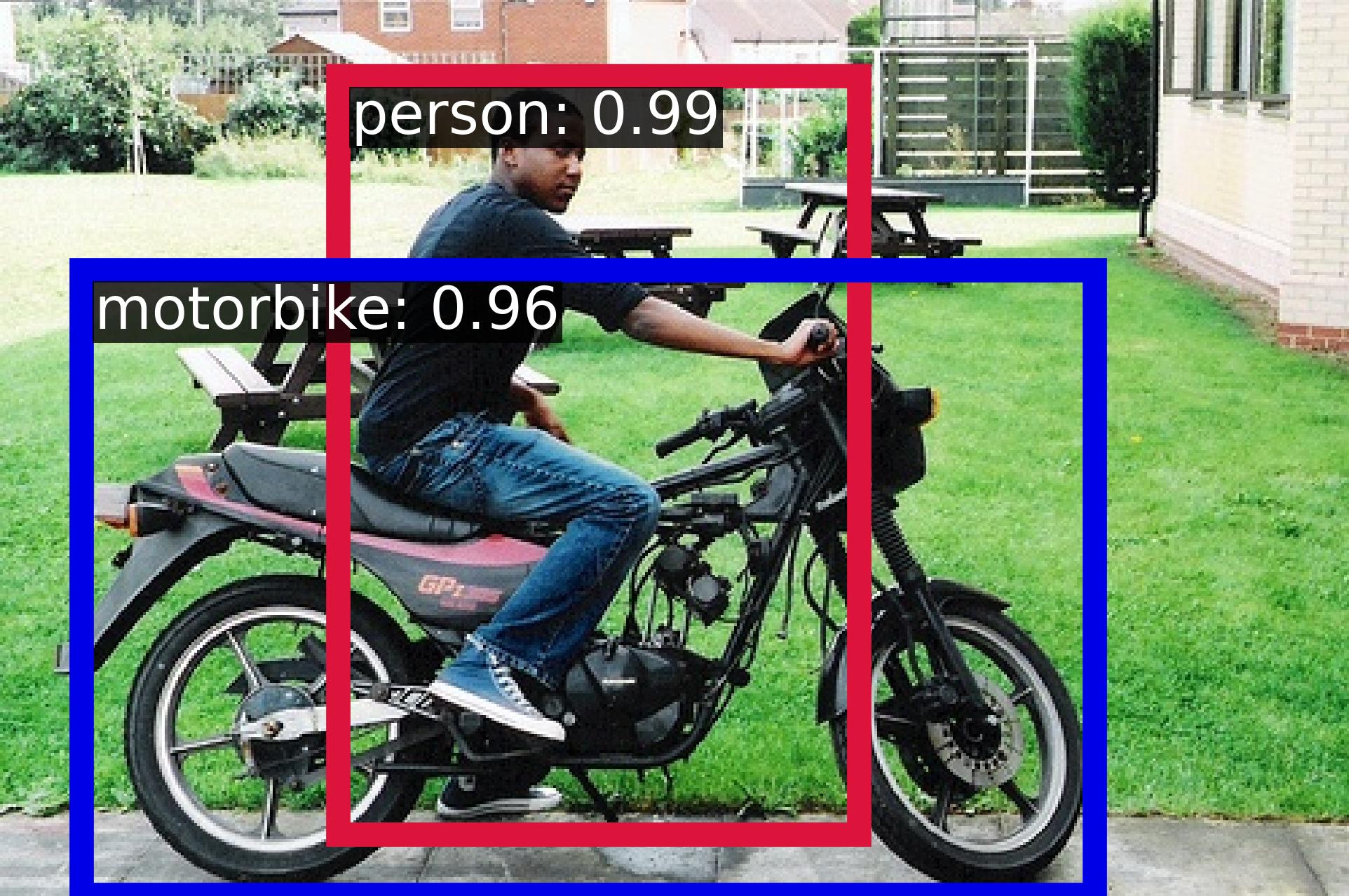} & 
				\includegraphics[width=0.15\textwidth]{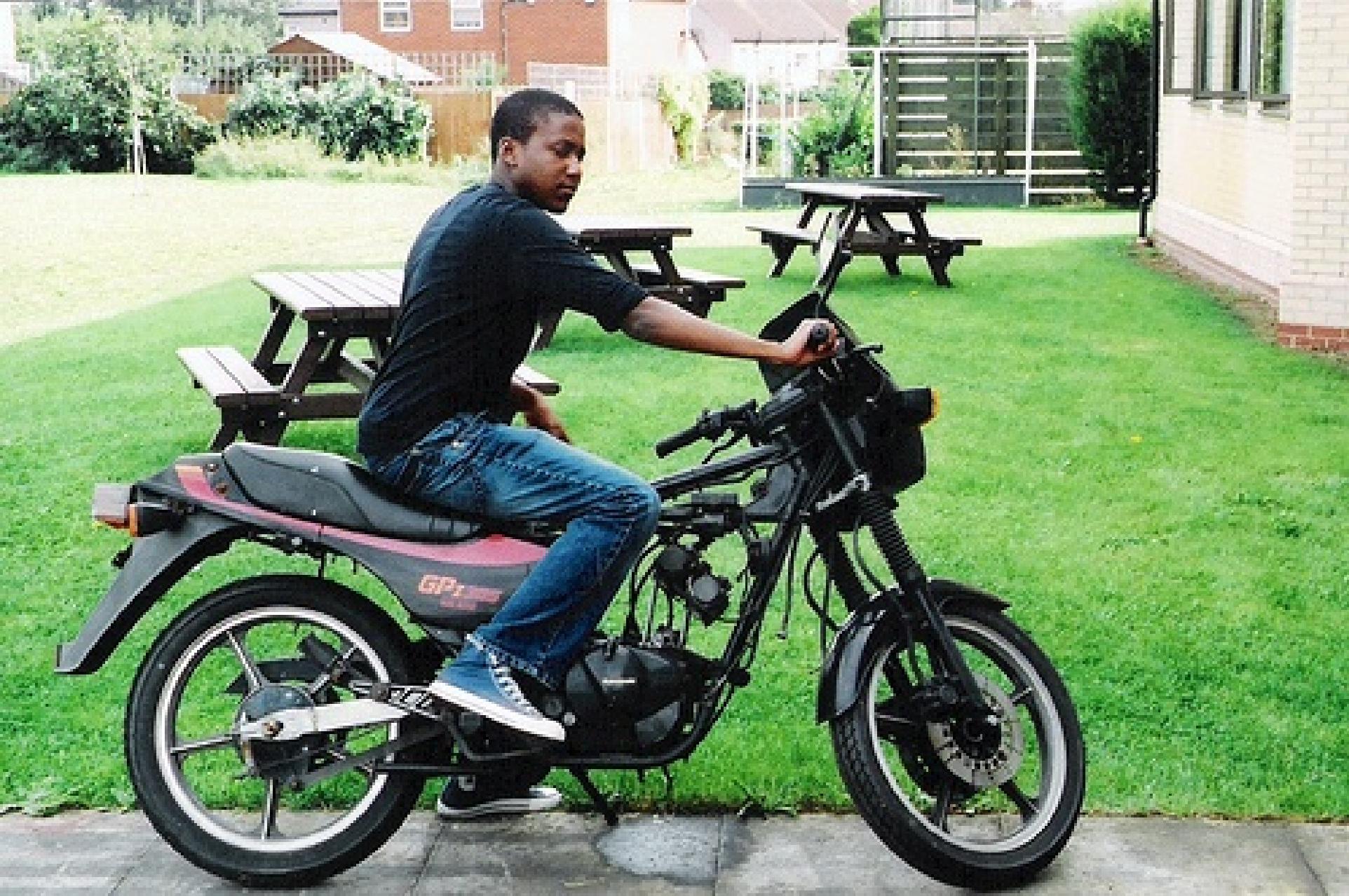} & 
				\includegraphics[width=0.15\textwidth]{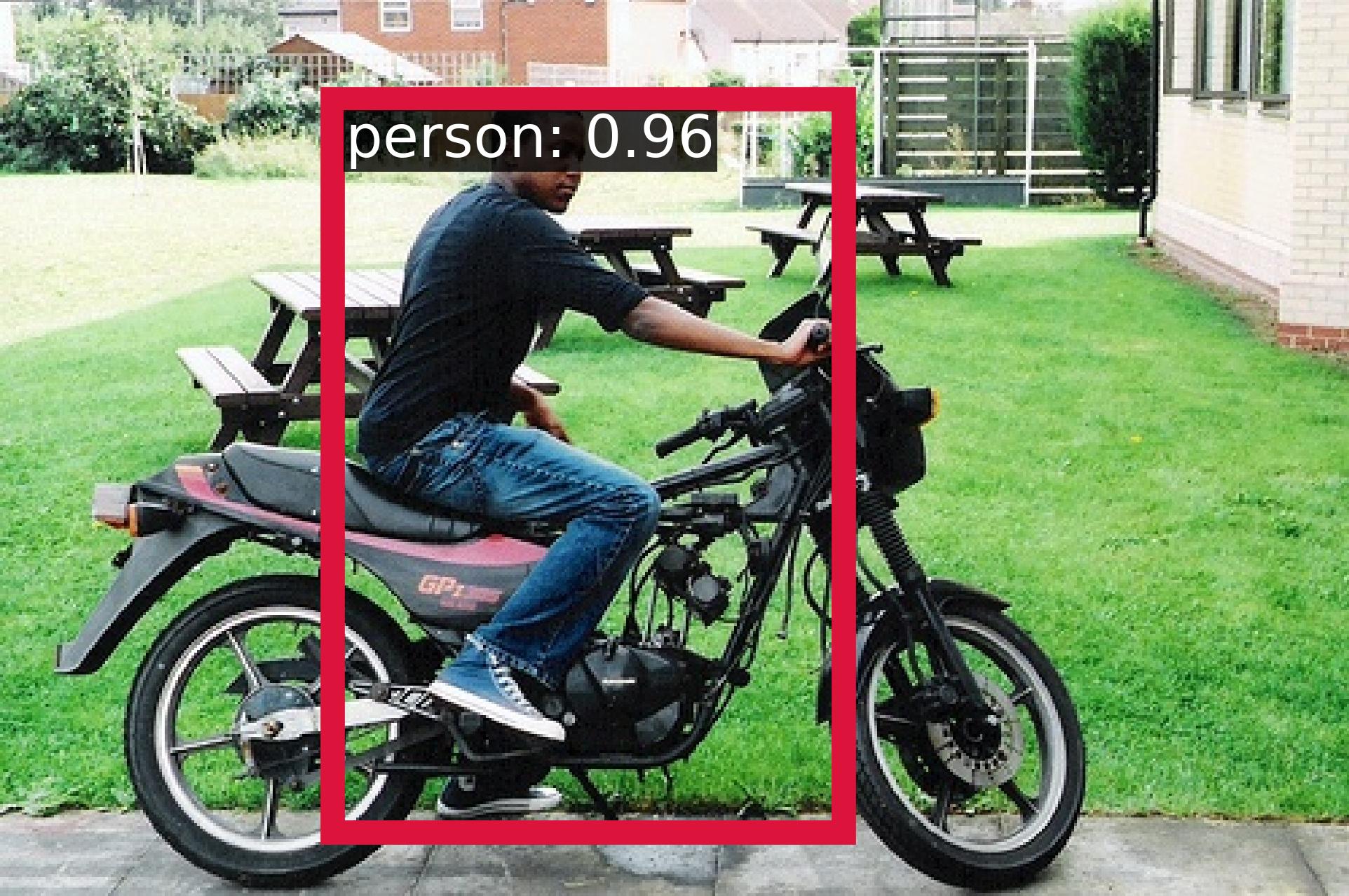} & 
				\includegraphics[width=0.15\textwidth]{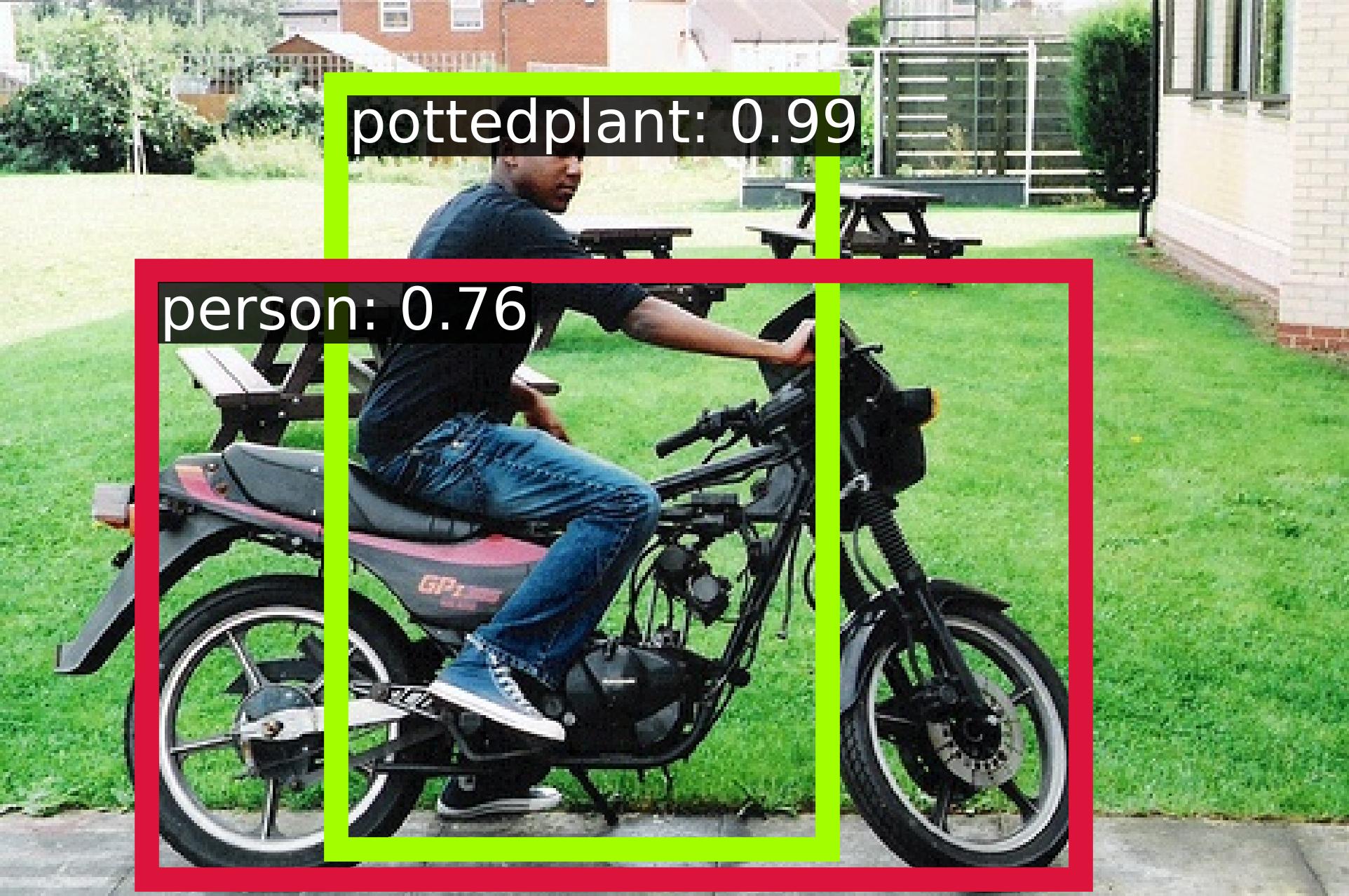} & 
				\includegraphics[width=0.15\textwidth]{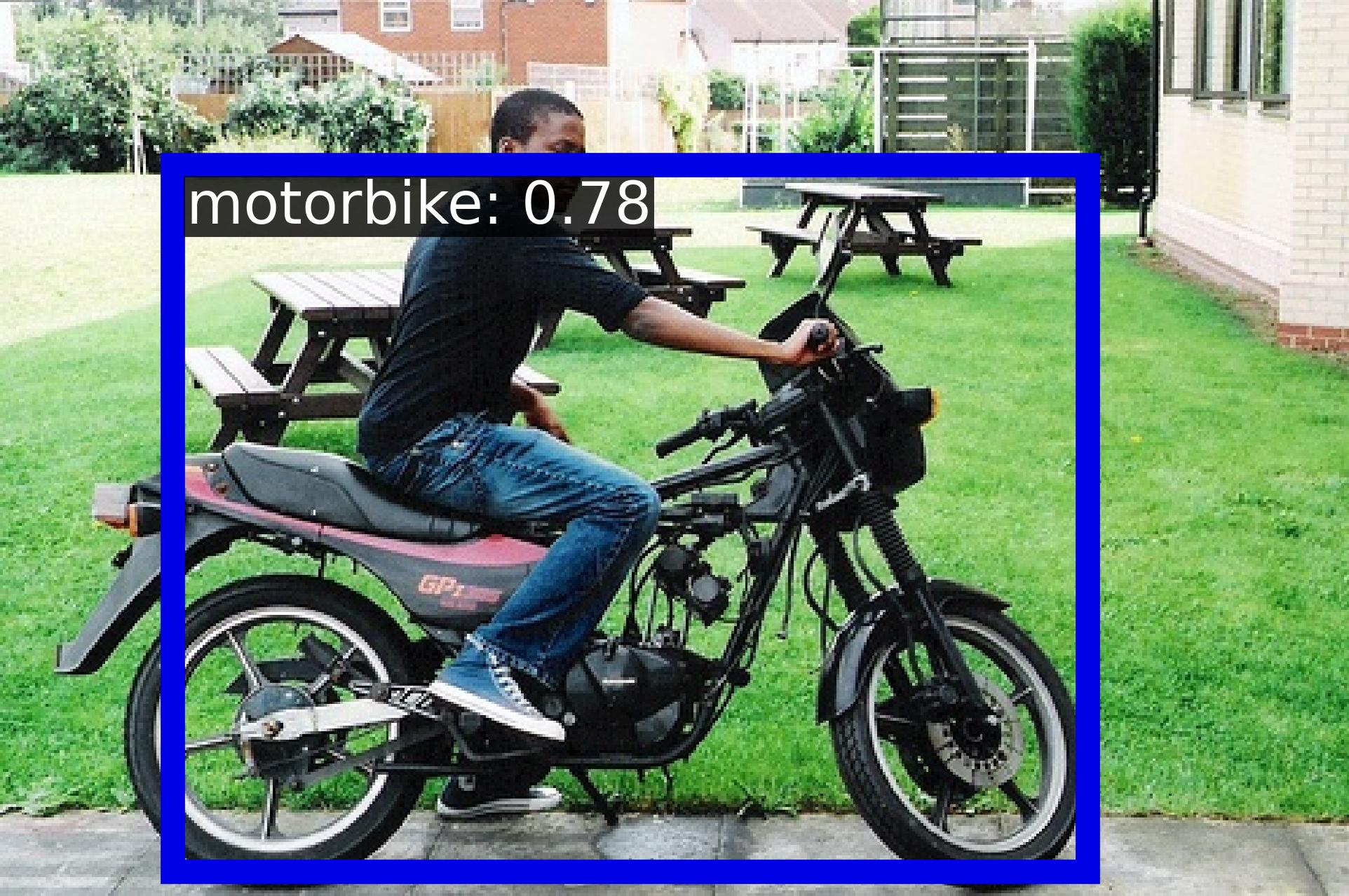} & 
				\includegraphics[width=0.15\textwidth]{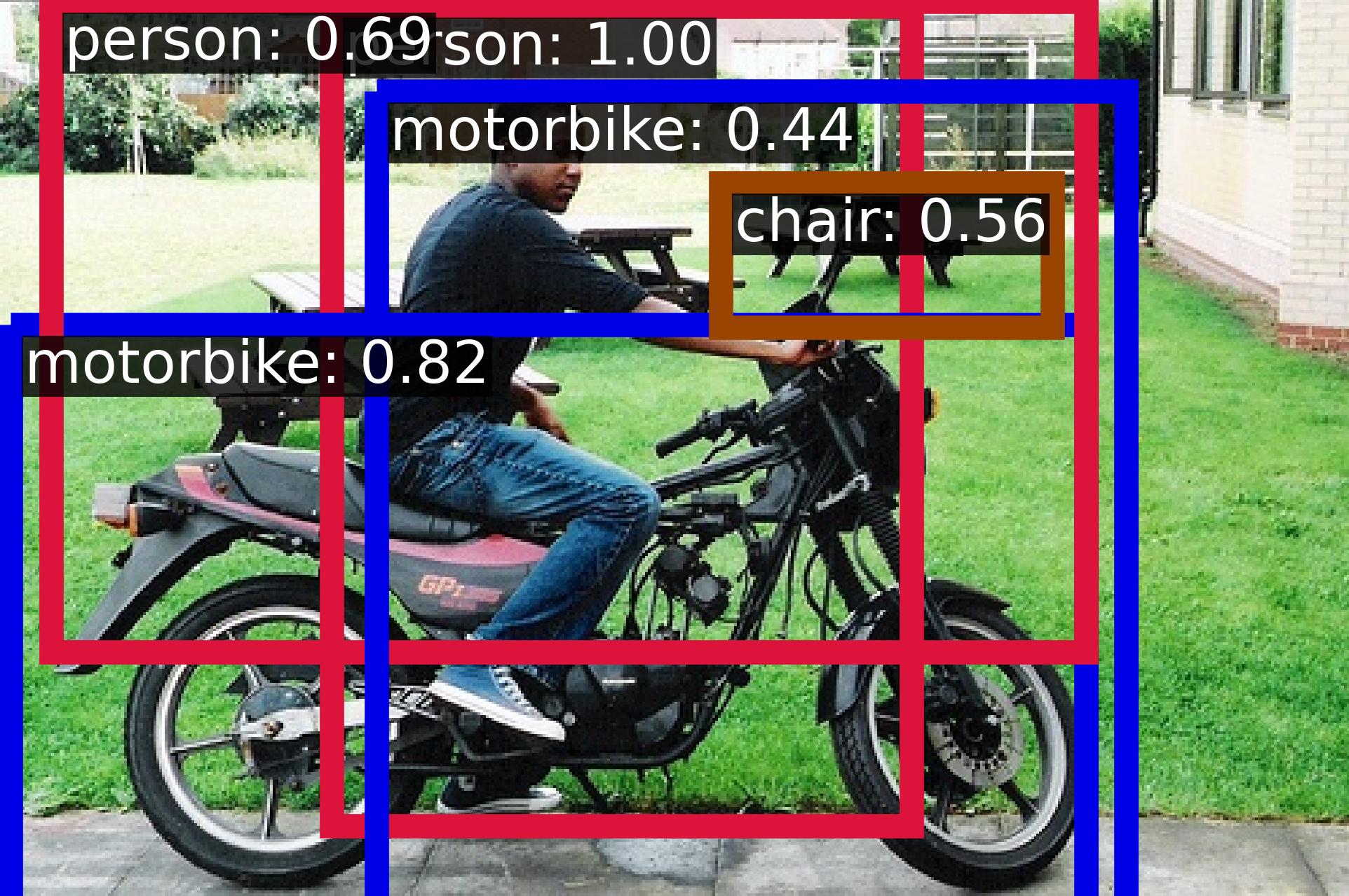} \\
				\bottomrule
			\end{tabular}
		}
		\caption{Visual samples from both PASCAL VOC07+12 and MSCOCO datasets.}
		\label{tab:visualization_samples}
	\end{table*}
	
	\section{Experimental Setup}
	\label{supp:setup}
	
	\subsection*{Datasets}
	
	The experimental configuration employs two standard object detection benchmarks: PASCAL VOC07+12 and MSCOCO. For PASCAL VOC07+12, we combine the VOC2007 \textit{train} set (2,501 images) with the VOC2012 \textit{trainval} set (11,540 images) to form the training data, while using the VOC2007 \textit{val} set (2,510 images) for testing. The MSCOCO dataset utilizes its full training set of 118k images (80 classes) for model training, with evaluation conducted on the validation set containing 5k images. For targeted misclassification attack, we select a 5-class subset consisting of \textit{person}, \textit{car}, \textit{bus}, \textit{bicycle}, and \textit{motorbike} (equivalent to \textit{motorcycle} in MSCOCO annotations) from both datasets. 
	
	\subsection*{Implementation Details}
	
	We utilize models pretrained on MSCOCO provided by the mmdetection~\cite{chen2019open}. The pretraining epochs are set as 12 for Faster R-CNN, 150 for DETR, and 273 for YOLOv3. The optimization framework employs distinct configurations for each architecture. Faster R-CNN utilizes SGD with learning rate 0.02, momentum 0.9, and weight decay 0.0001. DETR adopts AdamW optimizer ($\beta_1$ = 0.9, $\beta_2$ = 0.999) with a lower learning rate 0.0001 and equivalent weight decay regularization (0.0001). YOLOv3 implements an SGD optimizer with learning rate 0.0001 and weight decay 0.0005, combined with a cosine learning rate scheduler to accommodate its multi-scale detection mechanism. The trigger generator for all models is trained with Adam and has a learning rate of 0.1.
	
	We use input resolutions of $(1000, 600)$ with batch size 8 for VOC and $(500, 300)$ with batch size 16 for COCO, maintaining a fixed poisoning rate $p=0.5$ (4 samples per VOC batch and 8 per COCO batch). The adversarial trigger is implemented as a $3\times30\times30$ RGB patch with $\ell_\infty$-norm perturbation constraint $\epsilon=0.05$, generated by a 1-layer fully connected network optimized via Adam ($\beta_1=0.5$, $\beta_2=0.999$).

\end{document}